\documentclass[
    aps,
    prl,
    reprint,
    superscriptaddress,
    amsmath,
    amssymb
]{revtex4-2}

\usepackage{graphicx}
\usepackage{bm}
\usepackage{mathtools}
\usepackage{xcolor}

\usepackage{mathptmx}

\usepackage[
    colorlinks=true,
    linkcolor=blue,
    citecolor=blue,
    urlcolor=blue
]{hyperref}

\newcommand{\dd}{\mathrm{d}}

\begin{document}
\title{Anomalous front propagation in many-body quantum transport with memory-dependent resetting}

\author{Gabriele de Mauro}
\affiliation{LPTMS, CNRS, Universit\'e Paris-Saclay, 91405 Orsay, France}

\author{Manas Kulkarni}
\affiliation{International Centre for Theoretical Sciences, Tata Institute of Fundamental Research,
Bengaluru 560089, India}

\author{Satya N. Majumdar}
\affiliation{LPTMS, CNRS, Universit\'e Paris-Saclay, 91405 Orsay, France}

\date{\today}

\begin{abstract}
We introduce an exactly solvable framework for non-Markovian quantum many-body dynamics based on stochastic resetting with memory. Applicable to arbitrary time-independent Hamiltonians, including interacting systems, our framework provides an exact representation of the dynamics with memory in terms of the corresponding reset-free evolution, which can be accessed either analytically or numerically. We demonstrate the power of this approach by studying domain-wall melting in a one-dimensional chain of noninteracting fermions, obtaining exact analytical results for the density profile, the local structure of the front edge, the extreme-value statistics of the rightmost fermion, and the integrated current. Remarkably, memory transforms ballistic transport into logarithmically slow propagation without arresting spreading. While the macroscopic density profile retains its original scaling function, the front edge broadens as $\sqrt{\ln t}$ rather than $t^{1/3}$ at late times and develops a qualitatively different edge structure. Furthermore, the extreme-value statistics change from Tracy--Widom to Gaussian, while both the mean and variance of the integrated current grow logarithmically in time. Our predictions are in excellent agreement with numerical simulations. These results reveal how memory can fundamentally alter quantum many-body transport and its fluctuations.
\end{abstract}

\maketitle

Realistic quantum systems are inevitably coupled to their surroundings, such as thermal reservoirs, external fields, or measurement apparatuses, giving rise to nonunitary dynamics characterized by decoherence, dissipation, and relaxation. These interactions can profoundly affect quantum many-body dynamics, as illustrated by measurement-induced phase transitions, where the entanglement structure changes qualitatively with the measurement rate~\cite{LiChenFisher2018,SkinnerRuhmanNahum2019}. While open quantum dynamics is often described within the Markovian approximation~\cite{GKSL1976,Lindblad1976,BreuerPetruccione2006,Carmichael2002}, memory effects become crucial when the separation between system and environment timescales breaks down~\cite{Breuer2016,deVegaAlonso2017,Xu2026,MilzModi2021,Ciccarello2022,Keeling2026}.

A standard approach consists of deriving an effective equation for the reduced system dynamics by tracing out the environmental degrees of freedom. Several techniques achieve this, including non-Markovian master equations~\cite{Nakajima1958,Zwanzig1960}, Feynman--Vernon influence-functional methods~\cite{FeynmanVernon1963}, stochastic Schr\"odinger equations~\cite{Diosi1998}, and hierarchical equations of motion~\cite{TanimuraKubo1989}. 
In particular, different frameworks based on classical stochastic processes and their quantum generalizations have been developed to construct non-Markovian quantum master equations that guarantee completely positive and trace-preserving dynamics~\cite{Budini2004,Vacchini2016,Chenu2017}, i.e., a dynamics that preserves the positivity and normalization of the density matrix.
More general mathematical constructions have also established sufficient conditions for completely positive dynamics governed by memory-kernel master equations, including time-inhomogeneous formulations~\cite{Wudarski2015,ChruscinskiKossakowski2016,Chruscinski2022}. Although exact solutions exist for some classes of non-Markovian quantum dynamics~\cite{Goychuk2004,BudiniGarrahan2021}, obtaining explicit analytical results for nontrivial observables, in the presence of memory, remains a formidable challenge, particularly in quantum many-body systems~\cite{deVegaAlonso2017,Xu2026,Keeling2026}.

In this Letter, we pursue a different approach based on a stochastic protocol that explicitly couples the quantum dynamics to its own past history. Rather than modeling the microscopic structure of the environment, we describe its influence through an externally imposed stochastic protocol with memory, which can be interpreted as a simplified effective environment. While this approach sacrifices the microscopic details of the environment, it offers three major advantages. First, despite its non-Markovian nature, the resulting dynamics admits a simple exact solution at all times for any time-independent many-body Hamiltonian, including interacting systems. Remarkably, this representation enables the analytical investigation of nontrivial many-body observables, revealing profound memory-induced effects and qualitatively new dynamical behavior. Second, the framework naturally allows for extensions to a broad class of memory kernels, potentially capturing different environmental effects. Finally, recent advances in the experimental control of quantum many-body systems suggest possible routes toward implementing the protocol in practice. Together, these features provide a versatile and analytically tractable framework for investigating the effects of memory on quantum many-body dynamics.

Our protocol interrupts the unitary evolution of a many-body system at random Poissonian times, resetting the entire state to one visited at a uniformly chosen time in its past history. It is inspired by the preferential relocation model of classical random walks~\cite{BSS2014,BRC2014,BP2016,BEM2017,FCBGM2017,MUB2019,BFCGM2019,MCM2019,CM2019,VCMLNA2022,BociMailler2023,BM2024Active,BM2024,BociMailler2024,Sepehrinia2025,Sepehrinia2026,BEM2025,BM2026,ArmfieldMailler2026,BM2026a}, in which a walker occasionally returns to a previously visited position. Originally motivated by animal foraging, this model reproduces features of free-ranging capuchin monkey movement patterns~\cite{BSS2014} and exhibits unusual dynamics, including ultraslow spreading~\cite{BSS2014,BEM2017}. Extensions to multiple classical Brownian particles~\cite{BM2026a} and single quantum particles~\cite{dMKM2026} have also been investigated. Extending the protocol to quantum many-body systems, however, requires a description in terms of many-body wave functions even for noninteracting particles, introducing additional complexity and, as we show, qualitatively new collective phenomena. Although reminiscent of conventional quantum stochastic resetting~\cite{EM2011,EM2011JPA,EMS2020,Perfetto2021,MSM2018,RTLG2018}, our protocol is fundamentally different. Conventional resetting repeatedly returns the system to a fixed state, typically leading to a stationary state and arresting the spreading. In sharp contrast, our protocol resets the system to states selected from its entire past evolution, allowing it to spread indefinitely. The resulting dynamics reflects a competition between unitary evolution, which drives spreading, and memory-based resets, which counteract it. Ultimately, unitary spreading prevails, but at the cost of a dramatic slowdown.

Recent experimental advances provide further motivation for this approach. Rydberg-atom arrays~\cite{Bernien2017,BL2020,BCYL2026} and programmable quantum processors~\cite{KRMZG2023,Kim2023,Haghshenas2026,Hartnett2026,Fischer2026} offer increasingly precise control over many-body quantum dynamics. In particular, our protocol could be implemented through controlled forward and backward unitary evolution, allowing the system to return to previously visited states. The possibility of reversing quantum evolution dates back to the pioneering spin-echo experiments, in which the effects of spin dephasing can be reversed~\cite{Hahn1950,Carr1954}. More recently, time reversal has been demonstrated in interacting quantum many-body systems~\cite{Colombo2022,Linnemann2016,Widera2008,Garttner2017,Wei2018,Gilmore2021,Rovny2018,Mi2022,Braumuller2022,Li2023,Geier2024}. In particular, reversing the sign of the Hamiltonian, $\hat{\mathcal H}\to-\hat{\mathcal H}$, reverses the unitary evolution, $e^{-i\hat{\mathcal H}t}\to e^{i\hat{\mathcal H}t}$ (setting $\hbar=1$), and experiments have demonstrated the recovery of magnetization in interacting spin systems even after substantial relaxation~\cite{Geier2024}.  In our protocol, the clock is paused during this operation and resumes once the selected past state has been recovered.

Our protocol can be applied, in principle, to any quantum many-body Hamiltonian $\hat{\mathcal H}$. To illustrate the resulting nontrivial non-Markovian dynamics, in this Letter we will focus on a specific many-body model of noninteracting spinless fermions on an infinite one-dimensional lattice, initially prepared in a domain-wall state with one half-chain fully occupied and the other half empty. Through a Jordan--Wigner transformation, this model maps onto the one-dimensional XX spin chain, making it experimentally accessible. The pure unitary dynamics of this model already exhibits nontrivial transport and fluctuation properties, and it has been studied extensively~\cite{ARRS1999,Hunyadi2004,AKR2008,Eisler2013}. We will show that resetting with memory affects these properties drastically. 
\begin{figure}[ht]
    \centering
    \includegraphics[width=0.45\textwidth]{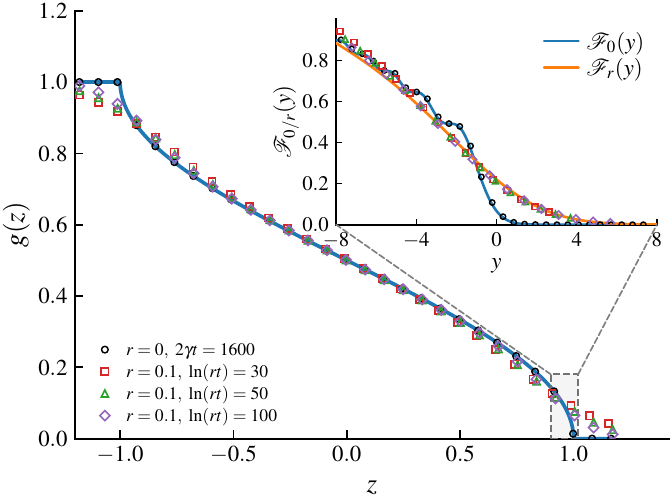}
    
\caption{
Macroscopic density profile and local density at the edge of the propagating front.
The solid line shows the analytical scaling function
$g(z)$ given in Eq.~\eqref{eq:averagedensitynoreset}, while symbols represent
numerical results for the reset-free case ($r=0$)
and memory-dependent resetting ($r=0.1$) at different
times, as indicated in the legend.
The scaling variable is $z=x/(2\gamma t)$ for $r=0$
and $z=x/\Lambda_r(t)$ for $r>0$, with
$\Lambda_r(t)=2\gamma\ln(rt)/r$.
The inset shows the rescaled local density at the front edge,
compared with the analytical scaling functions
$\mathcal F_0(y)$ and $\mathcal F_r(y)$ appearing
in Eqs.~\eqref{eq:edge_density} and \eqref{eq:memory_edge_final}, respectively. Details of the numerical simulations are provided
in the SM~\cite{SM}.
}
    \label{fig:front}
\end{figure}

Applying our framework to a noninteracting fermionic chain reveals rich many-body dynamics. We obtain exact analytical results for the density profile, the local structure of the front edge, the extreme-value statistics (EVS) of the rightmost fermion, and the integrated current. Memory transforms ballistic transport into logarithmically slow spreading, but leaves the macroscopic density scaling function unchanged. However, the local structure of the front edge and collective fluctuations change qualitatively: the front edge broadens as $\sqrt{\ln t}$ rather than $t^{1/3}$ at late times, the EVS become Gaussian instead of Tracy--Widom, and the mean integrated current grows logarithmically rather than linearly in time.

\paragraph{The model.}
We consider $N$ noninteracting spinless fermions on an infinite one-dimensional lattice, where each site $|x\rangle$ can accommodate at most one fermion due to the Pauli exclusion principle. The system is described by the many-body Hamiltonian $\hat{\mathcal H}=\sum_{m=1}^N\hat H_m$, where $\hat H_m$ acts on the $m$-th particle through the single-particle tight-binding Hamiltonian
\begin{equation}\label{eq:singlepartHamilt}
    \hat H=-\gamma\sum_{x\in\mathbb Z}\left(|x\rangle\langle x+1|+|x+1\rangle\langle x|\right).
\end{equation}
We consider a domain-wall initial state $|\Psi_N(0)\rangle$, constructed as a Slater determinant of the occupied orbitals $|\phi_n(0)\rangle=|-(n-1)\rangle$, with $n=1,\ldots,N$ (see Eq.~\eqref{eq:domainwallinitialstate_EM}). For $N\to\infty$, all sites $x\leq0$ are occupied, while those with $x>0$ are empty. Without resetting, the density matrix evolves unitarily as
\begin{equation}\label{eq:eqforrho0}
    \hat\rho_0^{(N)}(t)=e^{-i\hat{\mathcal H}t}\hat\rho^{(N)}(0)e^{i\hat{\mathcal H}t},
\end{equation}
where the initial condition $\hat\rho^{(N)}(0)$ reads $\hat\rho^{(N)}(0)=|\Psi_N(0)\rangle\langle\Psi_N(0)|$. This domain-wall problem has been extensively studied using second quantization in Refs.~\cite{ARRS1999,Hunyadi2004,AKR2008,Eisler2013}. For completeness and to establish the notation used throughout this work, we briefly revisit the reset-free dynamics using first quantization in the Supplementary Material (SM)~\cite{SM}.

We now introduce our protocol, which combines deterministic unitary evolution with a classical stochastic resetting process. Under this protocol, a state $|\psi(t)\rangle$ at time $t$ evolves over a small time interval $dt$ as
\begin{equation}\label{eq:timeevolpsi}
|\psi(t+dt)\rangle=
\    \begin{cases}
        |\psi(t)\rangle-i\hat{\mathcal H}|\psi(t)\rangle dt, & \text{with prob.} \; 1-rdt,\\
        |\psi(\tau)\rangle, & \text{with prob.} \; rdt,
    \end{cases}
\end{equation}
where $\tau\sim U[0,t]$ is uniformly distributed over $[0,t]$, and $r$ is the resetting rate. Here, $\hat{\mathcal H}$ is an arbitrary time-independent Hamiltonian. The first line in Eq.~\eqref{eq:timeevolpsi} refers to unitary evolution, while the second line refers to resetting to a previously occupied state. This is exactly equivalent to evolving a state from time $t'$ unitarily up to a random time $t'+u$, where $u$ is chosen from the exponential distribution $p(u)=re^{-ru}$, followed by an instantaneous resetting to a past state $|\psi(\tau)\rangle$, where $\tau\sim U[0, t'+u]$. 
Although the resetting events occur according to a Poisson process with constant rate $r$, the resulting quantum dynamics is non-Markovian: the state selected at each reset depends on the entire past trajectory, rather than on the current state alone. A more detailed description of this dynamics is provided in the SM~\cite{SM}.

Denoting by $\hat\rho_r^{(N)}(t)$ the density matrix averaged over all realizations of the resetting process, its evolution is governed by
\begin{equation}\label{eq:memory_masterequation}
    \frac{\dd \hat\rho_r^{(N)}(t)}{\dd t}
    =
    -i[\hat{\mathcal H},\hat\rho_r^{(N)}(t)]
    -r\hat\rho_r^{(N)}(t)
    +\frac{r}{t}\int_0^t\dd\tau\,\hat\rho_r^{(N)}(\tau),
\end{equation}
as shown in the Appendix~\ref{app:master_effective_time}. 
Equation~\eqref{eq:memory_masterequation} was previously derived for a single particle in Ref.~\cite{dMKM2026}. In this Letter, we extend this result to arbitrary $N$-particle systems and derive a new exact solution valid for any time-independent Hamiltonian, including interacting and noninteracting systems. This represents a nontrivial generalization of the single-particle result and provides a clearer physical interpretation and a convenient framework for calculating many-body observables. Furthermore, we show that, in the noninteracting case, tracing out $N-1$ particles recovers the single-particle equation of Ref.~\cite{dMKM2026} (see the Appendix~\ref{app:master_effective_time} and the SM~\cite{SM}).

Since each reset returns the system to a previously visited state, the dynamics remains on the same unitary (reset-free) orbit and can therefore be parametrized by a stochastic effective time $q(t)$. As shown in the Appendix~\ref{app:master_effective_time}, this allows us to express the solution of Eq.~\eqref{eq:memory_masterequation} exactly for any time-independent many-body Hamiltonian $\hat{\mathcal H}$, including interacting systems, in the form
\begin{equation}\label{eq:formasintegraloverq}
    \hat\rho_r^{(N)}(t)=\int_0^t\mathrm{d}q\,P(q,t)\,\hat\rho_0^{(N)}(q).
\end{equation}
Here, $\hat\rho_0^{(N)}(q)$ is the reset-free density matrix defined in Eq.~\eqref{eq:eqforrho0}, while $P(q,t)$ is the probability distribution of the effective time $q(t)$. This representation separates the quantum dynamics, encoded in $\hat\rho_0^{(N)}(q)$, from the classical stochastic protocol, described by $P(q,t)$. Interpreting the resetting protocol as an effective environment, $P(q,t)$ thus fully characterizes its influence on the reduced system dynamics. As shown in Appendix~\ref{app:master_effective_time} (see also SM~\cite{SM}), $P(q,t)$ obeys a Fokker--Planck-like equation
\begin{equation}\label{eq:Pq_final}
\frac{\partial P(q,t)}{\partial t}=-\frac{\partial P(q,t)}{\partial q}-rP(q,t)+\frac{r}{t}\int_0^t\mathrm{d}\tau\,P(q,\tau),
\end{equation}
where the first term describes advection and the last two account for resetting. Thus, $q(t)$ can be interpreted as the position of a particle moving ballistically at unit velocity and undergoing memory-dependent resetting with rate $r$ to a previously visited position, selected at a uniformly sampled past time $\tau\in[0,t]$. Equation~\eqref{eq:Pq_final} can be solved exactly in Laplace space (see Appendix~\ref{app:master_effective_time}). Defining $f(s,t):=\int_0^\infty \mathrm{d}q\,e^{-sq}P(q,t)$, we obtain $f(s,t)=M\left(\frac{s}{r+s};1;-(r+s)t\right)$, where $M(a;b;z)$ is Kummer's confluent hypergeometric function of the first kind, whose series expansion is given in Eq.~\eqref{eq:series_expansion_M}.
This solution coincides with that obtained for classical diffusion with memory-dependent resetting using a completely different method that relies on the diffusive nature of the underlying dynamics~\cite{BM2026a}. Our derivation is more general, as it does not require knowledge of the underlying dynamics and applies to both classical and quantum systems. Moreover, the stochastic dynamics of the effective time $q(t)$, governed by Eq.~\eqref{eq:Pq_final}, has not been explored before. At long times, $P(q,t)$ approaches a Gaussian distribution with mean $T(t)=\ln(rt)/r$ and variance $2T(t)/r$, as shown in the SM~\cite{SM}. Consequently, $\langle q\rangle\simeq T(t)$ and $\operatorname{Var}(q)\simeq 2T(t)/r$. This asymptotic result was also obtained in the mathematical literature through a mapping to weighted random recursive trees~\cite{MUB2019}.

The representation in Eq.~\eqref{eq:formasintegraloverq} is particularly useful, as it allows us to compute, at late times, exact results for any observable directly from its reset-free counterpart. In the following, for any observable $\hat O$, we use the subscripts $0$ and $r$ to distinguish expectation values computed without and with resetting, respectively, defined as $\langle\hat O(t)\rangle_\alpha\equiv\operatorname{Tr}[\hat\rho_\alpha^{(N)}(t)\hat O]$, with $\alpha=0,r$. For the initial condition in Eq.~\eqref{eq:domainwallinitialstate_EM}, we have $\langle\hat O(t)\rangle_0
=\operatorname{Tr}[\hat\rho_0^{(N)}(t)\hat O]
=\langle\Psi_N(t)|\hat O|\Psi_N(t)\rangle$, while Eq.~\eqref{eq:formasintegraloverq} yields 
\begin{equation}\label{eq:observablerelation}
   \langle\hat O(t)\rangle_r=\int_0^t\mathrm{d}q\,P(q,t)\langle\hat O(q)\rangle_0. 
\end{equation}
This equation directly connects observables in the presence of memory to their reset-free counterparts. It also provides an efficient numerical approach, as discussed in the SM~\cite{SM}, by reducing the simulation of quantum many-body dynamics with memory-dependent resetting to sampling the one-dimensional classical stochastic process $q(t)$, provided the reset-free behavior of the observable $\langle\hat O(q)\rangle_0$ in Eq.~\eqref{eq:observablerelation}
is known either analytically or numerically.

\paragraph{Average density profile.}
We first consider the average local particle density, defined as the average of the operator
\begin{equation}\label{eq:operator_density}
    \hat n_x=\sum_{m=1}^N(|x\rangle\langle x|)_m,
\end{equation}
where $(|x\rangle\langle x|)_m=\hat I_1\otimes\cdots\otimes\hat I_{m-1}\otimes|x\rangle\langle x|\otimes\hat I_{m+1}\otimes\cdots\otimes\hat I_N$. In the absence of resetting, it was shown in Ref.~\cite{ARRS1999} that, at large times, the density profile spreads ballistically and takes the scaling form
\begin{equation}\label{eq:averagedensitynoreset}
    \langle \hat n_x(t)\rangle_{0}
    =
    g\left(\frac{x}{2\gamma t}\right),
    \qquad
    g(z)=
    \begin{cases}
        1, & z<-1,\\[1mm]
        \displaystyle\frac{1}{\pi}\arccos z, & |z|<1,\\[2mm]
        0, & z>1.
    \end{cases}
\end{equation}
The function $g(z)$ is shown in Fig.~\ref{fig:front}. Thus, at large times, the fermions form a front that propagates ballistically, with a characteristic length scale $\Lambda_0(t)\simeq 2\gamma t$, while retaining a time-independent shape when positions are rescaled by $2\gamma t$.

In the presence of resetting with memory, we instead obtain (see Appendix~\ref{eq:sec_averagenwithr_APP})
\begin{equation}\label{eq:mx_memory_bulk}
    \langle\hat n_x(t)\rangle_r
    \simeq
    g\left(\frac{x}{\Lambda_r(t)}\right),
    \qquad
    \Lambda_r(t)=\frac{2\gamma}{r}\ln(rt).
\end{equation}
Remarkably, the scaling function $g(z)$ remains unchanged from the reset-free case, while the ballistic scale $2\gamma t$ is replaced by the slowly growing logarithmic scale $\Lambda_r(t)$. This prediction is confirmed by the numerical simulations shown in Fig.~\ref{fig:front}.
The front therefore continues to propagate, but logarithmically slowly rather than ballistically. This slowdown is considerably stronger than that induced by dephasing noise~\cite{Eisler2011} or strong continuous monitoring~\cite{BernardJinShpielberg2018}, both of which turn ballistic transport into diffusive transport, with a characteristic spreading length growing as $\sqrt{t}$.
Furthermore, unlike conventional resetting to a fixed state, which leads to a stationary density profile~\cite{MSM2018}, our protocol produces an indefinitely spreading front.
Interestingly, the logarithmic growth also differs from classical random walks under resetting with uniform-memory, where the characteristic spreading length grows as $\sqrt{\ln t}$~\cite{BSS2014,BRC2014}.

\paragraph{Local density at the edge of the front.}

We now focus on the region near the edge of the propagating front and study the local shape of the density profile. For $r=0$, the right edge of the front is located at $\Lambda_0(t)=2\gamma t$, while for $r>0$ it lies at $\Lambda_r(t)=2\gamma\ln(rt)/r$. To resolve the density profile in this region, we consider positions $x=\Lambda_{0/r}(t)+\delta x$, where $\delta x$ is of the order of the front edge width. Without resetting, the front edge width grows as $t^{1/3}$~\cite{Hunyadi2004}, and the local density takes the scaling form
\begin{equation}\label{eq:edge_density}
    \langle \hat n_x(t)\rangle_0
    \simeq
    \frac{1}{(\gamma t)^{1/3}}
    \mathcal{F}_0\left(
    \frac{x-2\gamma t}{(\gamma t)^{1/3}}
    \right),
\end{equation}
where $\mathcal{F}_0(y)=\int_y^\infty \dd s\,\operatorname{Ai}^2(s)$, with $\operatorname{Ai}(s)$ denoting the Airy function. 
The scaling function $\mathcal{F}_0(y)$ is shown in blue in the inset of Fig.~\ref{fig:front}, where it is compared with numerical simulations. Its asymptotic behaviors are given in Eq.~\eqref{eq:asymptF0_EM}.

We now investigate how this local density profile is modified by 
resetting with memory. Using Eqs.~\eqref{eq:observablerelation} 
and~\eqref{eq:edge_density}, as shown in the SM~\cite{SM}, we find
\begin{equation}\label{eq:memory_edge_final}
    \langle \hat n_x(t)\rangle_r
    \simeq
    \left(\frac{2\gamma}{r\Lambda_r(t)}\right)^{1/4}
    \mathcal{F}_r\left(
    \frac{x-\Lambda_r(t)}
    {\sqrt{\gamma\Lambda_r(t)/(2r)}}
    \right),
\end{equation}
where
\begin{equation}\label{eq:Fmemory_integral}
    \mathcal{F}_r(y)
    =
    \frac{1}{2\pi^{3/2}}
    \int_{y/2}^{\infty}\dd u\,
    e^{-u^2/4}\sqrt{2u-y}.
\end{equation}
The scaling function $\mathcal{F}_r(y)$ is shown in orange in the inset of Fig.~\ref{fig:front}, compared with numerical simulations. Its asymptotic behaviors are given in Eq.~\eqref{eq:asymptF0_EM}.
In particular, $\mathcal{F}_r(y)\sim\sqrt{-y}/\pi$ as $y\to-\infty$ ensures that the far-left tail of the edge density matches the far-right tail of the macroscopic density profile in Eq.~\eqref{eq:mx_memory_bulk} (see the SM~\cite{SM}). For $r>0$, the front edge width thus grows as $\sqrt{\ln(rt)}$, rather than $t^{1/3}$ without resetting, and the edge is governed by a different scaling function.

Beyond this change in scaling, the two cases also differ in the number of particles near the front edge. Without resetting, the front edge width grows as $(\gamma t)^{1/3}$, while the local density at $x=2\gamma t$ decreases as $(\gamma t)^{-1/3}$, as follows from Eq.~\eqref{eq:edge_density}. Their product remains $O(1)$, implying that the edge region contains only an order-one number of particles. With uniform-memory resetting, instead, the front edge broadens as $\sqrt{\ln t}$ while the local density decreases only as $(\ln t)^{-1/4}$. Consequently, the number of particles in the edge region grows as $(\ln t)^{1/4}$.

\paragraph{Extreme-value statistics.}
We now consider the position $X_{\max}(t)$ of the rightmost fermion. In the absence of resetting, the fluctuations of $X_{\max}(t)$ around the ballistic front edge $2\gamma t$ occur on the same $t^{1/3}$ scale as the density edge and are governed by the Tracy--Widom distribution,
\begin{equation}\label{eq:TW_density}
    \mathbb P_0\left[X_{\max}(t)=x\right]
    \simeq
    \frac{1}{(\gamma t)^{1/3}}
    f_0\left(
    \frac{x-2\gamma t}{(\gamma t)^{1/3}}
    \right),
\end{equation}
where $f_0(s)$ is the Tracy--Widom probability density with Dyson index $\beta=2$~\cite{Tracy1994,Tracy1996}. Originally derived in Ref.~\cite{Eisler2013}, this result is re-derived in the SM~\cite{SM} using our notation in the first quantization formalism.

In the presence of uniform-memory resetting, 
Eqs.~\eqref{eq:observablerelation} and~\eqref{eq:TW_density} yield
\begin{equation}
    \mathbb P_r\left[X_{\max}(t)=x\right]
    =\int_0^t\dd q\,P(q,t)\,
    \mathbb P_0\left[X_{\max}(q)=x\right].
\end{equation}
At fixed $q$, $\mathbb P_0[X_{\max}(q)=x]$ has Tracy--Widom 
fluctuations of width $O(q^{1/3})$, as follows from 
Eq.~\eqref{eq:TW_density}. However, at large times the effective 
time $q$ itself fluctuates around $T(t)=\Lambda_r(t)/(2\gamma)$ over 
a scale $\sqrt{\Lambda_r(t)/(2\gamma r)}$, which induces 
fluctuations of the front edge position of order $\Lambda_r^{1/2}(t)$.
Since $\Lambda_r^{1/2}(t)\gg\Lambda_r^{1/3}(t)$, these 
externally induced fluctuations asymptotically dominate the 
Tracy--Widom ones. As a result,
\begin{equation}\label{eq:Xmax_gaussian}
    \mathbb P_r\left[X_{\max}(t)=x\right]
    \simeq
    \frac{1}{2\sqrt{\gamma\Lambda_r(t)/r}}
    f_r\left(
    \frac{x-\Lambda_r(t)}
    {2\sqrt{\gamma\Lambda_r(t)/r}}
    \right),
\end{equation}
where $f_r(y)=e^{-y^2/2}/\sqrt{2\pi}$.
Thus, for any nonzero $r$, the asymptotic fluctuations
of the rightmost fermion become Gaussian instead of Tracy--Widom:
memory resetting changes both the width and universality class
of the edge fluctuations. This prediction is confirmed by
numerical simulations in the SM~\cite{SM}.
A similar fragility of Tracy–Widom statistics was found in Ref.~\cite{CDV2018}, where interactions in the XXZ chain were argued to prevent Tracy–Widom scaling.

\paragraph{Integrated current.}
We finally consider the integrated current $\hat C=\sum_{x=1}^{\infty}\hat n_x$, which counts the number of fermions that at time $t$ are in the initially empty half-chain. In the absence of resetting, its mean grows ballistically, while its variance grows only logarithmically,
\begin{equation}
    \langle\hat C(t)\rangle_0
    \simeq
    \frac{2\gamma t}{\pi},
    \quad
    \operatorname{Var}_0[\hat C(t)]
    \simeq
    \frac{1}{2\pi^2}
    \left[\ln(2\gamma t)+\bar c\right],
\end{equation}
with $\bar c\simeq2.963$, as shown in Ref.~\cite{AKR2008}.

In the presence of uniform-memory resetting, instead, using 
Eq.~\eqref{eq:formasintegraloverq} we obtain 
$\langle\hat C(t)\rangle_r\simeq\Lambda_r(t)/\pi$, where $\Lambda_r(t)$ is given in Eq.~\eqref{eq:mx_memory_bulk}.
The variance contains two contributions: the intrinsic reset-free 
fluctuations at fixed $q$ and those induced by the randomness of the effective time $q$. At long times the 
latter dominate, giving 
$\operatorname{Var}_r[\hat C(t)]\simeq
4\gamma\Lambda_r(t)/(\pi^2r)$.
Thus, both the mean and variance grow linearly with $\Lambda_r(t)$,
i.e., logarithmically in time, with the leading fluctuations 
dominated by the randomness of the effective time. These predictions 
are confirmed by numerical simulations presented in the SM~\cite{SM}. 
Consequently, current fluctuations are much larger relative to the 
mean than in the reset-free case.

\paragraph{Conclusions and outlook.}
In this Letter, we have introduced a general framework for memory-dependent resetting in quantum many-body systems, admitting an exact solution for arbitrary time-independent Hamiltonians, including interacting systems. The dynamics can be expressed in terms of the reset-free unitary evolution evaluated at a stochastic effective time. Applying this framework to a chain of noninteracting fermions initially prepared in a domain-wall state, we have shown that memory transforms ballistic transport into logarithmically slow spreading. While the macroscopic density profile retains its original scaling function, the local structure of the front edge and the collective fluctuations change qualitatively. In particular, the local density near the front edge and the extreme-value statistics change completely.

Our framework suggests several research directions. Since the exact representation in Eq.~\eqref{eq:formasintegraloverq} holds for arbitrary time-independent Hamiltonians and $P(q,t)$ is known exactly, the dynamics under memory-dependent resetting can be determined whenever the corresponding reset-free dynamics is accessible, either analytically or numerically. This opens the possibility of investigating memory-induced effects in interacting many-body systems beyond the noninteracting fermionic chain studied here. Another interesting direction is to explore whether memory-dependent resetting can generate dynamically emergent correlations~\cite{MS2026_DEC} between noninteracting quantum particles, as previously observed in classical diffusing systems~\cite{BM2026a}.

Our framework can also be extended to more general memory kernels. The integral term in Eq.~\eqref{eq:memory_masterequation} corresponds to a uniform-memory kernel $K(\tau,t)=1/t$, which could be generalized to favor resets to either distant or recent states in the system's history, as previously explored in classical systems~\cite{BEM2017,BEM2025}. Different kernels could lead to qualitatively different dynamical regimes, potentially allowing control over quantum spreading. Another important question is whether our effective dynamics can be derived microscopically from system--environment interactions or measurement-and-feedback mechanisms, providing a physical foundation for the protocol. Finally, recent experimental advances in reversing many-body unitary evolution~\cite{Colombo2022,Geier2024} suggest possible routes to implementing memory-dependent resetting in programmable quantum platforms.

\textit{Acknowledgments.} We acknowledge support from ANR Grant No. ANR-23-CE30-0020-01 EDIPS. M.K. acknowledges support from the Department of Atomic Energy, Government of India, under Project No. RTI4001. M. K. thanks the hospitality of Laboratoire de Physique Théorique et Modèles Statistiques (LPTMS), University Paris-Saclay and Collège de France, PSL Research University where a major part of the work took place. We thank Leonardo Mazza for useful discussions.

\clearpage
\onecolumngrid

\appendix
\setcounter{secnumdepth}{1}

\section{The model}
\label{app:model}

The $N$-fermions Hamiltonian introduced in the main Letter reads
\begin{equation}\label{eq:NbodyHam}
    \hat{\mathcal H}=\sum_{m=1}^N\hat H_m,
    \qquad
    \hat H_m=\hat I_1\otimes\cdots\otimes
    \underset{m\text{-th factor}}{\hat H}
    \otimes\cdots\otimes\hat I_N,
\end{equation}
where $\hat I_m$ is the identity operator acting on the $m$-th particle, while the single-particle tight-binding Hamiltonian $\hat H$ is given in Eq.~\eqref{eq:singlepartHamilt} in the main text. The eigenstates $|k\rangle$ of $\hat H$ are plane waves, $\langle x|k\rangle=e^{ikx}/\sqrt{2\pi}$, with eigenenergies $E(k)=-2\gamma\cos k$, where $k\in[-\pi,\pi]$. The corresponding single-particle propagator is $\langle x|e^{-i\hat Ht}|y\rangle=i^{x-y}J_{x-y}(2\gamma t)$, where $J_n$ denotes the Bessel function of the first kind.

We consider a domain-wall initial condition in which the
$N$ sites $x=0,-1,\ldots,-(N-1)$ are initially occupied,
with the corresponding single-particle orbitals given by
$|\phi_n(0)\rangle=|x=-(n-1)\rangle$, for $n=1,\ldots,N$.
The resulting many-body initial state is the Slater determinant
\begin{equation}\label{eq:domainwallinitialstate_EM}
    |\Psi_N(0)\rangle
    =
    \frac{1}{\sqrt{N!}}
    \sum_{\pi\in S_N}
    \operatorname{sgn}(\pi)
    \bigotimes_{m=1}^N
    |\phi_{\pi(m)}(0)\rangle,
\end{equation}
where $S_N$ is the permutation group of $N$ elements and $\operatorname{sgn}(\pi)$ is the signature of the permutation $\pi$. In the absence of resetting, each orbital evolves independently as $|\phi_n(t)\rangle=e^{-i\hat Ht}|\phi_n(0)\rangle$. Using the single-particle propagator given above, we obtain $\phi_n(x,t)\equiv\langle x|\phi_n(t)\rangle=i^{x+n-1}J_{x+n-1}(2\gamma t)$. Since the particles are noninteracting, the many-body state remains a Slater determinant of the time-evolved orbitals,
\begin{equation}
    |\Psi_N(t)\rangle=e^{-i\hat{\mathcal H}t}|\Psi_N(0)\rangle=\frac{1}{\sqrt{N!}}\sum_{\pi\in S_N}\operatorname{sgn}(\pi)\bigotimes_{m=1}^N|\phi_{\pi(m)}(t)\rangle.
\end{equation}
Further details on the derivation of the reset-free many-body
dynamics are provided in the SM~\cite{SM}.

\section{Master equation and effective-time representation}
\label{app:master_effective_time}

In this section, we derive Eqs.~\eqref{eq:memory_masterequation},~\eqref{eq:formasintegraloverq} and~\eqref{eq:Pq_final} of the main text for an arbitrary time-independent $N$-body Hamiltonian $\hat{\mathcal H}$. Let $\hat\rho_0^{(N)}(q)=e^{-i\hat{\mathcal H}q}\hat\rho^{(N)}(0)e^{i\hat{\mathcal H}q}$ denote the reset-free density matrix, which satisfies the von Neumann equation 
\begin{equation}\label{eq:vonNewmaneq_EM}
    \partial_q\hat\rho_0^{(N)}=-i[\hat{\mathcal H},\hat\rho_0^{(N)}]
\end{equation}
Since each reset returns the system to a previously visited state, the dynamics always remains on the same unitary orbit and can be represented as $\hat\rho_0^{(N)}(q(t))$, where $q(t)$ is a stochastic effective time. We define the density matrix in the presence of resetting with memory as the average over all realizations of the stochastic protocol, $\hat\rho_r^{(N)}(t)=\langle\hat\rho_0^{(N)}(q(t))\rangle$, where the average is taken over both the random resetting times and the past times selected at each resetting event.

We now consider the evolution over a short time interval $\dd t$. According to the protocol introduced in the main text, two possibilities arise. With probability $1-r\,\dd t$, no reset occurs, and the system evolves unitarily, so that its effective time increases from $q(t)$ to $q(t)+\dd t$. With probability $r\,\dd t$, a reset occurs to a state visited at a past time $\tau$, chosen uniformly from $[0,t]$. Averaging over these two possibilities gives
\begin{equation}
\hat\rho_r^{(N)}(t+\dd t)
\simeq (1-r\,\dd t)\big\langle
\hat\rho_0^{(N)}(q(t)+\dd t)\big\rangle
+\frac{r\,\dd t}{t}\int_0^t\dd\tau\,
\hat\rho_r^{(N)}(\tau).
\label{app:step}
\end{equation}
Expanding to first order in $\dd t$ and using Eq.~\eqref{eq:vonNewmaneq_EM} directly yields Eq.~\eqref{eq:memory_masterequation} of the main text.
We now show that Eq.~\eqref{eq:formasintegraloverq} in the main text is a valid ansatz for the solution of the master equation~\eqref{eq:memory_masterequation}, with a uniquely determined effective-time distribution $P(q,t)$, given by Eq.~\eqref{eq:Pq_final}. Substituting this ansatz into the master equation, we obtain the following equation for $P(q,t)$ (see also the SM~\cite{SM}):
\begin{equation}\label{eq:Pq_final_EM}
\frac{\partial P(q,t)}{\partial t}=-\frac{\partial P(q,t)}{\partial q}-rP(q,t)+\frac{r}{t}\int_0^t\mathrm{d}\tau\,P(q,\tau).
\end{equation}
Multiplying both sides by $e^{-s q}$ and integrating over $q$, and recalling the definition of the Laplace transform $f(s,t)\equiv\int_0^\infty \mathrm{d}q\,e^{-s q}P(q,t)$, we obtain
\begin{equation}\label{eq:f_memory_EM}
\frac{\partial f(s,t)}{\partial t}=-(r+s)f(s,t)+\frac{r}{t}\int_0^t\mathrm{d}\tau\,f(s,\tau).
\end{equation}
This equation can be solved by multiplying both sides by $t$ and then differentiating with respect to $t$, which gives
\begin{equation}\label{eq:f_kummer_EM}
t\frac{\partial^2 f(s,t)}{\partial t^2}+\left[1+(r+s)t\right]\frac{\partial f(s,t)}{\partial t}+s f(s,t)=0.
\end{equation}
The equation is supplemented by the initial conditions 
$f(s,0)=1$ and $\left.\frac{\partial f(s,t)}{\partial t}\right|_{t=0}=-s$, where the second condition follows from Eq.~\eqref{eq:f_memory_EM} in the limit $t\to0$. Equation~\eqref{eq:f_kummer_EM} is Kummer's confluent hypergeometric equation, whose standard form is $z\,y''+(b-z)y'-ay=0$. In our case, $a=s/(r+s)$, $b=1$, and $z=-(r+s)t$. Imposing the initial conditions then gives 
\begin{equation}
\label{eq:f_app}
f(s,t)=M\left(\frac{s}{r+s};1;-(r+s)t\right). 
\end{equation}
The function $M(a,b,z)$ is Kummer's confluent hypergeometric function of the first kind, and can be written as the power series
\begin{equation}\label{eq:series_expansion_M}
    M(a,b,z)=\sum_{n=0}^{\infty}\frac{(a)_n}{(b)_n}\frac{z^n}{n!}
\end{equation}
where $(a)_n=a(a+1)\cdots(a+n-1)$ is the Pochhammer symbol, with $(a)_0=1$. See also Eq. (13.2.2) in Ref.~\cite{NIST}.

\section{Derivation of the average density profile}
\label{eq:sec_averagenwithr_APP}

We now derive the large-time average density profile in the 
presence of resetting with memory, given in 
Eq.~\eqref{eq:mx_memory_bulk} in the main text. Using the 
effective-time representation in Eq.~\eqref{eq:formasintegraloverq} 
in the main text, we have
\begin{equation}\label{eq:density_integral_EM}
\langle\hat n_x(t)\rangle_r
=\int_0^t\dd q\,P(q,t)\langle\hat n_x(q)\rangle_0.
\end{equation}
At large times, the reset-free density takes the scaling form 
given in Eq.~\eqref{eq:averagedensitynoreset} of the main text, 
while $P(q,t)$ approaches the asymptotic Gaussian distribution 
derived in the SM~\cite{SM}. Substituting these results into 
Eq.~\eqref{eq:density_integral_EM}, we obtain
\begin{equation}
\langle\hat n_x(t)\rangle_r
\simeq\int_0^t\frac{\dd q}{\sqrt{4\pi T(t)/r}}
\exp\left[-\frac{(q-T(t))^2}{4T(t)/r}\right]
g\left(\frac{x}{2\gamma q}\right).
\end{equation}
We now introduce the variable $u=(q-T(t))/\sqrt{T(t)/r}$ 
and consider the scaling limit $t\to\infty$ with 
$z=x/\Lambda_r(t)$ fixed, where 
$\Lambda_r(t)=2\gamma T(t)=2\gamma\ln(rt)/r$.
The integral becomes
\begin{equation}
\langle\hat n_x(t)\rangle_r
\simeq\int_{-\infty}^{\infty}\frac{\dd u}{\sqrt{4\pi}}
e^{-u^2/4}
g\left(\frac{z}{1+u\sqrt{2\gamma/[r\Lambda_r(t)]}}\right),
\end{equation}
where the integration limits have been extended to infinity 
in the large-time limit. Since $\Lambda_r(t)\to\infty$, 
the fluctuations of $q$ become negligible compared to its mean, 
and the argument of $g$ approaches $z$. Using the normalization 
of the Gaussian distribution, we finally obtain 
$\langle\hat n_x(t)\rangle_r\simeq 
g\left(x/\Lambda_r(t)\right)$, as given in 
Eq.~\eqref{eq:mx_memory_bulk} of the main text. Further details 
of this derivation are provided in the SM~\cite{SM}.

\section{Asymptotics of $\mathcal{F}_0(y)$ and $\mathcal{F}_r(y)$}

The asymptotic behaviors of $\mathcal{F}_0(y)$ and $\mathcal{F}_r(y)$ read, respectively,
\begin{equation}\label{eq:asymptF0_EM}
    \mathcal{F}_0(y)\simeq
    \begin{cases}
        \displaystyle \frac{\sqrt{-y}}{\pi},
        & y\to-\infty,\\[2mm]
        \displaystyle \frac{e^{-4y^{3/2}/3}}{8\pi y},
        & y\to+\infty,
    \end{cases}
    \qquad
    \mathcal{F}_r(y)\simeq
    \begin{cases}
        \displaystyle \frac{\sqrt{-y}}{\pi},
        & y\to-\infty,\\[2mm]
        \displaystyle \frac{2^{3/2}}{\pi y^{3/2}}
        e^{-y^2/16},
        & y\to+\infty.
    \end{cases}
\end{equation}
The two functions have the same behavior for $y\to-\infty$, needed for the matching with the bulk density profile, but differ significantly for $y\to+\infty$.

\end{document}


\begin{center}

{\large \underline{Supplementary Material}}

\vspace{0.15cm}

{\large\bfseries
Anomalous front propagation in many-body quantum transport with memory-dependent resetting
\par}

\vspace{0.45cm}

Gabriele de Mauro$^{1}$,
Manas Kulkarni$^{2}$,
and Satya N. Majumdar$^{1}$

\vspace{0.15cm}

{\small\itshape
$^{1}$LPTMS, CNRS, Universit\'e Paris-Saclay,
91405 Orsay, France
\par}

{\small\itshape
$^{2}$International Centre for Theoretical Sciences,
Tata Institute of Fundamental Research,\\
Bengaluru 560089, India
\par}

\end{center}

\renewcommand{\theequation}{S\arabic{equation}}
\renewcommand{\thefigure}{S\arabic{figure}}
\setcounter{equation}{0}
\setcounter{figure}{0}

\vspace{1.5cm}

\renewcommand{\contentsname}{Contents}
\tableofcontents

\clearpage

In this Supplementary Material, we first re-derive several well-known results for a one-dimensional system of noninteracting fermions evolving unitarily from a domain-wall initial condition. While these results were originally derived using a second-quantized formalism~\cite{ARRS1999,Hunyadi2004,AKR2008,Eisler2013}, we reformulate them here in the first quantization formalism, which is more convenient for implementing our quantum resetting protocol with memory. This also allows us to establish the notation and formalism needed for the subsequent analysis. We then use these results to derive the corresponding properties of the system in the presence of quantum resetting with memory.

\section{Reset-free dynamics}\label{sec:onepartnomemory}

We describe here the reset-free unitary evolution of a one-dimensional tight-binding system of $N$ noninteracting spinless fermions, starting from a domain-wall initial state. As given in the main Letter, the many-body Hamiltonian is
\begin{equation}
    \hat{\mathcal H}
    =
    \sum_{m=1}^N \hat H_m,
    \qquad
    \hat H_m
    =
    \hat I_1\otimes\cdots\otimes
    \underset{m\text{-th factor}}{\hat H}
    \otimes\cdots\otimes\hat I_N,
    \label{eq:HN}
\end{equation}
where $\hat H_m$ denotes the single-particle Hamiltonian acting on the
$m$-th particle and as the identity on all the remaining particles.
The single-particle tight-binding Hamiltonian is
\begin{equation}
    \hat H
    =
    -\gamma\sum_{x\in\mathbb Z}
    \left(
    |x\rangle\langle x+1|
    +
    |x+1\rangle\langle x|
    \right),
    \label{eq:H1}
\end{equation}
and describes a particle hopping between nearest-neighbor sites on an infinite one-dimensional lattice, with hopping amplitude $\gamma$.
The initial many-body state, which we refer to as a domain-wall
state, is the Slater determinant
\begin{equation}
    |\Psi_N(0)\rangle
    =
    \frac{1}{\sqrt{N!}}
    \sum_{\pi\in S_N}
    \operatorname{sgn}(\pi)
    \bigotimes_{m=1}^N
    |\phi_{\pi(m)}(0)\rangle,
    \label{eq:Psi0}
\end{equation}
where the $N$ single-particle orbitals are defined as
\begin{equation}\label{eq:phinofzero}
    |\phi_n(0)\rangle:=
    |x=-(n-1)\rangle,
    \qquad
    n=1,\ldots,N,
\end{equation}
corresponding to particles initially localized at the
$N$ lattice sites $x=0,-1,\ldots,-(N-1)$.
In Eq.~\eqref{eq:Psi0}, $\pi\in S_N$ denotes a permutation of the $N$ orbital labels
$\{1,\ldots,N\}$, and $\operatorname{sgn}(\pi)$ denotes its sign. In
Eq.~\eqref{eq:Psi0}, the index $m$ labels the particle Hilbert-space
factor, whereas $\pi(m)$ specifies which single-particle orbital
occupies that factor in a given term of the antisymmetrized sum.
The corresponding initial $N$-body density matrix is
\begin{equation}
    \hat\rho^{(N)}(0)
    =
    |\Psi_N(0)\rangle\langle\Psi_N(0)|
    =
    \frac{1}{N!}
    \sum_{\pi,\sigma\in S_N}
    \operatorname{sgn}(\pi)\operatorname{sgn}(\sigma)
    \bigotimes_{m=1}^N
    |\phi_{\pi(m)}(0)\rangle
    \langle\phi_{\sigma(m)}(0)|,
    \label{eq:rhoN_initial}
\end{equation}
where the
superscript $N$ indicates that the density matrix describes the full
$N$-body system.

The many-body state in Eq.~\eqref{eq:Psi0} evolves unitarily according to
\begin{equation}
    |\Psi_N(t)\rangle
    =
    e^{-i\hat{\mathcal H}t}
    |\Psi_N(0)\rangle.
    \label{eq:Psi_unitary}
\end{equation}
Since the particles do not interact, the many-body evolution operator
factorizes as $e^{-i\hat{\mathcal H}t}=\bigotimes_{m=1}^N e^{-i\hat Ht}$.
Therefore, the time-evolved many-body state reads
\begin{equation}
    |\Psi_N(t)\rangle
    =
    \frac{1}{\sqrt{N!}}
    \sum_{\pi\in S_N}
    \operatorname{sgn}(\pi)
    \bigotimes_{m=1}^N
    |\phi_{\pi(m)}(t)\rangle,
    \label{eq:Psi_t}
\end{equation}
where the time-evolved single-particle orbitals are
\begin{equation}
    |\phi_n(t)\rangle
    =
    e^{-i\hat Ht}
    |\phi_n(0)\rangle,
    \qquad
    n=1,\ldots,N,
    \label{eq:phi_t}
\end{equation}
with $\hat H$ given in Eq.~\eqref{eq:H1} and $|\phi_n(0)\rangle$
given in Eq.~\eqref{eq:phinofzero}. Hence, the reset-free unitary
evolution preserves the Slater-determinant structure of the many-body state.
The corresponding $N$-body density matrix at time $t$ is
\begin{equation}
    \hat\rho_0^{(N)}(t)
    =
    |\Psi_N(t)\rangle\langle\Psi_N(t)|
    =
    e^{-i\hat{\mathcal H}t}
    \hat\rho^{(N)}(0)
    e^{i\hat{\mathcal H}t}.
    \label{eq:rhoN_t}
\end{equation}
Substituting Eq.~\eqref{eq:Psi_t} into Eq.~\eqref{eq:rhoN_t}, we obtain
the explicit expression
\begin{equation}
    \hat\rho_0^{(N)}(t)
    =
    \frac{1}{N!}
    \sum_{\pi,\sigma\in S_N}
    \operatorname{sgn}(\pi)\operatorname{sgn}(\sigma)
    \bigotimes_{m=1}^N
    |\phi_{\pi(m)}(t)\rangle
    \langle\phi_{\sigma(m)}(t)|.
    \label{eq:rhoN_expand}
\end{equation}
It therefore remains only to determine the time-evolved single-particle orbitals $|\phi_n(t)\rangle=e^{-i\hat H t}|\phi_n(0)\rangle$, where $\hat H$ is given in Eq.~\eqref{eq:H1}. This is done in the following subsection (Sec.~\ref{sec:singleparticledynam}).

\subsection{Single-particle dynamics}\label{sec:singleparticledynam}

To determine the time-evolved single-particle orbitals $|\phi_n(t)\rangle$, we first determine the energy spectrum of the single-particle Hamiltonian $\hat H$ given in Eq.~\eqref{eq:H1}. Its eigenstates satisfy
\begin{equation}\label{eq:spectrumdefinition}
    \hat H|\psi\rangle=E|\psi\rangle.
\end{equation}
Consider a generic eigenstate $|\psi\rangle$ of $\hat H$ and expand it on the $|x\rangle$ basis as $ |\psi\rangle=\sum_{x\in\mathbb Z}\psi_x|x\rangle$.
Acting with the Hamiltonian gives
\begin{align}
    \hat H|\psi\rangle
    &=
    -\gamma\sum_{x\in\mathbb Z}
    \left(
    |x\rangle\langle x+1|
    +
    |x+1\rangle\langle x|
    \right)
    \sum_{y\in\mathbb Z}\psi_y|y\rangle=
    -\gamma\sum_{x\in\mathbb Z}
    \left(
    \psi_{x+1}|x\rangle
    +
    \psi_x|x+1\rangle
    \right).
    \label{eq:Hpsi_1}
\end{align}
Relabeling $x+1\to x$ in the second term, we obtain
\begin{equation}
    \hat H|\psi\rangle
    =
    -\gamma\sum_{x\in\mathbb Z}
    \left(
    \psi_{x-1}
    +
    \psi_{x+1}
    \right)|x\rangle.
\end{equation}
Comparing this expression with Eq.~\eqref{eq:spectrumdefinition} gives
\begin{equation}
    -\gamma\left(
    \psi_{x-1}
    +
    \psi_{x+1}
    \right)
    =
    E\psi_x.
    \label{eq:discrete_sch}
\end{equation}
Because of translational invariance, we look for plane-wave solutions of the form $\psi_x=Ae^{ikx}$.
Substituting this ansatz into Eq.~\eqref{eq:discrete_sch} gives $-\gamma\left(e^{-ik}+e^{ik}\right)Ae^{ikx}=EAe^{ikx}$,
and therefore the dispersion relation is
\begin{equation}
    E(k)
    =
    -2\gamma \cos k,
    \qquad
    k\in[-\pi,\pi].
    \label{eq:dispersion}
\end{equation}
The normalization constant $A$ is fixed by imposing $\langle k|k'\rangle=\delta(k-k')$. Using the position basis, we have
\begin{align}
\langle k|k'\rangle
&=
|A|^2
\sum_{x\in\mathbb Z}
e^{i(k'-k)x}.
\end{align}
Using the identity $\sum_{x\in\mathbb Z}e^{iqx}=2\pi\sum_{\ell\in\mathbb Z}\delta(q-2\pi\ell)$,
and restricting $k,k'\in[-\pi,\pi]$, only the term $\ell=0$ contributes, so that $\langle k|k'\rangle=2\pi |A|^2\delta(k-k')$.
Therefore, imposing $\langle k|k'\rangle=\delta(k-k')$ gives $|A|^2=1/(2\pi)$, so that one can choose
\begin{equation}
    \langle x|k\rangle
    =
    \frac{e^{ikx}}{\sqrt{2\pi}},
    \qquad
    k\in[-\pi,\pi].
    \label{eq:momentum_normalization}
\end{equation}
We now determine explicitly the time evolution of the orbitals entering the many-body state in Eq.~\eqref{eq:rhoN_expand}. From Eq.~\eqref{eq:phinofzero}, the $n$-th initially
occupied orbital is $|\phi_n(0)\rangle=|-(n-1)\rangle$.
Its wave function at time $t$ is therefore
\begin{equation}
    \phi_n(x,t)
    \equiv
    \langle x|\phi_n(t)\rangle
    =
    \langle x|
    e^{-i\hat Ht}
    |-(n-1)\rangle.
    \label{eq:phi_position_def}
\end{equation}
Inserting the momentum resolution of the identity $\hat I=\int_{-\pi}^{\pi}\mathrm d k\,|k\rangle\langle k|$
and using the dispersion relation in Eq.~\eqref{eq:dispersion}, we obtain
\begin{align}
    \phi_n(x,t)
    &=
    \int_{-\pi}^{\pi}\mathrm d k\,
    \langle x|k\rangle
    e^{-iE(k)t}
    \langle k|-(n-1)\rangle=
    \int_{-\pi}^{\pi}
    \frac{\mathrm d k}{2\pi}\,
    e^{ik(x+n-1)}
    e^{2i\gamma t\cos k}.
    \label{eq:orbital_integral}
\end{align}
Using the standard integral representation of the Bessel function of the first kind of order $\nu$, $J_{\nu}(z)$ [see Eq.~(10.9.2) in Ref.~\cite{NIST}], this becomes
\begin{equation}
    \phi_n(x,t)
    =
    i^{x+n-1}
    J_{x+n-1}(2\gamma t).
    \label{eq:orbital_bessel}
\end{equation}
More generally, the single-particle propagator from an arbitrary initial
site $y$ to a final site $x$ is
\begin{equation}
    \langle x|e^{-i\hat Ht}|y\rangle
    =
    i^{x-y}
    J_{x-y}(2\gamma t).
    \label{eq:propagator}
\end{equation}
By setting $y=-(n-1)$ in Eq.~\eqref{eq:propagator} one immediately recovers 
Eq.~\eqref{eq:orbital_bessel}.

Eqs.~\eqref{eq:rhoN_expand} and \eqref{eq:orbital_bessel} together provide the exact density matrix of the full many-body system under unitary evolution at any time $t$.

\subsection{One-body reduced density matrix}

For later convenience, it is useful to compute the one-body reduced density matrix associated with $\hat\rho_0^{(N)}(t)$, given in Eq.~\eqref{eq:rhoN_expand}. As we will show below, the matrix elements of all higher-order reduced density matrices can be expressed entirely in terms of those of the one-body reduced density matrix. This property is crucial for the exact solvability of the system. We therefore begin by considering the one-body reduced density matrix, defined as
\begin{equation}
    \hat\rho_0^{(1)}(t)
    =
    \operatorname{Tr}_{2,\ldots,N}\hat\rho_0^{(N)}(t),
    \label{eq:rho1_def}
\end{equation}
where $\hat\rho_0^{(N)}(t)$ is given in Eq.~\eqref{eq:rhoN_expand} and $\operatorname{Tr}_{2,\ldots,N}$ denotes the partial trace over all single-particle Hilbert spaces except the first one.
To perform the partial trace explicitly, we consider the position basis $\{|x\rangle\}_{x\in\mathbb Z}$ in each of the traced single-particle spaces:
\begin{align}
    \hat\rho_0^{(1)}(t)
    &=
    \sum_{x_2\in\mathbb Z}\cdots\sum_{x_N\in\mathbb Z}
    \Bigl(
    \hat I_1\otimes
    \langle x_2|\otimes\cdots\otimes\langle x_N|
    \Bigr)
    \hat\rho_0^{(N)}(t)
    \Bigl(
    \hat I_1\otimes
    |x_2\rangle\otimes\cdots\otimes|x_N\rangle
    \Bigr).
    \label{eq:rho1_partial_trace_explicit}
\end{align}
Substituting Eq.~\eqref{eq:rhoN_expand} into Eq.~\eqref{eq:rho1_partial_trace_explicit} gives
\begin{align}
    \hat\rho_0^{(1)}(t)
    &=
    \frac{1}{N!}
    \sum_{\pi,\sigma\in S_N}
    \operatorname{sgn}(\pi)\operatorname{sgn}(\sigma)
    \sum_{x_2\in\mathbb Z}\cdots\sum_{x_N\in\mathbb Z}
    \Bigl(
    \hat I_1\otimes
    \langle x_2|\otimes\cdots\otimes\langle x_N|
    \Bigr)
    \nonumber\\
    &\qquad\qquad\times
    \left[
    \bigotimes_{m=1}^N
    |\phi_{\pi(m)}(t)\rangle
    \langle\phi_{\sigma(m)}(t)|
    \right]
    \Bigl(
    \hat I_1\otimes
    |x_2\rangle\otimes\cdots\otimes|x_N\rangle
    \Bigr).
    \label{eq:rho1_partial_trace_insert}
\end{align}
The operator $|\phi_{\pi(1)}(t)\rangle\langle\phi_{\sigma(1)}(t)|$ acting on the first Hilbert space is left untouched because of the identity matrices, while for
each $m=2,\ldots,N$ the trace produces a scalar matrix element. Hence
\begin{align}
    \hat\rho_0^{(1)}(t)
    &=
    \frac{1}{N!}
    \sum_{\pi,\sigma\in S_N}
    \operatorname{sgn}(\pi)\operatorname{sgn}(\sigma)
    |\phi_{\pi(1)}(t)\rangle
    \langle\phi_{\sigma(1)}(t)|
    \sum_{x_2\in\mathbb Z}\cdots\sum_{x_N\in\mathbb Z}
    \prod_{m=2}^N
    \langle x_m|\phi_{\pi(m)}(t)\rangle
    \langle\phi_{\sigma(m)}(t)|x_m\rangle.
    \label{eq:rho1_trace_all_sums}
\end{align}
Since each $x_m$ appears only in the corresponding factor, the sums factorize:
\begin{align}
    &\sum_{x_2\in\mathbb Z}\cdots\sum_{x_N\in\mathbb Z}
    \prod_{m=2}^N
    \langle x_m|\phi_{\pi(m)}(t)\rangle
    \langle\phi_{\sigma(m)}(t)|x_m\rangle=
    \prod_{m=2}^N
    \left[
    \sum_{x_m\in\mathbb Z}
    \langle\phi_{\sigma(m)}(t)|x_m\rangle
    \langle x_m|\phi_{\pi(m)}(t)\rangle
    \right].
    \label{eq:rho1_sums_factorize}
\end{align}
Using the completeness relation
$\sum_{x_m\in\mathbb Z}|x_m\rangle\langle x_m|=\hat I_m$, each sum becomes
\begin{equation}
    \sum_{x_m\in\mathbb Z}
    \langle\phi_{\sigma(m)}(t)|x_m\rangle
    \langle x_m|\phi_{\pi(m)}(t)\rangle
    =
    \langle\phi_{\sigma(m)}(t)|
    \phi_{\pi(m)}(t)\rangle.
    \label{eq:rho1_completeness}
\end{equation}
The single-particle states are orthonormal at $t=0$, and unitary
evolution preserves their scalar products:
\begin{align}
    \langle\phi_n(t)|\phi_{n'}(t)\rangle
    &=
    \langle\phi_n(0)|
    e^{i\hat Ht}e^{-i\hat Ht}
    |\phi_{n'}(0)\rangle=
    \langle\phi_n(0)|\phi_{n'}(0)\rangle
    =
    \delta_{nn'}.
    \label{eq:orbital_orthogonality}
\end{align}
Therefore Eq.~\eqref{eq:rho1_trace_all_sums} reduces to
\begin{equation}
    \hat\rho_0^{(1)}(t)
    =
    \frac{1}{N!}
    \sum_{\pi,\sigma\in S_N}
    \operatorname{sgn}(\pi)\operatorname{sgn}(\sigma)
    |\phi_{\pi(1)}(t)\rangle
    \langle\phi_{\sigma(1)}(t)|
    \prod_{m=2}^N
    \delta_{\sigma(m),\pi(m)}.
    \label{eq:rho1_delta_product}
\end{equation}
The product of Kronecker deltas imposes
$\sigma(m)=\pi(m)$ for every $m=2,\ldots,N$. Since $\pi$ and $\sigma$
are both permutations of the same $N$ labels, once their last $N-1$
entries coincide, the remaining entry must also coincide,
$\sigma(1)=\pi(1)$. Hence $\sigma=\pi$, and therefore
$\operatorname{sgn}(\pi)\operatorname{sgn}(\sigma)=\operatorname{sgn}^2(\pi)=1$. We then obtain
\begin{equation}
    \hat\rho_0^{(1)}(t)
    =
    \frac{1}{N!}
    \sum_{\pi\in S_N}
    |\phi_{\pi(1)}(t)\rangle
    \langle\phi_{\pi(1)}(t)|.
    \label{eq:rho1_perm}
\end{equation}
Now fix the value of the first entry, $\pi(1)=m$. For each fixed $m$,
the remaining $N-1$ labels can be arranged in $(N-1)!$ different ways.
Therefore
\begin{align}
    \sum_{\pi\in S_N}
    |\phi_{\pi(1)}(t)\rangle
    \langle\phi_{\pi(1)}(t)|
    &=
    (N-1)!
    \sum_{m=1}^N
    |\phi_m(t)\rangle
    \langle\phi_m(t)|.
    \label{eq:rho1_perm_count}
\end{align}
Using $(N-1)!/N!=1/N$, we finally obtain
\begin{equation}
    \hat\rho_0^{(1)}(t)
    =
    \frac{1}{N}
    \sum_{m=1}^N
    |\phi_m(t)\rangle\langle\phi_m(t)|,
    \qquad
    |\phi_m (t)\rangle=e^{-i\hat Ht}|\phi_m(0)\rangle,
    \label{eq:rho1_result}
\end{equation}
where $\hat H$ is given in Eq.~\eqref{eq:H1}.
The matrix elements of $\hat\rho_0^{(1)}(t)$ in position space read
\begin{equation}
    \rho_0^{(1)}(x,y;t)
    \equiv
    \langle x|\hat\rho_0^{(1)}(t)|y\rangle=\frac{1}{N}
    \sum_{m=1}^N
    \phi_m(x,t)\phi_m^*(y,t),
    \label{eq:rho1_xy}
\end{equation}
where we recall (see Eq.~\eqref{eq:phi_position_def}) $\phi_m(x,t)=\langle x|\phi_m(t)\rangle$,
so that $\phi_m(x,t)$ represents the time-evolved wave function of a single fermion initially (at $t=0$) localized at position $x=-(m-1)$ along the
chain, namely in the state $|\phi_m(0)\rangle=|-(m-1)\rangle$. 

The above derivation is valid also at time $t=0$, for which it follows from Eq.~\eqref{eq:rho1_result} that
\begin{equation}
    \hat\rho_0^{(1)}(0)=\frac{1}{N}\sum_{m=1}^N
    |\phi_m(0)\rangle\langle\phi_m(0)|,
    \qquad
    |\phi_m(0)\rangle=|-(m-1)\rangle.
    \label{eq:rho1_initial}
\end{equation}

\subsection{Two-body reduced density matrix}

We now consider the two-body reduced density matrix. Tracing over particles $3,\ldots,N$, we define
\begin{equation}
    \hat\rho_0^{(2)}(t)
    =
    \operatorname{Tr}_{3,\ldots,N}\hat\rho_0^{(N)}(t).
    \label{eq:rho2_def}
\end{equation}
Using Eq.~\eqref{eq:rhoN_expand} and repeating the same steps as in the
derivation of the one-body reduced density matrix, from
Eq.~\eqref{eq:rho1_partial_trace_explicit} to
Eq.~\eqref{eq:orbital_orthogonality}, but now tracing only over the
single-particle spaces $3,\ldots,N$, we obtain
\begin{align}
    \hat\rho_0^{(2)}(t)
    &=
    \frac{1}{N!}
    \sum_{\pi,\sigma\in S_N}
    \operatorname{sgn}(\pi)\operatorname{sgn}(\sigma)
    |\phi_{\pi(1)}(t)\rangle
    \langle\phi_{\sigma(1)}(t)|
    \otimes
    |\phi_{\pi(2)}(t)\rangle
    \langle\phi_{\sigma(2)}(t)|
    \prod_{m=3}^N
    \delta_{\sigma(m),\pi(m)}.
    \label{eq:rho2_trace_step}
\end{align}
The trace fixes $\sigma(m)=\pi(m)$ for $m=3,\ldots,N$. For fixed $\pi$, only two possibilities remain for the first two entries of $\sigma$:
\begin{equation}
    \bigl(\sigma(1),\sigma(2)\bigr)
    =\bigl(\pi(1),\pi(2)\bigr)
    \quad\text{or}\quad
    \bigl(\pi(2),\pi(1)\bigr).
    \label{eq:two_perms}
\end{equation}
The first choice in Eq.~\eqref{eq:two_perms} has the same parity as $\pi$, whereas the second differs by one transposition and therefore carries the opposite sign. Consequently,
\begin{align}
    \hat\rho_0^{(2)}(t)
    =\frac{1}{N!}\sum_{\pi\in S_N}
    \Big[&|\phi_{\pi(1)}\rangle\langle\phi_{\pi(1)}|
    \otimes|\phi_{\pi(2)}\rangle\langle\phi_{\pi(2)}|-|\phi_{\pi(1)}\rangle\langle\phi_{\pi(2)}|
    \otimes|\phi_{\pi(2)}\rangle\langle\phi_{\pi(1)}|\Big].
    \label{eq:rho2_perm}
\end{align}
For any fixed ordered pair $(m,m')$ with $m\neq m'$, the remaining $N-2$ labels can be permuted in $(N-2)!$ ways. Thus
\begin{align}
    \hat\rho_0^{(2)}(t)
    &=
    \frac{1}{N(N-1)}
    \sum_{m\neq m'}
    \Big[
    |\phi_m\rangle\langle\phi_m|
    \otimes
    |\phi_{m'}\rangle\langle\phi_{m'}|
    -
    |\phi_m\rangle\langle\phi_{m'}|
    \otimes
    |\phi_{m'}\rangle\langle\phi_m|
    \Big].
    \label{eq:rho2_result}
\end{align}
The restriction $m\neq m'$ can be dropped because the two terms cancel identically for $m=m'$. Taking a matrix element and using Eq.~\eqref{eq:rho1_xy}, the first term on the right-hand side of Eq.~\eqref{eq:rho2_result} gives
\begin{align}
    \sum_{m,m'}
    \langle x_1|\phi_m\rangle
    \langle\phi_m|y_1\rangle
    \langle x_2|\phi_{m'}\rangle
    \langle\phi_{m'}|y_2\rangle
    =
    N^2
    \rho_0^{(1)}(x_1,y_1;t)
    \rho_0^{(1)}(x_2,y_2;t),
    \label{eq:rho2_firstterm}
\end{align}
where $\rho_0^{(1)}(x,y;t)$ is defined in Eq.~\eqref{eq:rho1_xy}, whereas the second term gives
\begin{align}
    \sum_{m,m'}
    \langle x_1|\phi_m\rangle
    \langle\phi_{m'}|y_1\rangle
    \langle x_2|\phi_{m'}\rangle
    \langle\phi_m|y_2\rangle
    =
    N^2
    \rho_0^{(1)}(x_1,y_2;t)
    \rho_0^{(1)}(x_2,y_1;t).
    \label{eq:rho2_secondterm}
\end{align}
Substituting these two expressions into Eq.~\eqref{eq:rho2_result},
\begin{equation}
    \langle x_1,x_2|\hat\rho_0^{(2)}(t)|y_1,y_2\rangle
    =
    \frac{N}{N-1}
    \left[
    \rho_0^{(1)}(x_1,y_1;t)\rho_0^{(1)}(x_2,y_2;t)
    -
    \rho_0^{(1)}(x_1,y_2;t)\rho_0^{(1)}(x_2,y_1;t)
    \right].
    \label{eq:rho2_rho1}
\end{equation}
Therefore, the two-body reduced density matrix can be expressed entirely in terms of the one-body reduced density matrix. More generally, for a Slater-determinant state, all higher-order correlation functions can be expressed in terms of one-body correlations through Wick's theorem~\cite{Wick1950,Borodin2011}. We demonstrate this property explicitly for arbitrary $p$ in Sec.~\ref{sec:generalp}.

\subsection{Average density profile}\label{sec:domainwallnoreset}

In this section, we derive the large-time average density profile of the system within our first-quantized framework. This result was originally obtained in Ref.~\cite{ARRS1999} using a second-quantized formalism. We begin by considering the local occupation operator
\begin{equation}
    \hat n_x
    =
    \sum_{m=1}^N (|x\rangle\langle x|)_m,
    \qquad
    (|x\rangle\langle x|)_m=
    \hat I_1\otimes\cdots\otimes
    \underset{m\text{-th factor}}{|x\rangle\langle x|}
    \otimes\cdots\otimes\hat I_N,
    \label{eq:mx_def}
\end{equation}
where $(|x\rangle\langle x|)_m$ denotes the projector onto site $x$ acting on the Hilbert space of particle $m$ and as the identity on all the other particles.
Its expectation value gives the average number of fermions occupying
site $x$ at time $t$:
\begin{align}
    \langle \hat n_x(t)\rangle_0
    &=
    \operatorname{Tr}_{1,\ldots,N}
    \left[
    \hat\rho_0^{(N)}(t)\hat n_x
    \right]=
    \sum_{m=1}^N
    \operatorname{Tr}_{1,\ldots,N}
    \left[
    \hat\rho_0^{(N)}(t)
    (|x\rangle\langle x|)_m
    \right],
    \label{eq:nx_sum_particles}
\end{align}
where the subscript $0$ indicates the absence of resetting.
Because the many-body state is invariant under particle relabeling,
all $N$ terms in the sum in Eq.~\eqref{eq:nx_sum_particles} give the same contribution. We can therefore
choose, for instance, particle $1$ and multiply by $N$:
\begin{align}
    \langle \hat n_x(t)\rangle_0
    &=
    N\,
    \operatorname{Tr}_{1,\ldots,N}
    \left[
    \hat\rho_0^{(N)}(t)
    (|x\rangle\langle x|)_1
    \right].
    \label{eq:nx_perm_sym}
\end{align}
Using the definition of the one-body reduced density matrix,
$\hat\rho_0^{(1)}(t)=\operatorname{Tr}_{2,\ldots,N}\hat\rho_0^{(N)}(t)$,
the trace over the remaining particles can be performed first:
\begin{align}
    \langle \hat n_x(t)\rangle_0
    &=
    N\,
    \operatorname{Tr}_{1}
    \left[
    \hat\rho_0^{(1)}(t)|x\rangle\langle x|
    \right]=
    N\langle x|\hat\rho_0^{(1)}(t)|x\rangle.
    \label{eq:mx_rho1}
\end{align}
Note that the derivation of Eq.~\eqref{eq:mx_rho1} relies only on the definition of the partial trace and not on the specific form of the one-body reduced density matrix. Therefore, the same relation remains valid in the presence of resetting with memory, upon replacing $\hat\rho_0^{(1)}(t)$ with $\hat\rho_r^{(1)}(t)$.

We now insert Eq.~\eqref{eq:rho1_result} into Eq.~\eqref{eq:mx_rho1}, which gives
\begin{align}
    \langle \hat n_x(t)\rangle_0=N\langle x|\hat\rho_0^{(1)}(t)|x\rangle
    &=
    \sum_{m=1}^N
    \langle x|\phi_m(t)\rangle
    \langle\phi_m(t)|x\rangle=
    \sum_{m=1}^N
    |\phi_m(x,t)|^2.
\end{align}
Therefore, using Eq.~\eqref{eq:orbital_bessel}, i.e.
$\phi_m(x,t)=i^{x+m-1}J_{x+m-1}(2\gamma t)$, we obtain
\begin{equation}
    \langle \hat n_x(t)\rangle_0
    =
    \sum_{m=1}^N
    J_{x+m-1}^2(2\gamma t).
    \label{eq:mx_finiteN}
\end{equation}

In the limit $N\to\infty$, setting $\ell=x+m-1$ gives
\begin{equation}
    \langle \hat n_x(t)\rangle_0=\sum_{\ell=x}^{\infty}J_\ell^2(2\gamma t).
    \label{eq:mx_bessel}
\end{equation}
At $t=0$, $J_\ell(0)=\delta_{\ell0}$, and therefore
\begin{equation}
    \langle \hat n_x(0)\rangle_0=
    \begin{cases}
        1,&x\leq0,\\
        0,&x>0,
    \end{cases}
    \label{eq:domainwall_IC}
\end{equation}
as required from the domain-wall initial condition.

We now derive the large-time scaling form of the density
in Eq.~\eqref{eq:mx_bessel}. To this end, we consider the limit
$|\ell|,t\to\infty$ with $\ell/(2\gamma t)$ fixed.
For $z=x/(2\gamma t)>1$, all terms in Eq.~\eqref{eq:mx_bessel} are exponentially small, since $|\ell|>2\gamma t$ and the Bessel functions decay exponentially in this regime [see Eq.~(10.19.3) of Ref.~\cite{NIST}]. Therefore, the density vanishes outside the ballistic front, $\langle\hat n_x(t)\rangle_0\to 0$.
For $z<-1$, instead, we can write
\begin{equation}
    \langle\hat n_x(t)\rangle_0
    =
    \sum_{\ell=-\infty}^{\infty}J_\ell^2(2\gamma t)
    -
    \sum_{\ell=-\infty}^{x-1}J_\ell^2(2\gamma t).
\end{equation}
Using the identity
$\sum_{\ell=-\infty}^{\infty}J_\ell^2(2\gamma t)=1$,
and noting that the second sum is again exponentially
suppressed since $\ell<x<-2\gamma t$, we obtain
$\langle\hat n_x(t)\rangle_0\to 1$.

We now consider the bulk region $|z|<1$. As discussed above, the contributions from $\ell>2\gamma t$ are negligible, so Eq.~\eqref{eq:mx_bessel} reduces to
\begin{equation}
    \langle \hat n_x(t)\rangle_0
    \simeq
    \sum_{\ell=x}^{\ell\lesssim 2\gamma t}
    J_\ell^2(2\gamma t).
    \label{eq:mx_bessel_truncated}
\end{equation}
For $|\ell|/(2\gamma t)<1$, when both $\ell$ and $2\gamma t$ are large, the Bessel functions
exhibit oscillatory behavior, with the asymptotic form [see Eq. (10.19.6)~\cite{NIST}]
\begin{equation}
    J_\ell(2\gamma t)
    \simeq
    \sqrt{\frac{2}{\pi\sqrt{(2\gamma t)^2-\ell^2}}}
    \cos\!\left[
        \sqrt{(2\gamma t)^2-\ell^2}
        -\ell\arccos\!\left(\frac{\ell}{2\gamma t}\right)
        -\frac{\pi}{4}
    \right].
    \label{eq:bessel_bulk_asympt}
\end{equation}
Squaring this expression and using
$2\cos^2\theta=1+\cos(2\theta)$, we obtain
\begin{equation}
    J_\ell^2(2\gamma t)
    \simeq
    \frac{1}{\pi\sqrt{(2\gamma t)^2-\ell^2}}
    \left[
    1+
    \cos\!\left(
        2\sqrt{(2\gamma t)^2-\ell^2}
        -2\ell\arccos\!\left(\frac{\ell}{2\gamma t}\right)
        -\frac{\pi}{2}
    \right)
    \right].
    \label{eq:bessel_bulk_squared}
\end{equation}
Substituting Eq.~\eqref{eq:bessel_bulk_squared} into
Eq.~\eqref{eq:mx_bessel_truncated}, we can separate
the density into a non-oscillatory and an oscillatory
contribution:
\begin{align}
    \langle \hat n_x(t)\rangle_0
    \simeq {}&
    \frac{1}{\pi}
    \sum_{\ell=x}^{\ell<2\gamma t}
    \frac{1}{\sqrt{(2\gamma t)^2-\ell^2}}+
    \frac{1}{\pi}
    \sum_{\ell=x}^{\ell<2\gamma t}
    \frac{
    \cos\!\left[
        2\sqrt{(2\gamma t)^2-\ell^2}
        -2\ell\arccos\!\left(\frac{\ell}{2\gamma t}\right)
        -\frac{\pi}{2}
    \right]
    }{
    \sqrt{(2\gamma t)^2-\ell^2}
    }.
    \label{eq:density_smooth_osc}
\end{align}
In the large-time scaling limit, the second term
on the right-hand side of Eq.~\eqref{eq:density_smooth_osc}
is subleading, as the rapidly oscillating contributions
cancel upon summation over $\ell$.
Consequently, the leading behavior of the density
is determined by the first term. We can now replace the sum by an integral, yielding
\begin{equation}
    \langle \hat n_x(t)\rangle_0
    \simeq
    \frac{1}{\pi}
    \int_x^{2\gamma t}
    \frac{\mathrm{d}\ell}
    {\sqrt{(2\gamma t)^2-\ell^2}}.
    \label{eq:density_bulk_integral}
\end{equation}
Introducing the scaling variable $z=x/(2\gamma t)$
and changing variables to $u=\ell/(2\gamma t)$,
we obtain
\begin{align}
    \langle \hat n_x(t)\rangle_0
    &\simeq
    \frac{1}{\pi}
    \int_z^1\frac{\mathrm{d}u}{\sqrt{1-u^2}}
    =
    \frac{1}{\pi}\arccos z,
    \qquad |z|<1.
    \label{eq:density_bulk_arccos}
\end{align}

Combining the results for the three regions,
we find that, in the limit $x,t\to\infty$ with
$z=x/(2\gamma t)$ fixed, the density takes the scaling form
\begin{equation}
    \langle \hat n_x(t)\rangle_0\simeq g\left( \frac{x}{2\gamma t}\right),
    \label{eq:bulk_scaling}
\end{equation}
where
\begin{equation}
    g(z)=
    \begin{cases}
        1,&z<-1,\\[1mm]
        \displaystyle\frac{1}{\pi}\arccos z,&|z|<1,\\[2mm]
        0,&z>1.
    \end{cases}
    \label{eq:g_bulk}
\end{equation}
Equations~\eqref{eq:bulk_scaling} and~\eqref{eq:g_bulk}
thus reproduce Eq.~(9) of the main text.

\subsection{Local density at the edge of the front}
\label{sec:finitetimecorrnoreset}

We now revisit the derivation of the local density near the edge
of the propagating front, first obtained in Ref.~\cite{Hunyadi2004} with a second quantization formalism.
From Eq.~\eqref{eq:g_bulk}, the bulk density vanishes for
$x>2\gamma t$, so that the right edge is located at
$x\simeq 2\gamma t$. Starting from Eq.~\eqref{eq:mx_bessel},
we therefore need the asymptotic behavior of
$J_\ell(2\gamma t)$ for large $t$, with $\ell$ close to
$2\gamma t$.

We start from the integral representation of the Bessel function,
\begin{equation}
    J_\ell(2\gamma t)
    =
    \frac{1}{2\pi}
    \int_{-\pi}^{\pi}\mathrm d\theta\,
    e^{i(\ell\theta-2\gamma t\sin\theta)}.
    \label{eq:bessel_integral}
\end{equation}
Since we are interested in the region near the front edge, we write
$\ell=2\gamma t+\delta\ell$, where $\delta\ell$ is subleading
with respect to $t$. The phase of the integrand is then
\begin{equation}
    \Phi(\theta)
    =
    \delta\ell\,\theta
    +2\gamma t(\theta-\sin\theta).
\end{equation}
For $\gamma t\gg1$, the integrand oscillates rapidly everywhere except near $\theta=0$. Indeed, the derivative of the phase is
$\Phi'(\theta)=\delta\ell+2\gamma t(1-\cos\theta)$, which is of order $t$ for any fixed $\theta\neq0$. The leading contribution therefore comes from a shrinking neighborhood of $\theta=0$, where we can expand $\sin\theta=\theta-\theta^3/6+O(\theta^5)$. This gives
\begin{equation}
    \Phi(\theta)
    \simeq
    \delta\ell\,\theta+\frac{\gamma t}{3}\theta^3.
    \label{eq:bessel_phase_expand}
\end{equation}

We now determine the scaling of $\delta\ell$ for which both terms in the right-hand side of Eq.~\eqref{eq:bessel_phase_expand} contribute to the phase. The cubic term is of order one when $\theta\sim(\gamma t)^{-1/3}$. Requiring the linear term to remain of the same order then gives
$\delta\ell\sim(\gamma t)^{1/3}$.
This identifies the characteristic scale near the front edge. We therefore introduce a scaling variable $s=O(1)$ by writing
\begin{equation}
    \ell=2\gamma t+(\gamma t)^{1/3}s.
    \label{eq:l_edge}
\end{equation}
With this choice, Eq.~\eqref{eq:bessel_phase_expand} becomes
\begin{equation}
    \Phi(\theta)
    \simeq
    (\gamma t)^{1/3}s\theta
    +\frac{\gamma t}{3}\theta^3.
    \label{eq:bessel_phase_edge}
\end{equation}
Rescaling the integration variable as $u=(\gamma t)^{1/3}\theta$, we obtain, in the large-time limit,
\begin{equation}
    J_\ell(2\gamma t)
    \simeq
    \frac{1}{(\gamma t)^{1/3}}
    \frac{1}{2\pi}
    \int_{-\infty}^{\infty}
    \mathrm d u\,
    e^{i(u^3/3+su)}.
    \label{eq:bessel_rescaled}
\end{equation}
Here, the integration limits have been extended to infinity since $t$ is large. 
Using the integral representation of the Airy function [Eq. (9.5.1) in Ref.~\cite{NIST}], namely
\begin{equation}
    \operatorname{Ai}(s)
    =
    \frac{1}{2\pi}
    \int_{-\infty}^{\infty}
    \mathrm d u\,
    e^{i(u^3/3+su)},
    \label{eq:airy_integral}
\end{equation}
we obtain
\begin{equation}
    J_\ell(2\gamma t)
    \simeq
    \frac{1}{(\gamma t)^{1/3}}
    \operatorname{Ai}\left(
    \frac{\ell-2\gamma t}{(\gamma t)^{1/3}}
    \right).
    \label{eq:bessel_airy}
\end{equation}

We can now determine the density near the front edge. Substituting Eq.~\eqref{eq:bessel_airy} into Eq.~\eqref{eq:mx_bessel}, we find
\begin{equation}
    \langle\hat n_x(t)\rangle_0
    \simeq
    \frac{1}{(\gamma t)^{2/3}}
    \sum_{\ell=x}^{\infty}
    \operatorname{Ai}^2\left(
    \frac{\ell-2\gamma t}{(\gamma t)^{1/3}}
    \right).
    \label{eq:mx_edge_sum}
\end{equation}
We consider positions $x$ such that
$x-2\gamma t=O((\gamma t)^{1/3})$. Since the terms sufficiently far to the right of the front edge are exponentially suppressed, the leading contribution to the sum comes from the edge scaling region where Eq.~\eqref{eq:bessel_airy} applies.
Introducing $s_\ell=(\ell-2\gamma t)/(\gamma t)^{1/3}$, consecutive values of $\ell$ correspond to increments $\Delta s=(\gamma t)^{-1/3}$. Thus, as $t\to\infty$, the sum becomes an integral,
\begin{equation}
    \sum_{\ell=x}^{\infty}
    \operatorname{Ai}^2(s_\ell)
    \simeq
    (\gamma t)^{1/3}
    \int_{s_x}^{\infty}
    \mathrm d s\,\operatorname{Ai}^2(s).
    \label{eq:sum_to_integral_airy}
\end{equation}
We therefore obtain the scaling form of the density near the front edge,
\begin{equation}
    \langle\hat n_x(t)\rangle_0
    \simeq
    \frac{1}{(\gamma t)^{1/3}}
    \mathcal{F}_0\left(
    \frac{x-2\gamma t}{(\gamma t)^{1/3}}
    \right),
    \qquad
    \mathcal{F}_0(y)
    =
    \int_y^\infty
    \mathrm d s\,\operatorname{Ai}^2(s).
    \label{eq:edge_density}
\end{equation}
This reproduces Eq.~(11) of the main text.
Thus, the density near the right edge is described by the scaling function $\mathcal{F}_0(y)$, with a characteristic width growing as $(\gamma t)^{1/3}$ and a density amplitude decaying as $(\gamma t)^{-1/3}$.
The asymptotic behaviors of $\mathcal{F}_0(y)$ read
\begin{equation}
    \mathcal{F}_0(y)
    \simeq
    \begin{cases}
        \displaystyle
        \frac{\sqrt{-y}}{\pi},
        & y\to-\infty,\\[2mm]
        \displaystyle
        \frac{1}{8\pi y}
        \exp\left(-\frac{4}{3}y^{3/2}\right),
        & y\to+\infty.
    \end{cases}
    \label{eq:F_airy_asympt}
\end{equation}

Finally, we verify that the local density at the edge, given in Eq.~\eqref{eq:edge_density}, matches the bulk density profile in Eq.~\eqref{eq:g_bulk}. Approaching the front edge from the left, we write
$x=2\gamma t+(\gamma t)^{1/3}y$, with $y<0$, so that
$x/(2\gamma t)=1+y/[2(\gamma t)^{2/3}]$.
Substituting this expression into Eq.~\eqref{eq:bulk_scaling} and using the expansion
$\arccos(1-\epsilon)\simeq\sqrt{2\epsilon}$ for small positive $\epsilon$, we obtain
\begin{equation}
    g\left(\frac{x}{2\gamma t}\right)
    \simeq
    \frac{1}{(\gamma t)^{1/3}}
    \frac{\sqrt{-y}}{\pi}.
    \label{eq:bulk_edge_matching}
\end{equation}
This agrees with the $y\to-\infty$ asymptotic behavior of the edge scaling function in Eq.~\eqref{eq:F_airy_asympt}, confirming that the edge and bulk density profiles match in their common regime of validity.

\subsection{General \texorpdfstring{$p$}{p}-body correlations and determinantal structure}\label{sec:generalp}

In the next section, we will study the extreme-value statistics of the rightmost fermion, for which we will need the equal-time $p$-point correlation functions of the local occupation operators. To compute these correlations, we first generalize the result obtained for the two-body reduced density matrix in Eq.~\eqref{eq:rho2_rho1} to arbitrary $p$-body reduced density matrices, with $p=1,\ldots,N$, showing that they can be expressed entirely in terms of the one-body reduced density matrix, provided that the full $N$-body state is a Slater determinant.

We now introduce the one-body correlation kernel $\Gamma_{xy}(t)$, commonly denoted by $K(x,y;t)$ in the literature on determinantal point processes (DPPs)~\cite{Hough2009,Borodin2011,Dean2016}. In terms of the one-body reduced density matrix defined in Eq.~\eqref{eq:rho1_def}, this kernel reads
\begin{equation}
    \Gamma_{xy}(t)
    \equiv
    N\langle x|\hat\rho_0^{(1)}(t)|y\rangle
    =
    \sum_{m=1}^N
    \phi_m(x,t)\phi_m^*(y,t).
    \label{eq:Gamma_def}
\end{equation}
We now consider the $p$-body reduced density matrix, obtained by tracing
over the last $N-p$ single-particle Hilbert spaces,
\begin{equation}
    \hat\rho_0^{(p)}(t)
    =
    \operatorname{Tr}_{p+1,\ldots,N}\hat\rho_0^{(N)}(t).
    \label{eq:rhop_def}
\end{equation}
We will show that its matrix elements
$\langle x_1,\ldots,x_p|\hat\rho_0^{(p)}(t)|y_1,\ldots,y_p\rangle$
can be expressed entirely in terms of the matrix elements of the one-body reduced density matrix, or equivalently, the one-body correlation matrix $\Gamma_{xy}(t)$. Specifically, we will prove the identity
\begin{equation}
    \frac{N!}{(N-p)!}
    \langle x_1,\ldots,x_p|
    \hat\rho_0^{(p)}(t)
    |y_1,\ldots,y_p\rangle
    =
    \det_{1\leq \ell,\ell'\leq p}
    \left[\Gamma_{x_\ell y_{\ell'}}(t)\right].
    \label{eq:pbody_det}
\end{equation}
Here, the determinant is taken over the $p\times p$ matrix
whose $(\ell,\ell')$ entry is $\Gamma_{x_\ell y_{\ell'}}(t)$.

Starting from the Slater determinant in Eq.~\eqref{eq:rhoN_expand}, we can repeat exactly the same steps used in the derivation of $\hat\rho_0^{(1)}(t)$, from
Eq.~\eqref{eq:rho1_partial_trace_explicit} to
Eq.~\eqref{eq:orbital_orthogonality}, but now tracing only over the
single-particle spaces $p+1,\ldots,N$. We obtain
\begin{align}
    \hat\rho_0^{(p)}(t)
    &=
    \frac{1}{N!}
    \sum_{\pi,\sigma\in S_N}
    \operatorname{sgn}(\pi)\operatorname{sgn}(\sigma)
    \left[
    \bigotimes_{\ell=1}^p
    |\phi_{\pi(\ell)}(t)\rangle
    \langle\phi_{\sigma(\ell)}(t)|
    \right]
    \prod_{\ell=p+1}^N
    \delta_{\sigma(\ell),\pi(\ell)}.
    \label{eq:rhop_trace_step}
\end{align}
The product of Kronecker deltas imposes
$\sigma(\ell)=\pi(\ell)$ for every $\ell=p+1,\ldots,N$. Therefore $\pi$ and
$\sigma$ have exactly the same last $N-p$ entries. Since both are
permutations of the same $N$ labels, their first $p$ entries must also
contain the same set of labels, but possibly in a different order.
Hence, for fixed $\pi$, there exists a permutation $\tau\in S_p$ such that
\begin{equation}
    \sigma(\ell)=\pi(\tau(\ell)),
    \qquad
    \ell=1,\ldots,p.
    \label{eq:sigma_pi_r}
\end{equation}
The permutation $\sigma$ is therefore obtained from $\pi$ by applying
$\tau$ to its first $p$ entries, so that $\operatorname{sgn}(\sigma)=\operatorname{sgn}(\pi)\operatorname{sgn}(\tau)$.
Consequently,
$\operatorname{sgn}(\pi)\operatorname{sgn}(\sigma)=\operatorname{sgn}(\tau)$, and
Eq.~\eqref{eq:rhop_trace_step} becomes
\begin{equation}
    \hat\rho_0^{(p)}(t)
    =
    \frac{1}{N!}
    \sum_{\pi\in S_N}
    \sum_{\tau\in S_p}
    \operatorname{sgn}(\tau)
    \bigotimes_{\ell=1}^p
    |\phi_{\pi(\ell)}(t)\rangle
    \langle\phi_{\pi(\tau(\ell))}(t)|.
    \label{eq:rhop_perm_pi}
\end{equation}
We now denote the first $p$ entries of $\pi$ by
$m_\ell=\pi(\ell)$, with $\ell=1,\ldots,p$. Since $\pi$ is a permutation, the
indices $m_1,\ldots,m_p$ are all distinct. Once these first $p$ entries
are fixed, the remaining $N-p$ labels can be arranged in $(N-p)!$
different ways. Therefore the sum over $\pi$ can be rewritten as
\begin{equation}
    \hat\rho_0^{(p)}(t)
    =
    \frac{(N-p)!}{N!}
    \sum_{\substack{m_1,\ldots,m_p=1\\\text{all distinct}}}^N
    \sum_{\tau\in S_p}
    \operatorname{sgn}(\tau)
    \bigotimes_{\ell=1}^p
    |\phi_{m_\ell}(t)\rangle
    \langle\phi_{m_{\tau(\ell)}}(t)|.
    \label{eq:rhop_explicit}
\end{equation}
We now take the matrix element of Eq.~\eqref{eq:rhop_explicit} between
$|x_1,\ldots,x_p\rangle$ and $|y_1,\ldots,y_p\rangle$. Using
$\phi_m(x,t)=\langle x|\phi_m(t)\rangle$, we find
\begin{align}
    \frac{N!}{(N-p)!}
    &\langle x_1,\ldots,x_p|
    \hat\rho_0^{(p)}(t)
    |y_1,\ldots,y_p\rangle=
    \sum_{\substack{m_1,\ldots,m_p=1\\\text{all distinct}}}^N
    \sum_{\tau\in S_p}
    \operatorname{sgn}(\tau)
    \prod_{\ell=1}^p
    \phi_{m_\ell}(x_\ell,t)
    \phi_{m_{\tau(\ell)}}^*(y_\ell,t).
    \label{eq:rhop_matrix_step}
\end{align}
The restriction that the indices $m_1,\ldots,m_p$ are all distinct can
now be removed. Indeed, if two indices coincide, for example
$m_\ell=m_{\ell'}$ with $\ell\neq \ell'$, the antisymmetrized sum over $\tau$ vanishes:
interchanging the two corresponding indices changes the sign of the
permutation while leaving the product unchanged. We can thus write
\begin{align}
    \frac{N!}{(N-p)!}
    &\langle x_1,\ldots,x_p|
    \hat\rho_0^{(p)}(t)
    |y_1,\ldots,y_p\rangle=
    \sum_{m_1=1}^N\cdots\sum_{m_p=1}^N
    \sum_{\tau\in S_p}
    \operatorname{sgn}(\tau)
    \prod_{\ell=1}^p
    \phi_{m_\ell}(x_\ell,t)
    \phi_{m_{\tau(\ell)}}^*(y_\ell,t).
    \label{eq:rhop_independent_sums}
\end{align}
To factorize the sums over the indices $m_\ell$, we first relabel the
indices in the second product. Since $\tau\in S_p$ is a permutation, the
set $\{\tau(1),\ldots,\tau(p)\}$ is again $\{1,\ldots,p\}$. Setting
$\ell'=\tau(\ell)$, or equivalently $\ell=\tau^{-1}(\ell')$, we have
\begin{equation}
    \prod_{\ell=1}^p
    \phi_{m_{\tau(\ell)}}^*(y_\ell,t)
    =
    \prod_{\ell'=1}^p
    \phi_{m_{\ell'}}^*(y_{\tau^{-1}(\ell')},t).
\end{equation}
Renaming the dummy index $\ell'\to\ell$,
\begin{equation}
    \prod_{\ell=1}^p
    \phi_{m_{\tau(\ell)}}^*(y_\ell,t)
    =
    \prod_{\ell=1}^p
    \phi_{m_\ell}^*(y_{\tau^{-1}(\ell)},t).
\end{equation}
Therefore,
\begin{equation}
    \prod_{\ell=1}^p
    \phi_{m_\ell}(x_\ell,t)
    \phi_{m_{\tau(\ell)}}^*(y_\ell,t)
    =
    \prod_{\ell=1}^p
    \phi_{m_\ell}(x_\ell,t)
    \phi_{m_\ell}^*(y_{\tau^{-1}(\ell)},t).
\end{equation}
Now each summation index $m_\ell$ appears only in the corresponding factor,
so the sums over $m_1,\ldots,m_p$ factorize independently.
Hence
\begin{align}
    \frac{N!}{(N-p)!}
    &\langle x_1,\ldots,x_p|
    \hat\rho_0^{(p)}(t)
    |y_1,\ldots,y_p\rangle=
    \sum_{\tau\in S_p}
    \operatorname{sgn}(\tau)
    \prod_{\ell=1}^p
    \left[
    \sum_{m_\ell=1}^N
    \phi_{m_\ell}(x_\ell,t)
    \phi_{m_\ell}^*(y_{\tau^{-1}(\ell)},t)
    \right].
    \label{eq:rhop_factorized}
\end{align}
Since the inverse of a permutation has the same sign,
$\operatorname{sgn}(\tau^{-1})=\operatorname{sgn}(\tau)$, and since $\tau^{-1}$ runs over the same set $S_p$
as $\tau$, we can simply relabel $\tau^{-1}\to\tau$. Using the definition of
the kernel in Eq.~\eqref{eq:Gamma_def}, we obtain
\begin{equation}
    \frac{N!}{(N-p)!}
    \langle x_1,\ldots,x_p|
    \hat\rho_0^{(p)}(t)
    |y_1,\ldots,y_p\rangle
    =
    \sum_{\tau\in S_p}
    \operatorname{sgn}(\tau)
    \prod_{\ell=1}^p
    \Gamma_{x_\ell y_{\tau(\ell)}}(t).
    \label{eq:rhop_det_expansion}
\end{equation}
The right-hand side is precisely the Leibniz expansion of a determinant,
which proves Eq.~\eqref{eq:pbody_det}. Setting $p=1$ in Eq.~\eqref{eq:rhop_det_expansion}, we recover
Eq.~\eqref{eq:Gamma_def}, while for $p=2$ we recover
Eq.~\eqref{eq:rho2_rho1}.

This result is the generalization of Eq.~\eqref{eq:rho2_rho1}: for a
Slater-determinant state, all higher-order reduced density matrices are
completely determined by the one-body correlation kernel.

\subsubsection{Equal-time occupation correlations}

To derive the extreme-value statistics of the rightmost fermion, we also need the equal-time $p$-point correlation functions of the local occupation operators at $p$ distinct lattice sites $x_1,\ldots,x_p$, namely
\begin{equation}
    \left\langle
    \hat n_{x_1}(t)\cdots\hat n_{x_p}(t)
    \right\rangle_0
    =
    \operatorname{Tr}_{1,\ldots,N}
    \left[
    \hat\rho_0^{(N)}(t)
    \hat n_{x_1}\cdots\hat n_{x_p}
    \right],
    \label{eq:p_point_occupation_def}
\end{equation}
where the subscript $0$ denotes the absence of resetting.
Using the definition in Eq.~\eqref{eq:mx_def}, each local occupation operator is $\hat n_{x_\ell}=\sum_{m_\ell=1}^N(|x_\ell\rangle\langle x_\ell|)_{m_\ell}$.
Therefore, their product can be written as
\begin{equation}
    \hat n_{x_1}\cdots\hat n_{x_p}
    =
    \sum_{m_1=1}^N\cdots\sum_{m_p=1}^N
    (|x_1\rangle\langle x_1|)_{m_1}
    \cdots
    (|x_p\rangle\langle x_p|)_{m_p}.
    \label{eq:product_n_expand}
\end{equation}
Substituting Eq.~\eqref{eq:product_n_expand} into
Eq.~\eqref{eq:p_point_occupation_def}, we obtain
\begin{align}
    \left\langle
    \hat n_{x_1}(t)\cdots\hat n_{x_p}(t)
    \right\rangle_0
    &=
    \sum_{m_1=1}^N\cdots\sum_{m_p=1}^N
    \operatorname{Tr}_{1,\ldots,N}
    \Big[
    \hat\rho_0^{(N)}(t)
    (|x_1\rangle\langle x_1|)_{m_1}
    \cdots
    (|x_p\rangle\langle x_p|)_{m_p}
    \Big].
    \label{eq:p_point_occupation_expand}
\end{align}
We now use the fact that the sites $x_1,\ldots,x_p$ must be all distinct.
In fact, if two particle labels coincide, for example $m_\ell=m_{\ell'}$ with $\ell\neq \ell'$,
then two different position projectors act on the same particle. Their
product is
\begin{equation}
    |x_\ell\rangle\langle x_\ell|
    |x_{\ell'}\rangle\langle x_{\ell'}|
    =
    \langle x_\ell|x_{\ell'}\rangle
    |x_\ell\rangle\langle x_{\ell'}|
    =
    0,
    \qquad x_\ell\neq x_{\ell'}.
    \label{eq:distinct_sites_projectors}
\end{equation}
Hence only terms with $m_1,\ldots,m_p$ all distinct contribute to
Eq.~\eqref{eq:p_point_occupation_expand}.
The number of ordered choices of $p$ distinct particle labels is
\begin{equation}
    N(N-1)\cdots(N-p+1)
    =
    \frac{N!}{(N-p)!}.
    \label{eq:number_distinct_particles}
\end{equation}
Moreover, because the many-body state is invariant under particle
relabeling, all these terms give the same contribution. We can therefore
choose the particular set $m_1=1,\ldots,m_p=p$ and multiply by the
number of equivalent terms:
\begin{align}
    \left\langle
    \hat n_{x_1}(t)\cdots\hat n_{x_p}(t)
    \right\rangle_0
    &=
    \frac{N!}{(N-p)!}
    \operatorname{Tr}_{1,\ldots,N}
    \Big[
    \hat\rho_0^{(N)}(t)
    (|x_1\rangle\langle x_1|)_1
    \cdots
    (|x_p\rangle\langle x_p|)_p
    \Big].
    \label{eq:occupation_full_trace}
\end{align}
We can now perform first the trace over particles $p+1,\ldots,N$.
Using the definition $\hat\rho_0^{(p)}(t)=\operatorname{Tr}_{p+1,\ldots,N}\hat\rho_0^{(N)}(t)$,
Eq.~\eqref{eq:occupation_full_trace} becomes
\begin{align}
    \left\langle
    \hat n_{x_1}(t)\cdots\hat n_{x_p}(t)
    \right\rangle_0
    &=
    \frac{N!}{(N-p)!}
    \operatorname{Tr}_{1,\ldots,p}
    \left[
    \hat\rho_0^{(p)}(t)
    \bigotimes_{\ell=1}^p
    |x_\ell\rangle\langle x_\ell|
    \right].
    \label{eq:occupation_rhop_trace}
\end{align}
The remaining trace is simply the diagonal matrix element of the
$p$-body reduced density matrix,
\begin{equation}
    \left\langle
    \hat n_{x_1}(t)\cdots\hat n_{x_p}(t)
    \right\rangle_0
    =
    \frac{N!}{(N-p)!}
    \langle x_1,\ldots,x_p|
    \hat\rho_0^{(p)}(t)
    |x_1,\ldots,x_p\rangle.
    \label{eq:occupation_rhop}
\end{equation}
We can now directly use
Eq.~\eqref{eq:pbody_det}. Setting $y_\ell=x_\ell$ for every
$\ell=1,\ldots,p$ in Eq.~\eqref{eq:pbody_det} and comparing it with Eq.~\eqref{eq:occupation_rhop}, we find
\begin{equation}
    \left\langle
    \hat n_{x_1}(t)\cdots\hat n_{x_p}(t)
    \right\rangle_0
    =
    \det_{1\leq \ell,\ell'\leq p}
    \left[
    \Gamma_{x_\ell x_{\ell'}}(t)
    \right],
    \qquad
    x_\ell\neq x_{\ell'}
    \quad\text{for }\ell\neq \ell',
    \label{eq:occupation_det}
\end{equation}
where $\Gamma_{x,y}(t)$ is given in Eq.~\eqref{eq:Gamma_def}.
Thus, all equal-time $p$-point occupation correlations are completely determined by the one-body correlation kernel $\Gamma_{xy}(t)$. This recovers the well-known result that the positions of noninteracting fermions at fixed time form a determinantal point process, with correlation kernel $\Gamma_{xy}(t)$~\cite{Hough2009,Borodin2011,Dean2016}.

\subsection{Extreme-value statistics}\label{sec:EVSnoreset}

We now study the position of the rightmost fermion. Let $X_{\max}(t)$
denote its position at time $t$. For an integer $M$, the cumulative
distribution of $X_{\max}(t)$ is
\begin{equation}
    \mathbb P_0\left[X_{\max}(t)\leq M\right].
    \label{eq:Xmax_cdf_def}
\end{equation}
The event $X_{\max}(t)\leq M$ means that there are no fermions on any
site to the right of $M$. We therefore define the set
\begin{equation}
    A_M=\{M+1,M+2,\ldots\}.
    \label{eq:AM_def}
\end{equation}
For spinless fermions, the local occupation number at a given site can only be $0$ or $1$. Therefore, on the fermionic Hilbert space, $\hat n_x^2=\hat n_x$, and $1-\hat n_x$ is the projector onto the state
in which site $x$ is empty. Consequently, the product
$\prod_{x>M}(1-\hat n_x)$ projects onto configurations in which every
site in $A_M$ is empty. Hence
\begin{equation}
    \mathbb P_0\left[X_{\max}(t)\leq M\right]
    =
    \left\langle
    \prod_{x=M+1}^{\infty}(1-\hat n_x)
    \right\rangle_0.
    \label{eq:max_projector}
\end{equation}
We now expand the product as
\begin{align}
    \mathbb P_0\left[X_{\max}(t)\leq M\right]
    &=
    1
    -\sum_{x_1>M}\langle\hat n_{x_1}\rangle_0
    +\sum_{M<x_1<x_2}
    \langle\hat n_{x_1}\hat n_{x_2}\rangle_0
    -\sum_{M<x_1<x_2<x_3}
    \langle\hat n_{x_1}\hat n_{x_2}\hat n_{x_3}\rangle_0
    +\cdots .
    \label{eq:max_expand}
\end{align}
Using Eq.~\eqref{eq:occupation_det}, each $p$-point correlation function
is a determinant,
\begin{equation}
    \left\langle
    \hat n_{x_1}\cdots\hat n_{x_p}
    \right\rangle_0
    =
    \det_{1\leq \ell,\ell'\leq p}
    \left[\Gamma_{x_\ell x_{\ell'}}(t)\right].
    \label{eq:max_pcorrelation}
\end{equation}
The determinant is unchanged under a simultaneous permutation of its
rows and columns. Therefore, for any permutation $\pi\in S_p$,
\begin{equation}
    \det\left[
    \Gamma_{x_{\pi(\ell)}x_{\pi(\ell')}}(t)
    \right]
    =
    \det\left[
    \Gamma_{x_\ell x_{\ell'}}(t)
    \right].
\end{equation}
Hence, each unordered set of $p$ distinct sites appears $p!$ times when
we sum independently over their possible orderings. We can therefore write
\begin{align}
    &\sum_{M<x_1<\cdots<x_p}
    \det_{1\leq \ell,\ell'\leq p}
    \left[
    \Gamma_{x_\ell x_{\ell'}}(t)
    \right]=
    \frac{1}{p!}
    \sum_{\substack{x_1,\ldots,x_p>M\\
    \text{all distinct}}}
    \det_{1\leq \ell,\ell'\leq p}
    \left[
    \Gamma_{x_\ell x_{\ell'}}(t)
    \right].
    \label{eq:ordered_to_unordered}
\end{align}
Moreover, the restriction that the sites be distinct can be removed.
Indeed, if $x_\ell=x_{\ell'}$ for some $\ell\neq\ell'$, the corresponding
matrix has two identical rows (and columns), and its determinant
vanishes. Therefore,
\begin{equation}
    \sum_{\substack{x_1,\ldots,x_p>M\\
    \text{all distinct}}}
    \det\left[
    \Gamma_{x_\ell x_{\ell'}}(t)
    \right]
    =
    \sum_{x_1,\ldots,x_p>M}
    \det\left[
    \Gamma_{x_\ell x_{\ell'}}(t)
    \right].
    \label{eq:remove_distinct_constraint}
\end{equation}
Using these two observations, Eq.~\eqref{eq:max_expand} can be rewritten as
\begin{equation}
    \mathbb P_0\left[X_{\max}(t)\leq M\right]
    =
    1+
    \sum_{p=1}^{\infty}
    \frac{(-1)^p}{p!}
    \sum_{x_1,\ldots,x_p>M}
    \det_{1\leq \ell,\ell'\leq p}
    \left[\Gamma_{x_\ell x_{\ell'}}(t)\right].
    \label{eq:fredholm_series}
\end{equation}
Equation~\eqref{eq:fredholm_series} is precisely the Fredholm expansion
of the determinant of the correlation kernel restricted to the set
$A_M$. Denoting by $\hat \Gamma_{A_M}(t)$ the operator with matrix elements
$\Gamma_{xy}(t)$ for $x,y\in A_M$, we obtain
\begin{equation}
    \mathbb P_0\left[X_{\max}(t)\leq M\right]
    =
    \det\left(I-\hat \Gamma_{A_M}(t)\right).
    \label{eq:fredholm}
\end{equation}
Thus the distribution of the rightmost fermion is completely
determined by the one-body correlation kernel.

We now study Eq.~\eqref{eq:fredholm} close to the ballistic edge. With a domain-wall initial condition, using Eqs.~\eqref{eq:Gamma_def} and
\eqref{eq:orbital_bessel}, the kernel can be written as
\begin{align}
    \Gamma_{xy}(t)
    &=
    \sum_{m=1}^{\infty}
    \phi_m(x,t)\phi_m^*(y,t)=
    i^{x-y}
    \sum_{\ell=0}^{\infty}
    J_{x+\ell}(2\gamma t)J_{y+\ell}(2\gamma t),
    \label{eq:Gamma_bessel_edge}
\end{align}
where we set $\ell=m-1$. Since we are interested in the position of the rightmost fermion, we focus on the region near the right edge of the propagating front, located at $x\simeq 2\gamma t$. As shown in Sec.~\ref{sec:finitetimecorrnoreset}, the front edge broadens over a characteristic scale $(\gamma t)^{1/3}$, which therefore sets the natural scale for the fluctuations of the rightmost fermion. We thus introduce the scaling variables
\begin{equation}
    x=2\gamma t+(\gamma t)^{1/3}u,
    \qquad
    y=2\gamma t+(\gamma t)^{1/3}v,
    \label{eq:edge_xy_scaling}
\end{equation}
where $u,v=O(1)$ are continuous variables which describe positions within the edge scaling region.
The sum in Eq.~\eqref{eq:Gamma_bessel_edge} is therefore dominated by
$\ell=O((\gamma t)^{1/3})$. We therefore write $\ell=(\gamma t)^{1/3}w$ and $\Delta w=\frac{1}{(\gamma t)^{1/3}}$.
In this regime we can use the asymptotic of the Bessel function derived in Sec.~\ref{sec:finitetimecorrnoreset} and given in Eq.~\eqref{eq:bessel_airy}, namely
\begin{equation}
    J_{x+\ell}(2\gamma t)
    \simeq
    \frac{1}{(\gamma t)^{1/3}}
    \operatorname{Ai}(u+w),
    \qquad
    J_{y+\ell}(2\gamma t)
    \simeq
    \frac{1}{(\gamma t)^{1/3}}
    \operatorname{Ai}(v+w).
    \label{eq:bessel_edge_kernel}
\end{equation}
Substituting these expressions into
Eq.~\eqref{eq:Gamma_bessel_edge} we obtain
\begin{align}
    \Gamma_{xy}(t)
    &\simeq
    \frac{i^{x-y}}{(\gamma t)^{2/3}}
    \sum_{\ell=0}^{\infty}
    \operatorname{Ai}\left(u+\frac{\ell}{(\gamma t)^{1/3}}\right)
    \operatorname{Ai}\left(v+\frac{\ell}{(\gamma t)^{1/3}}\right).
    \label{eq:Gamma_edge_sum}
\end{align}
For large $t$, the sum becomes an integral. Since
$\Delta w=(\gamma t)^{-1/3}$,
\begin{equation}
    \sum_{\ell=0}^{\infty}
    \longrightarrow
    (\gamma t)^{1/3}\int_0^\infty\mathrm d w,
\end{equation}
and therefore
\begin{equation}
    \Gamma_{xy}(t)
    \simeq
    \frac{i^{x-y}}{(\gamma t)^{1/3}}
    K_{\mathrm{Ai}}(u,v),
    \label{eq:Gamma_Airy}
\end{equation}
where
\begin{equation}
    K_{\mathrm{Ai}}(u,v)
    =
    \int_0^\infty\mathrm d w\,
    \operatorname{Ai}(u+w)\operatorname{Ai}(v+w)
    \label{eq:Airy_kernel}
\end{equation}
is the Airy kernel~\cite{TracyWidom1994,Eisler2013}. The phase factor $i^{x-y}$ in Eq.~\eqref{eq:Gamma_Airy} can be absorbed into a diagonal unitary
transformation, $\Gamma=UKU^\dagger$, where
$U_{xy}=i^x\delta_{xy}$. Since such a transformation leaves
the determinant invariant, this phase factor can be neglected.

To study the rightmost particle on the same scale, we set
\begin{equation}
    M=2\gamma t+(\gamma t)^{1/3}s.
    \label{eq:M_edge_scaling}
\end{equation}
Under the scaling in Eqs.~\eqref{eq:edge_xy_scaling} and
\eqref{eq:M_edge_scaling}, the discrete Fredholm determinant in
Eq.~\eqref{eq:fredholm} converges to the Fredholm determinant~\cite{Eisler2013,TracyWidom1994} of the
Airy kernel on the interval $(s,\infty)$,
\begin{equation}
    \mathbb P_0\left[
    X_{\max}(t)\leq
    2\gamma t+(\gamma t)^{1/3}s
    \right]
    \simeq
    \det\left(I-K_{\mathrm{Ai}}\right)_{L^2(s,\infty)}.
    \label{eq:Xmax_Airy_det}
\end{equation}
The Fredholm determinant on the right-hand side is the Tracy--Widom
distribution for $\beta=2$, conventionally denoted by $F_2(s)$:
\begin{equation}
    F_2(s)
    \equiv
    \det\left(I-K_{\mathrm{Ai}}\right)_{L^2(s,\infty)}.
    \label{eq:F2_fredholm}
\end{equation}
Hence
\begin{equation}
    \mathbb P_0\left[
    X_{\max}(t)\leq
    2\gamma t+(\gamma t)^{1/3}s
    \right]
    \simeq
    F_2(s).
    \label{eq:TW_cdf}
\end{equation}
Defining the Tracy--Widom probability density
$f_0(s)=F_2'(s)$, we finally obtain
\begin{equation}
    \mathbb P_0\left[X_{\max}(t)=x\right]
    \simeq
    \frac{1}{(\gamma t)^{1/3}}
    f_0\left(
    \frac{x-2\gamma t}{(\gamma t)^{1/3}}
    \right).
    \label{eq:TW_density}
\end{equation}
This reproduces Eq.~(14) of the main text.
The Tracy--Widom cumulative distribution can also be written in terms of the Hastings--McLeod solution of the Painlev\'e II equation,
\begin{equation}
    F_2(s)
    =
    \exp\left[
    -\int_s^\infty\mathrm d u\,(u-s)q^2(u)
    \right],
    \label{eq:TW_F2}
\end{equation}
where
\begin{equation}
    q''(u)=u q(u)+2q^3(u),
    \qquad
    q(u)\simeq\operatorname{Ai}(u)
    \quad\text{as }u\to+\infty.
    \label{eq:painleve}
\end{equation}
The corresponding asymptotic tails of $f_0(s)$ are
\begin{equation}
    f_0(s)\simeq
    \begin{cases}
        \displaystyle
        \frac{\tau_2}{4}|s|^{15/8}
        \exp\left(-\frac{|s|^3}{12}\right),
        & s\to-\infty,\\[3mm]
        \displaystyle
        \frac{1}{8\pi s}
        \exp\left(-\frac{4}{3}s^{3/2}\right),
        & s\to+\infty,
    \end{cases}
    \qquad
    \tau_2=2^{1/24}e^{\zeta'(-1)},
    \label{eq:TW_tails}
\end{equation}
where $\zeta'(-1)$ denotes the derivative of the Riemann zeta function
$\zeta(s)$ evaluated at $s=-1$.

\subsection{Integrated current}

In this section, we derive the mean and variance of the integrated current. These quantities were first studied using a second-quantized formalism in Ref.~\cite{AKR2008}. Here, we rederive the corresponding results within our first-quantized framework, which will also be useful below when studying the effects of uniform-memory resetting.

The integrated current measures the net number of fermions that have crossed a given bond up to time $t$, with crossings from left to right counted positively and those from right to left negatively. We consider the bond connecting sites $0$ and $1$. Since all sites $x\geq1$ are initially empty in the domain-wall configuration, the integrated current is simply equal to the total number of fermions found on the positive half-line at time $t$. It can therefore be written as
\begin{equation}
    \hat C=\sum_{x=1}^{\infty}\hat n_x.
    \label{eq:C0_def}
\end{equation}
Taking the expectation value with respect to the reset-free many-body density matrix $\hat\rho_0^{(N)}(t)$, we obtain
\begin{equation}
    \langle\hat C(t)\rangle_0
    =
    \operatorname{Tr}_{1,\ldots,N}
    \left[\hat\rho_0^{(N)}(t)\hat C\right]
    =
    \sum_{x=1}^{\infty}
    \langle\hat n_x(t)\rangle_0.
    \label{eq:C_mean_density}
\end{equation}
At large times, we can use the bulk density profile derived in Eq.~\eqref{eq:bulk_scaling},
$\langle\hat n_x(t)\rangle_0\approx g\left(\frac{x}{2\gamma t}\right)$,
where $g(z)$ is given in Eq.~\eqref{eq:g_bulk}. Since the density vanishes for $x>2\gamma t$, and the front scale becomes large as $t\to\infty$, the sum in Eq.~\eqref{eq:C_mean_density} can be replaced by an integral,
\begin{align}
    \langle\hat C(t)\rangle_0
    &\simeq
    \int_0^{2\gamma t}
    \mathrm dx\,
    g\left(\frac{x}{2\gamma t}\right)=
    2\gamma t
    \int_0^1\mathrm dz\,g(z).
\end{align}
Using $g(z)=\pi^{-1}\arccos z$ for $0<z<1$, we obtain
\begin{align}
    \langle\hat C(t)\rangle_0
    &\approx
    \frac{2\gamma t}{\pi}
    \int_0^1\mathrm dz\,\arccos z=
    \frac{2\gamma t}{\pi}.
    \label{eq:C_mean_asympt}
\end{align}
Thus the mean integrated current grows linearly with time. This agrees with the result obtained for the domain-wall XX chain in Ref.~\cite{AKR2008}.

\subsubsection{Variance}

We now compute the fluctuations of the integrated current. By
definition,
\begin{equation}
    \operatorname{Var}_0[\hat C(t)]
    =
    \langle\hat C^2(t)\rangle_0
    -
    \langle\hat C(t)\rangle_0^2.
    \label{eq:C_var_def}
\end{equation}
Using Eq.~\eqref{eq:C0_def},
\begin{equation}
    \hat C^2(t)
    =
    \sum_{x,y\geq1}
    \hat n_x(t)\hat n_y(t),
\end{equation}
where the operator $\hat n_x(t)$ is defined in Eq.~\eqref{eq:mx_def}.
We separate the terms with $x=y$ from those with $x\neq y$. Since the
local occupation operator satisfies $\hat n_x^2=\hat n_x$, we obtain
\begin{align}
    \langle\hat C^2(t)\rangle_0
    &=
    \sum_{x\geq1}
    \langle\hat n_x(t)\rangle_0
    +
    \sum_{\substack{x,y\geq1\\x\neq y}}
    \langle\hat n_x(t)\hat n_y(t)\rangle_0=
    \langle\hat C(t)\rangle_0
    +
    \sum_{\substack{x,y\geq1\\x\neq y}}
    \langle\hat n_x(t)\hat n_y(t)\rangle_0.
    \label{eq:C2_split}
\end{align}
For two distinct sites, Eq.~\eqref{eq:occupation_det} with $p=2$ gives
\begin{align}
    \langle\hat n_x(t)\hat n_y(t)\rangle_0
    &=
    \det
    \begin{pmatrix}
        \Gamma_{xx}(t) & \Gamma_{xy}(t)\\
        \Gamma_{yx}(t) & \Gamma_{yy}(t)
    \end{pmatrix}=
    \Gamma_{xx}(t)\Gamma_{yy}(t)
    -
    \Gamma_{xy}(t)\Gamma_{yx}(t),
\end{align}
where $\Gamma_{xy}(t)$ is given in Eq.~\eqref{eq:Gamma_def}. 
Since the correlation kernel $\Gamma_{xy}(t)$ is Hermitian,
$\Gamma_{yx}(t)=\Gamma_{xy}^*(t)$, and therefore
\begin{equation}
    \langle\hat n_x(t)\hat n_y(t)\rangle_0
    =
    \Gamma_{xx}(t)\Gamma_{yy}(t)
    -
    |\Gamma_{xy}(t)|^2,
    \qquad x\neq y.
    \label{eq:nxny_kernel}
\end{equation}
Moreover, from Eqs.~\eqref{eq:mx_rho1} and \eqref{eq:Gamma_def}, we have $\langle\hat n_x(t)\rangle_0=\Gamma_{xx}(t)$.
Substituting Eq.~\eqref{eq:nxny_kernel} into Eq.~\eqref{eq:C2_split},
\begin{equation}
    \langle\hat C^2(t)\rangle_0
    =
    \sum_{x\geq1}\Gamma_{xx}(t)
    +
    \sum_{\substack{x,y\geq1\\x\neq y}}
    \left[
    \Gamma_{xx}(t)\Gamma_{yy}(t)
    -
    |\Gamma_{xy}(t)|^2
    \right].
    \label{eq:C2_kernel}
\end{equation}
On the other hand,
\begin{align}
    \langle\hat C(t)\rangle_0^2
    &=
    \left[
    \sum_{x\geq1}\Gamma_{xx}(t)
    \right]^2=
    \sum_{x\geq1}\Gamma_{xx}^2(t)
    +
    \sum_{\substack{x,y\geq1\\x\neq y}}
    \Gamma_{xx}(t)\Gamma_{yy}(t).
    \label{eq:Cmean_square_kernel}
\end{align}
Subtracting Eq.~\eqref{eq:Cmean_square_kernel} from
Eq.~\eqref{eq:C2_kernel}, the terms
$\Gamma_{xx}\Gamma_{yy}$ with $x\neq y$ cancel, giving
\begin{align}
    \operatorname{Var}_0[\hat C(t)]
    &=
    \sum_{x\geq1}
    \left[
    \Gamma_{xx}(t)-\Gamma_{xx}^2(t)
    \right]
    -
    \sum_{\substack{x,y\geq1\\x\neq y}}
    |\Gamma_{xy}(t)|^2=
    \sum_{x\geq1}\Gamma_{xx}(t)
    -
    \sum_{x,y\geq1}|\Gamma_{xy}(t)|^2.
    \label{eq:C_var_intermediate}
\end{align}
We can simplify this expression using the fact that $\hat \Gamma$ is a
projector. Indeed, using Eqs.~\eqref{eq:rho1_result} and \eqref{eq:Gamma_def}, one has
\begin{equation}
    \hat\Gamma(t)=\sum_{m=1}^N
    |\phi_m(t)\rangle\langle\phi_m(t)|.
\end{equation}
Using the orthonormality relation in
Eq.~\eqref{eq:orbital_orthogonality}, we obtain
\begin{align}
    \hat\Gamma^2(t)
    &=
    \sum_{m,m'=1}^N
    |\phi_m(t)\rangle
    \langle\phi_m(t)|\phi_{m'}(t)\rangle
    \langle\phi_{m'}(t)|=
    \sum_{m=1}^N
    |\phi_m(t)\rangle\langle\phi_m(t)|
    =
    \hat\Gamma(t).
    \label{eq:Gamma_projector}
\end{align}
Taking the diagonal matrix element $\langle x|\cdots|x\rangle$ of Eq.~\eqref{eq:Gamma_projector} gives
\begin{align}
    \Gamma_{xx}(t)
    &=
    \langle x|\hat\Gamma^2(t)|x\rangle
    =
    \sum_{y\in\mathbb Z}
    \langle x|\hat\Gamma(t)|y\rangle
    \langle y|\hat\Gamma(t)|x\rangle=
    \sum_{y\in\mathbb Z}
    |\Gamma_{xy}(t)|^2.
    \label{eq:Gamma_square_matrix}
\end{align}
Therefore
\begin{equation}
    \sum_{x\geq1}\Gamma_{xx}(t)
    =
    \sum_{x\geq1}
    \sum_{y\in\mathbb Z}
    |\Gamma_{xy}(t)|^2.
    \label{eq:Cmean_kernel}
\end{equation}
Substituting Eq.~\eqref{eq:Cmean_kernel} into
Eq.~\eqref{eq:C_var_intermediate}, the contribution with $y\geq1$
cancels and we are left with
\begin{equation}
    \operatorname{Var}_0[\hat C(t)]
    =
    \sum_{x\geq1}
    \sum_{y\leq0}
    |\Gamma_{xy}(t)|^2.
    \label{eq:C_var_cross_Gamma}
\end{equation}
Since $\Gamma_{xy}(t)=N\rho_0^{(1)}(x,y;t)$ by
Eq.~\eqref{eq:Gamma_def}, this can equivalently be written as
\begin{equation}
    \operatorname{Var}_0[\hat C(t)]
    =
    N^2
    \sum_{x\geq1}
    \sum_{y\leq0}
    |\rho_0^{(1)}(x,y;t)|^2.
    \label{eq:C_var_cross}
\end{equation}
We now evaluate the kernel explicitly for the domain-wall initial condition. From Eq.~\eqref{eq:Gamma_def} and Eq.~\eqref{eq:orbital_bessel},
\begin{align}
    \Gamma_{xy}(t)
    &=
    \sum_{m=1}^{\infty}
    \phi_m(x,t)\phi_m^*(y,t)=
    i^{x-y}
    \sum_{m=1}^{\infty}
    J_{x+m-1}(2\gamma t)J_{y+m-1}(2\gamma t).
\end{align}
Setting $\ell=m-1$, we obtain
\begin{equation}
    \Gamma_{xy}(t)
    =
    N\rho_0^{(1)}(x,y;t)
    =
    i^{x-y}
    \sum_{\ell=0}^{\infty}
    J_{x+\ell}(2\gamma t)J_{y+\ell}(2\gamma t).
    \label{eq:kernel_bessel_sum}
\end{equation}
The sum can be evaluated using the  identity~\cite{AKR2008,GradshteynRyzhik2007}
\begin{equation}
    \sum_{\ell=0}^{\infty}
    J_{\ell+x}(z)J_{\ell+y}(z)
    =
    \frac{z}{2(x-y)}
    \left[
    J_{x-1}(z)J_y(z)
    -
    J_x(z)J_{y-1}(z)
    \right].
    \label{eq:lommel}
\end{equation}
To substitute this result into Eq.~\eqref{eq:C_var_cross_Gamma}, we
write $x=\ell$, $y=1-\ell'$ with $\ell,\ell'\geq1$,
so that $x\geq1$ and $y\leq0$. We then have
$x-y=\ell+\ell'-1$, and Eq.~\eqref{eq:lommel} gives
\begin{align}
    \Gamma_{\ell,1-\ell'}(t)
    &=
    i^{\ell+\ell'-1}
    \frac{2\gamma t}{2(\ell+\ell'-1)}
    \Big[
    J_{\ell-1}(2\gamma t)J_{1-\ell'}(2\gamma t)
    -
    J_\ell(2\gamma t)J_{-\ell'}(2\gamma t)
    \Big].
    \label{eq:kernel_lommel_step}
\end{align}
Using the identities $J_{1-\ell'}(z)=(-1)^{\ell'-1}J_{\ell'-1}(z)$ and $J_{-\ell'}(z)=(-1)^{\ell'}J_{\ell'}(z)$, (see 10.4.1 in Ref.~\cite{NIST})
the expression in square brackets becomes
\begin{equation}
    (-1)^{\ell'-1}
    \left[
    J_{\ell-1}(2\gamma t)J_{\ell'-1}(2\gamma t)
    +
    J_\ell(2\gamma t)J_{\ell'}(2\gamma t)
    \right].
\end{equation}
Considering the modulus squared yields
\begin{equation}
    |\Gamma_{\ell,1-\ell'}(t)|^2
    =
    \frac{(2\gamma t)^2}{4}
    \left[
    \frac{
    J_{\ell-1}(2\gamma t)J_{\ell'-1}(2\gamma t)
    +
    J_\ell(2\gamma t)J_{\ell'}(2\gamma t)
    }{\ell+\ell'-1}
    \right]^2.
    \label{eq:kernel_lommel}
\end{equation}
Substituting this expression into
Eq.~\eqref{eq:C_var_cross_Gamma}, we obtain
\begin{equation}
    \operatorname{Var}_0[\hat C(t)]
    =
    \frac{(2\gamma t)^2}{4}
    \sum_{\ell,\ell'\geq1}
    \left[
    \frac{
    J_{\ell-1}(2\gamma t)J_{\ell'-1}(2\gamma t)
    +
    J_\ell(2\gamma t)J_{\ell'}(2\gamma t)
    }{\ell+\ell'-1}
    \right]^2.
    \label{eq:C_var_bessel}
\end{equation}
In Ref.~\cite{AKR2008} it is shown that a detailed asymptotic analysis of
Eq.~\eqref{eq:C_var_bessel} gives, for $\gamma t\gg1$,
\begin{equation}
    \operatorname{Var}_0[\hat C(t)]
    \simeq
    \frac{1}{2\pi^2}
    \left[
    \ln(2\gamma t)+\bar c
    \right],
    \qquad
    \bar c\simeq2.963\ldots.
    \label{eq:C_var_log}
\end{equation}
Together with Eq.~\eqref{eq:C_mean_asympt}, this reproduces
Eq.~(17) of the main text.
Thus, while the mean integrated current grows linearly with time,
$\langle\hat C(t)\rangle_0\sim t$, its variance grows only
logarithmically, $\operatorname{Var}_0[\hat C(t)]\sim\ln t$.

\section{Dynamics with uniform-memory resetting}
\label{sec:uniform_memory_resetting}

We now introduce uniform-memory resetting into the $N$-fermion problem. As discussed in the main text, the reset-averaged many-body density matrix obeys
\begin{equation}
    \frac{\mathrm{d}\hat\rho_r^{(N)}(t)}{\mathrm{d}t}
    =-i[\hat{\mathcal H},\hat\rho_r^{(N)}(t)]
    -r\hat\rho_r^{(N)}(t)
    +\frac{r}{t}\int_0^t\mathrm{d}\tau\,\hat\rho_r^{(N)}(\tau).
    \label{eq:memory_master_recalled}
\end{equation}
As shown in the End Matter (EM), the solution of Eq.~\eqref{eq:memory_master_recalled} can be expressed as
\begin{equation}
    \hat\rho_r^{(N)}(t)
    =\int_0^t\mathrm{d}q\,P(q,t)\hat\rho_0^{(N)}(q),
    \label{eq:subordination}
\end{equation}
where $P(q,t)$ is the probability density of the effective time $q$, and $\hat\rho_0^{(N)}(q)$ is the density matrix of the corresponding reset-free system, evolving from the same initial condition. In particular, $\hat\rho_0^{(N)}(q)$ satisfies the von Neumann equation
\begin{equation}
    \frac{\mathrm{d}\hat\rho_0^{(N)}(q)}{\mathrm{d}q}
    =-i[\hat{\mathcal H},\hat\rho_0^{(N)}(q)],
    \label{eq:vNeq}
\end{equation}
which describes the purely unitary evolution obtained by setting $r=0$. 

The representation in Eq.~\eqref{eq:subordination} is a key property that makes the system analytically tractable. Indeed, computing many-body observables is often challenging even in the absence of resetting. However, once an observable is known for the reset-free system, its expectation value in the presence of uniform-memory resetting can be obtained exactly at any time $t$ by averaging the corresponding reset-free result over the effective-time distribution $P(q,t)$. In particular, for any observable $\hat O$, we have
\begin{equation}
    \langle\hat O(t)\rangle_r
    =
    \int_0^t \mathrm{d}q\,P(q,t)
    \langle\hat O(q)\rangle_0.
    \label{eq:observable_subordination}
\end{equation}
Thus, the problem with resetting can be reduced to the corresponding reset-free problem, followed by an average over the effective time.

Furthermore, taking the partial trace over the last $N-p$ particles in both Eqs.~\eqref{eq:memory_master_recalled} and \eqref{eq:subordination}, we find that the same relations hold for any $p$-body reduced density matrix. Indeed, since the particles are noninteracting, the partial trace of the Hamiltonian commutator involves only the Hamiltonian acting on the remaining $p$ particles. Defining $\hat\rho_r^{(p)}(t)=\operatorname{Tr}_{p+1,\ldots,N}\hat\rho_r^{(N)}(t)$, we obtain
\begin{equation}
    \frac{\mathrm{d}\hat\rho_r^{(p)}(t)}{\mathrm{d}t}
    =
    -i[\hat{\mathcal H}^{(p)},\hat\rho_r^{(p)}(t)]
    -r\hat\rho_r^{(p)}(t)
    +\frac{r}{t}\int_0^t\mathrm{d}\tau\,\hat\rho_r^{(p)}(\tau),
    \label{eq:memory_master_p}
\end{equation}
where $\hat{\mathcal H}^{(p)}=\sum_{m=1}^p\hat H_m$.
Similarly, the subordination relation becomes
\begin{equation}
    \hat\rho_r^{(p)}(t)=\int_0^t\mathrm{d}q\,P(q,t)\hat\rho_0^{(p)}(q).
    \label{eq:subordination_p}
\end{equation}
Thus, the same effective-time distribution $P(q,t)$ governs
the evolution of all reduced density matrices, independently
of $p$. A more detailed microscopic description of the effective-time dynamics is given in Sec.~\ref{sec:microscopic_q} (see also Fig.~\ref{fig:q_dynamics}).

As shown in the EM of the main Letter, the Laplace transform of $P(q,t)$, namely
\begin{equation}\label{eq:fdefinition}
    f(s,t)\equiv
    \int_0^\infty \mathrm{d}q\,
    e^{-s q}P(q,t),
\end{equation}
is known exactly at all times and reads
\begin{equation}
    f(s,t)
    =
    M\left(
    \frac{s}{r+s};
    1;
    -(r+s)t
    \right),
    \label{eq:f_characteristic}
\end{equation}
where $M(a;b;z)$ denotes Kummer's confluent hypergeometric function.

We now show that, as stated in the main Letter, at large times $P(q,t)$ approaches a Gaussian distribution with mean $\langle q\rangle\simeq T(t)$ and variance $\operatorname{Var}(q)\simeq\frac{2T(t)}{r}$, where
$T(t)=\frac{\ln(rt)}{r}$.
For $rt\gg1$, we use the large-argument asymptotic behavior [see (13.7.2) in Ref.~\cite{NIST}]
\begin{equation}
    M(a;1;-z)
    \simeq
    \frac{z^{-a}}{\Gamma(1-a)},
    \qquad
    |z|\to\infty,
\end{equation}
with $a=\frac{s}{r+s}$ and $z=(r+s)t$.
Equation~\eqref{eq:f_characteristic} therefore becomes
\begin{equation}
    f(s,t)
    \simeq
    \frac{
    \exp\left[
    -\dfrac{s}{r+s}
    \ln\left((r+s)t\right)
    \right]
    }{
    \Gamma\left(
    1-\dfrac{s}{r+s}
    \right)
    }.
    \label{eq:f_large_t}
\end{equation}
To determine the large $t$ behavior of $P(q,t)$, we expand the logarithm of the Laplace transform for small $s$. We first have
\begin{equation}
    \frac{s}{r+s}
    =
    \frac{s}{r}
    -
    \frac{s^2}{r^2}
    +
    O(s^3),
\end{equation}
and
\begin{equation}
    \ln\left[(r+s)t\right]\approx
    \ln(rt)
    +
    \frac{s}{r}
    -
    \frac{s^2}{2r^2}.
\end{equation}
Moreover, using
\begin{equation}
    \ln\Gamma(1-a)\approx
    \gamma_{\rm E}a
    +
    \frac{\pi^2}{12}a^2,
\end{equation}
where $\gamma_{\rm E}$ is the Euler--Mascheroni constant, we obtain
\begin{align}
    \ln f(s,t)\approx
    -\frac{s}{r}
    \left[
    \ln(rt)+\gamma_{\rm E}
    \right]+
    \frac{s^2}{r^2}
    \left[
    \ln(rt)-1+\gamma_{\rm E}
    -\frac{\pi^2}{12}
    \right].
    \label{eq:logf_expansion}
\end{align}
The mean and variance of $q$ can be obtained from the
derivatives of the logarithm of its Laplace transform,
\begin{equation}
    \langle q\rangle
    =-\left.\frac{\partial \ln f(s,t)}{\partial s}\right|_{s=0},
    \qquad
    \operatorname{Var}(q)
    =\left.\frac{\partial^2 \ln f(s,t)}{\partial s^2}\right|_{s=0}.
\end{equation}
Using Eq.~\eqref{eq:logf_expansion}, we find
\begin{equation}
    \langle q\rangle
    \simeq
    \frac{\ln(rt)+\gamma_{\rm E}}{r},
\end{equation}
and
\begin{equation}
    \operatorname{Var}(q)
    \simeq
    \frac{2}{r^2}
    \left[
    \ln(rt)-1+\gamma_{\rm E}
    -\frac{\pi^2}{12}
    \right].
\end{equation}
Keeping only the leading terms at large $t$, and recalling the definition $T(t)\equiv\frac{\ln(rt)}{r}$, we obtain
\begin{equation}
    \langle q\rangle\simeq T(t),
    \qquad
    \operatorname{Var}(q)\simeq\frac{2T(t)}{r}.
    \label{eq:q_moments}
\end{equation}
Correspondingly, on the scale of the typical fluctuations
$q-T(t)=O(\sqrt{T(t)/r})$, the Laplace transform becomes
\begin{equation}
    f(s,t)
    \simeq
    \exp\left[
    -s T(t)
    +\frac{s^2T(t)}{r}
    \right],
\end{equation}
which is the Laplace transform of a Gaussian random variable with mean $T(t)$ and variance $2T(t)/r$. Therefore,
\begin{equation}
    P(q,t)
    \simeq
    \frac{1}{\sqrt{4\pi T(t)/r}}
    \exp\left[
    -\frac{[q-T(t)]^2}{4T(t)/r}
    \right],
    \qquad rt\gg1.
    \label{eq:P_gaussian_T}
\end{equation}
Thus, at long times, the effective time grows only logarithmically with the physical time, $q\sim \ln(rt)/r$, while its fluctuations have a width of order $\sqrt{\ln(rt)}/r$.

\subsection{Microscopic dynamics of the effective time}
\label{sec:microscopic_q}

We now provide a more detailed description of the stochastic
dynamics underlying the effective time $q(t)$ introduced above.
As discussed in the main Letter, the system evolves unitarily
between resetting events, while at each reset it returns to a
state visited at a randomly selected time in its past history.
Since the Hamiltonian $\hat{\mathcal H}$ is time independent,
the state along any stochastic realization can always be
written as
\begin{equation}
    \hat\rho^{(N)}(t)
    =
    \hat\rho_0^{(N)}(q(t))
    =
    e^{-i\hat{\mathcal H}q(t)}
    \hat\rho^{(N)}(0)
    e^{i\hat{\mathcal H}q(t)},
\end{equation}
where $q(t)$ is a stochastic effective time and
$\hat\rho_0^{(N)}(q)$ denotes the reset-free density matrix
evolved for a time $q$ from the initial condition.

To understand the evolution of $q(t)$, consider a short
physical time interval $\mathrm{d}t$. With probability
$1-r\,\mathrm{d}t$, no resetting event occurs, and the
system evolves unitarily, so that its effective time
increases by $\mathrm{d}t$. With probability
$r\,\mathrm{d}t$, a resetting event occurs, and the system
returns to the state it occupied at a past physical time
$\tau=\eta t$, where $\eta$ is drawn uniformly from $[0,1]$.
The effective time therefore evolves according to
\begin{equation}
    q(t+\mathrm{d}t)
    =
    \begin{cases}
        q(t)+\mathrm{d}t,
        & \text{with probability }1-r\,\mathrm{d}t,\\[2mm]
        q(\eta t),
        & \text{with probability }r\,\mathrm{d}t,
    \end{cases}
    \label{eq:q_stochastic_dynamics}
\end{equation}
with the initial condition $q(0)=0$. Correspondingly, one can write a Fokker–Planck-like equation for the evolution of the probability density $P(q,t)$. In a small time $dt$, the distribution $P(q,t)$ evolves as
\begin{equation}
    \label{eq:Pevol}
    P(q, t+dt)= (1-r dt) P(q- dt, t) + r dt \int_0^1 P(q, \eta t) d\eta.
\end{equation}
The first term refers to the advection due to the first line in the Langevin equation~\eqref{eq:q_stochastic_dynamics}. It refers to the event that if the particle is at $q$ at time $t+dt$, it must have been at $q-dt$ at time $t$ and moved deterministically from $q-dt$ to $q$ in time interval $dt$. The second term in Eq.~\eqref{eq:Pevol} corresponds to the resetting event encoded in the second line of the Langevin equation~\eqref{eq:q_stochastic_dynamics}. If the particle is at $q$ at time $t+dt$ via memory resetting event, it must have been at $q$ in the past epoch $\eta t$ where $\eta$ is drawn uniformly from the interval $\eta \in (0,1)$. Making a change of variable $\eta t =\tau$ and taking the limit $dt\to 0 $ gives the Fokker–Planck-like equation
\begin{equation}\label{eq:supp:Pq_final}
\frac{\partial P(q,t)}{\partial t}=-\frac{\partial P(q,t)}{\partial q}-rP(q,t)+\frac{r}{t}\int_0^t\mathrm{d}\tau\,P(q,\tau)\,, 
\end{equation}
which is Eq.~(6) of the main text and was also derived in Eq.~(B3) of Appendix~B via a different method. The derivation presented here does not require the knowledge of the underlying dynamics of the process. It depends only on the classical stochastic resetting protocol and holds universally for any underlying dynamics, classical or quantum. 

Importantly, at a resetting event the effective time
becomes $q(\eta t)$ rather than $\eta t$, because the
state visited at the selected past time may itself have
already undergone previous resetting events. Thus, the
effective time is not simply the physical time elapsed
since the last reset. Between resets, $q(t)$ grows
linearly with unit slope, whereas at each reset it jumps
to a value previously attained along the same stochastic
realization. Depending on the selected past time, this
jump can be either upward or downward.
\begin{figure}[htbp]
  \centering
  \begin{subfigure}{0.48\textwidth}
    \includegraphics[width=\linewidth]{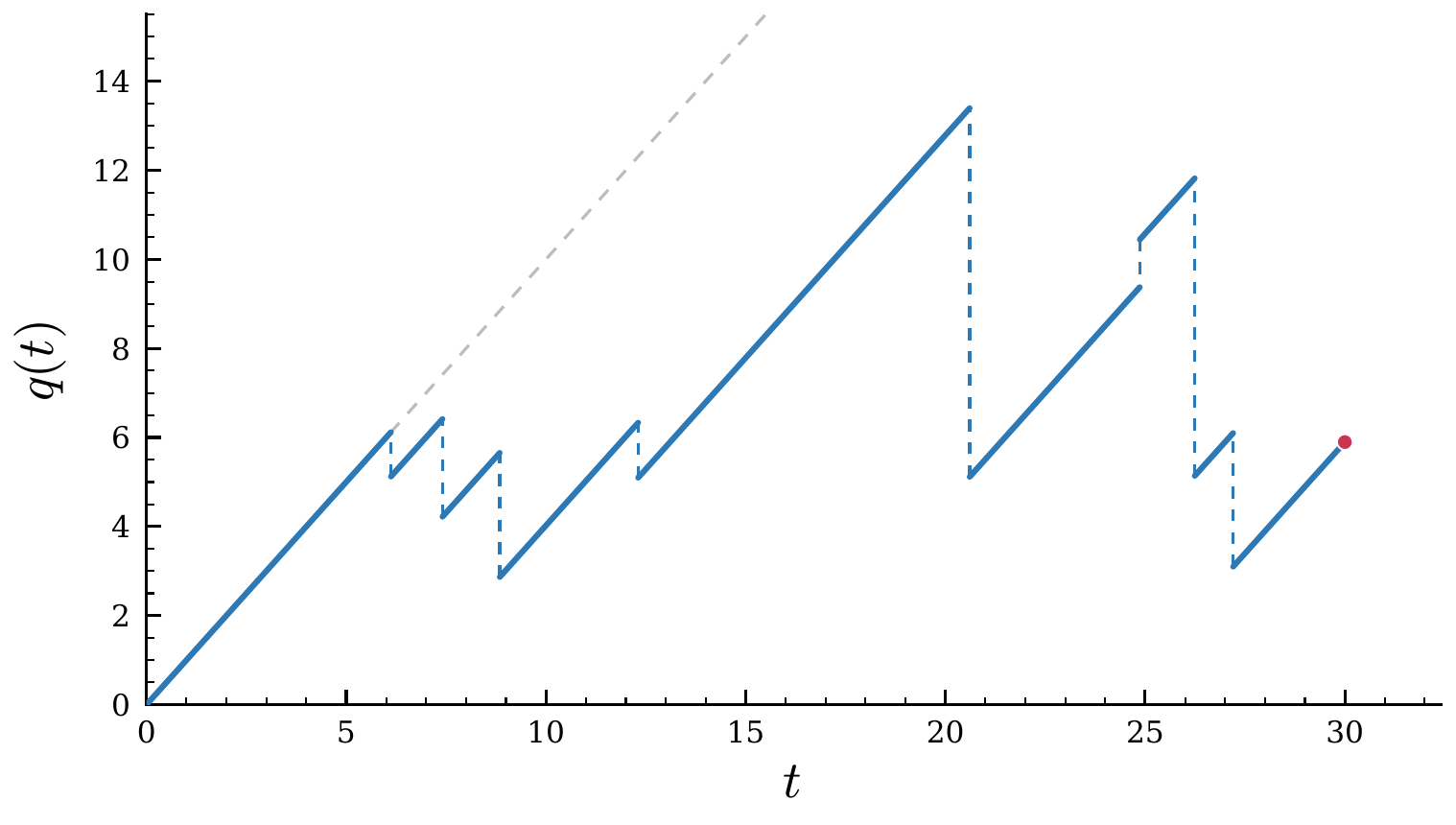}
    \caption{}
  \end{subfigure}\hfill
  \begin{subfigure}{0.48\textwidth}
    \includegraphics[width=\linewidth]{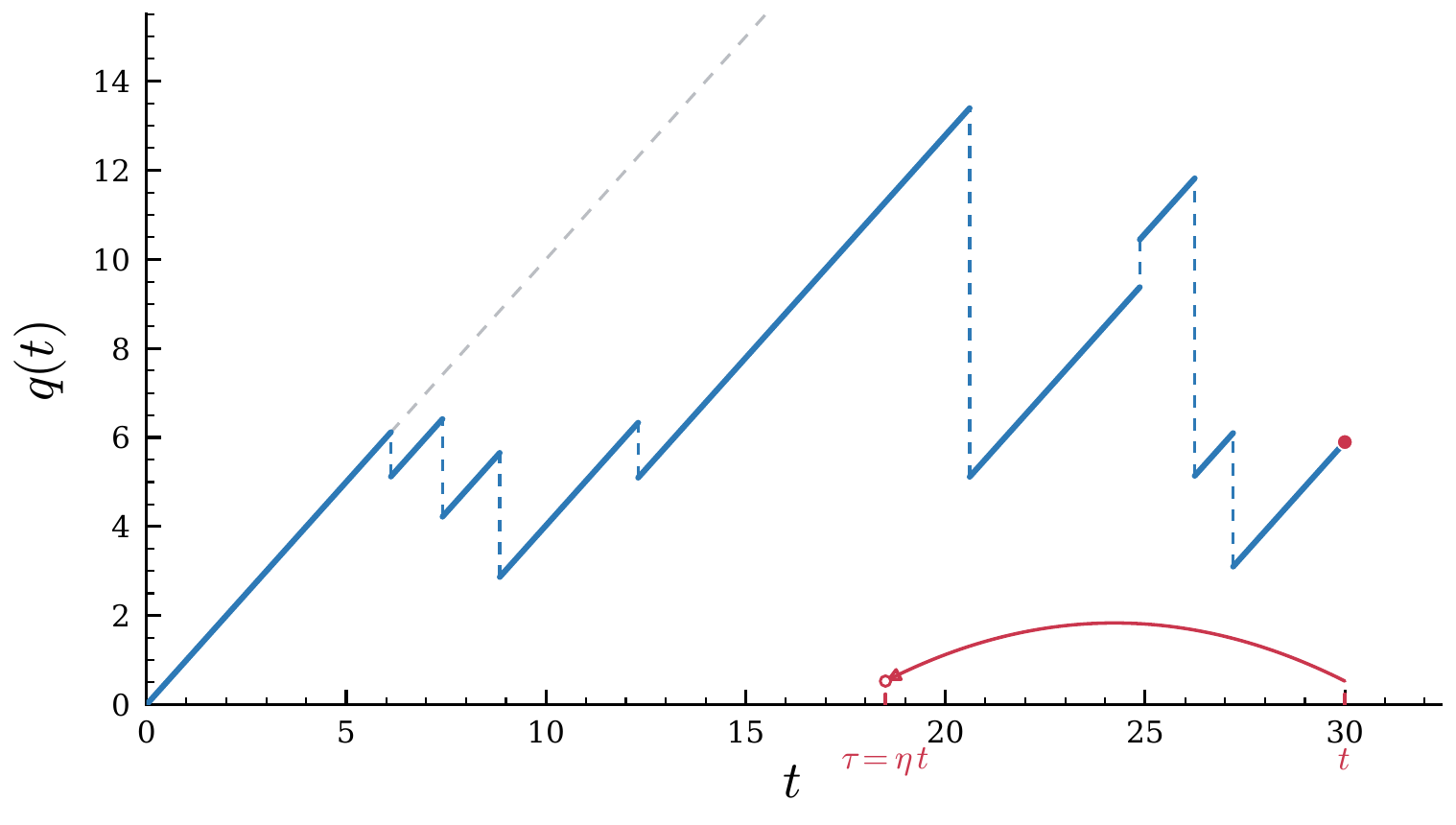}
    \caption{}
  \end{subfigure}

  \medskip
  \begin{subfigure}{0.48\textwidth}
    \includegraphics[width=\linewidth]{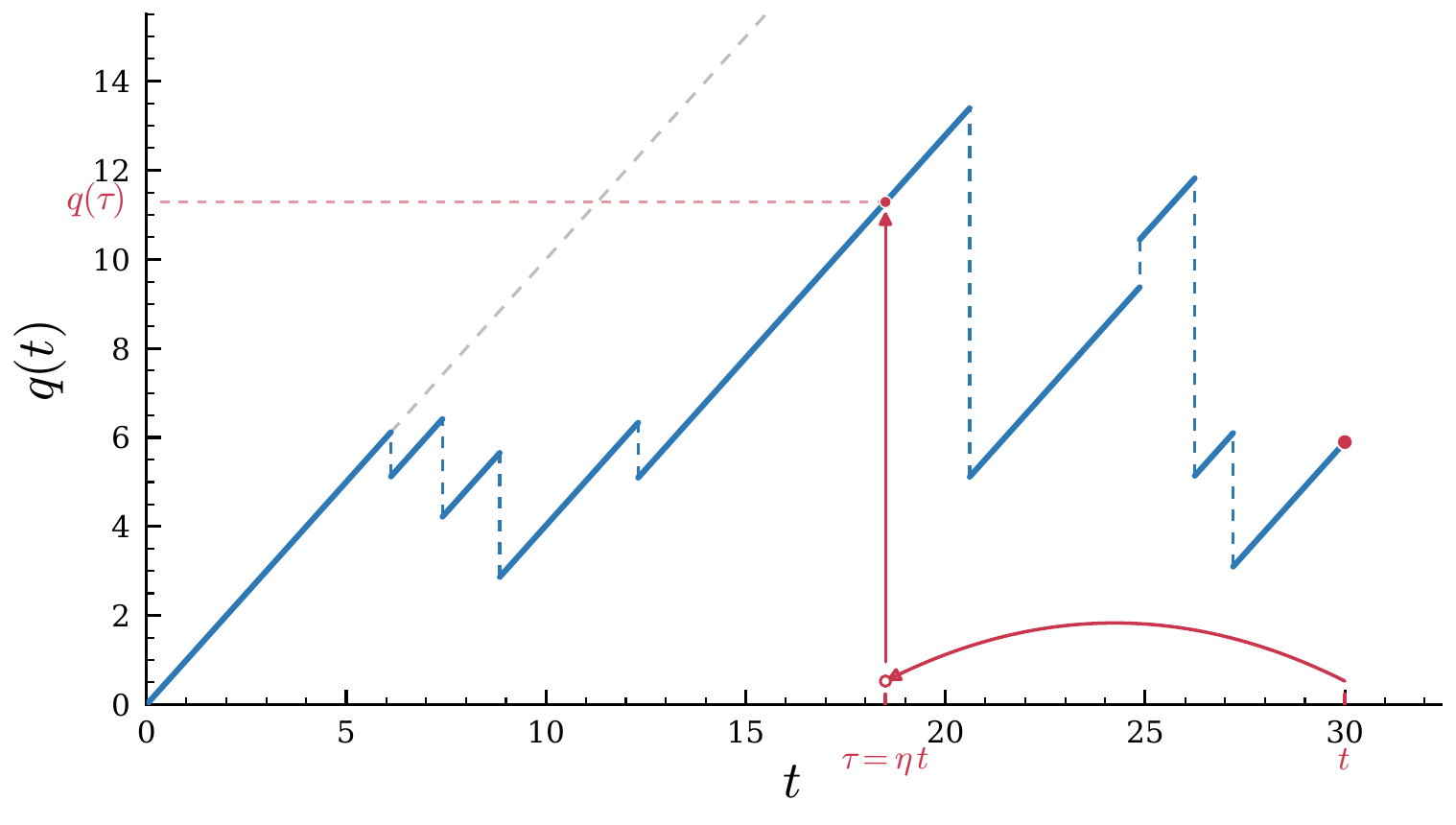}
    \caption{}
  \end{subfigure}\hfill
  \begin{subfigure}{0.48\textwidth}
    \includegraphics[width=\linewidth]{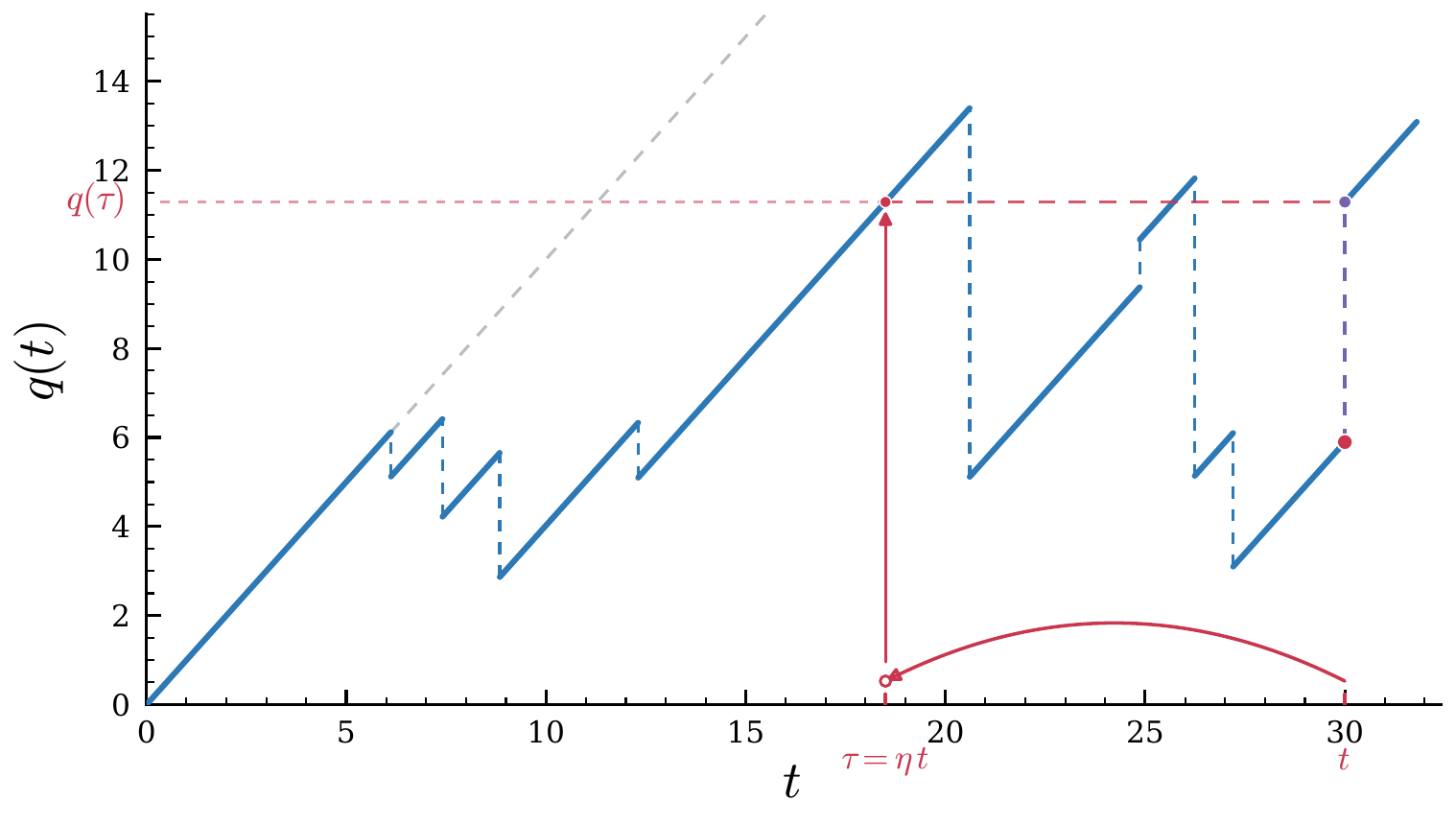}
    \caption{}
  \end{subfigure}
  
\caption{
Schematic illustration of the stochastic dynamics of the
effective time $q(t)$ under uniform-memory resetting.
\textbf{(a)} A typical realization of $q(t)$ as a function
of the physical time $t$. Between resetting events,
$q(t)$ increases linearly with unit slope, while at each
reset (vertical blue dashed lines) it jumps to a value
previously attained along the same trajectory.
The gray dashed line represents the reset-free evolution,
$q(t)=t$.
\textbf{(b)} At a resetting event occurring at time $t$
(red dot), a past physical time $\tau=\eta t$ is selected,
where $\eta$ is uniformly distributed in $[0,1]$.
\textbf{(c)} The effective time $q(\tau)$ corresponding
to the selected past time is identified on the trajectory.
\textbf{(d)} The effective time is instantaneously updated
to $q(t^+)=q(\tau)$, after which it resumes increasing
linearly until the next resetting event.
}
\label{fig:q_dynamics}
  \label{fig:quattro}
\end{figure}

A schematic illustration of this dynamics is shown in
Fig.~\ref{fig:q_dynamics}. Panel (a) shows a typical
realization of $q(t)$, with linear growth between
successive resetting events and instantaneous jumps
at the resetting times. The gray dashed line represents
the reset-free evolution, $q(t)=t$. In panel (b), a new
reset occurs at physical time $t$, and a past time
$\tau=\eta t$ is selected uniformly from $[0,t]$.
Panel (c) illustrates how the corresponding effective
time $q(\tau)$ is identified from the previously
visited trajectory. Finally, in panel (d), the system
is reset to this past state, so that
$q(t^+)=q(\tau)$, after which unitary evolution
resumes with unit slope.

The probability distribution $P(q,t)$ introduced in
Eq.~\eqref{eq:subordination} is obtained by averaging
over all possible realizations of this stochastic
process,
\begin{equation}
    P(q,t)
    =
    \left\langle
    \delta\bigl(q-q(t)\bigr)
    \right\rangle,
    \label{eq:Pq_stochastic_definition}
\end{equation}
where the average includes both the random resetting
times and the past times selected at each resetting
event. This provides the microscopic interpretation
of the effective-time representation used throughout
our analysis.

\subsection{Single-particle dynamics}

We first consider a single particle initially localized at the origin, $|x=0\rangle$, evolving under the tight-binding Hamiltonian in Eq.~\eqref{eq:H1} and subject to uniform-memory resetting at rate $r$. The corresponding initial density matrix is $ \hat\rho(0)=|0\rangle\langle0|$.
Using $\langle k|0\rangle=1/\sqrt{2\pi}$, where $|k\rangle$ is an eigenstate of the single particle Hamiltonian given in Eq.~\eqref{eq:H1} and described in Sec.~\ref{sec:singleparticledynam}, its matrix elements in the momentum basis are
\begin{equation}\label{eq:IC1partmomspace}
    \rho(k,k';0)=\langle k|\hat\rho(0)|k'\rangle
    =
    \frac{1}{2\pi}.
\end{equation}
More generally, for a particle initially localized at a general site $x_0$,
\begin{equation}
    \rho(k,k';0)
    =
    \frac{1}{2\pi}
    e^{i(k'-k)x_0}.
    \label{eq:initaldiagkspace}
\end{equation}
For a reset-free evolution of duration $q$, the density matrix
in the momentum basis evolves as
\begin{equation}
    \rho_0(k,k';q)
    =
    \rho(k,k';0)
    e^{-i[E(k)-E(k')]q},
\end{equation}
where $E(k)$ denotes the energy eigenvalue of the tight-binding Hamiltonian in Eq.~\eqref{eq:H1}.
Using the effective-time representation in Eq.~\eqref{eq:subordination},
we obtain the matrix elements of the density matrix in the presence
of resetting as
\begin{align}
    \rho_r(k,k';t)
    &=
    \int_0^t\mathrm d q\,P(q,t)\rho_0(k,k';q)=
    \rho_r(k,k';0)
    \int_0^t\mathrm d q\,P(q,t)
    e^{-i[E(k)-E(k')]q}
    \nonumber\\
    &=
    \rho_r(k,k';0)
    f\left(i[E(k)-E(k')],t\right),
    \label{eq:rho_k_memory}
\end{align}
where in the last equality we used the definition of the Laplace transform $f(s,t)$ given in Eq.~\eqref{eq:f_characteristic}.

For later convenience, we now compute the probability
$P_x^{(r)}(t):=\langle x|\hat\rho_r(t)|x\rangle$
of finding the particle at site $x$ at time $t$. For the initial condition in Eq.~\eqref{eq:IC1partmomspace}, we insert two resolutions of the identity in the momentum basis and use Eq.~\eqref{eq:rho_k_memory}, obtaining
\begin{align}
    P_x^{(r)}(t)
    &=
    \langle x|\hat\rho_r(t)|x\rangle=
    \int_{-\pi}^{\pi}\frac{\mathrm d k}{2\pi}
    \int_{-\pi}^{\pi}\frac{\mathrm d k'}{2\pi}
    e^{i(k-k')x}
    f\left(i[E(k)-E(k')],t\right).
    \label{eq:Px_memory_exact}
\end{align}
It was shown in Ref.~\cite{dMKM2026} that, for systems with a continuous energy spectrum, the large-time behavior of integrals of the form in Eq.~\eqref{eq:Px_memory_exact} is
\begin{equation}
    P_x^{(r)}(t)
    \simeq
    \frac{r}{\ln(rt)}
    \mathcal V_0\left(
    \frac{rx}{\ln(rt)}
    \right),
    \label{eq:single_memory_velocity}
\end{equation}
where
\begin{equation}
    \mathcal V_0(v)
    =
    \int_{-\pi}^{\pi}\mathrm d k\,
    \rho_0(k,k;0)
    \delta\left(
    v-\frac{\mathrm d E(k)}{\mathrm d k}
    \right)
    \label{eq:velocity_distribution}
\end{equation}
is the distribution of the group velocities in the initial state.

For a particle initially localized at the origin, the diagonal elements of the initial density matrix in momentum space are given by $\rho_0(k,k;0)=1/(2\pi)$. Moreover, from the dispersion relation in Eq.~\eqref{eq:dispersion}, the group velocity is
$\mathrm dE(k)/\mathrm dk=2\gamma\sin k$. Therefore,
\begin{align}
    \mathcal V_0(v)
    &=
    \frac{1}{2\pi}
    \int_{-\pi}^{\pi}\mathrm d k\,
    \delta\left(v-2\gamma\sin k\right)
    =
    \frac{\Theta(2\gamma-|v|)}
    {\pi\sqrt{4\gamma^2-v^2}},
    \label{eq:velocity_distribution_tb}
\end{align}
so that Eq.~\eqref{eq:single_memory_velocity} takes the form
\begin{equation}
    P_x^{(r)}(t)
    \simeq
    \frac{1}{\Lambda_r(t)}
    \mathcal{\tilde V}\left(
    \frac{x}{\Lambda_r(t)}
    \right),
    \qquad
    \mathcal{\tilde V}(z)
    =
    \frac{\Theta(1-|z|)}
    {\pi\sqrt{1-z^2}},
    \label{eq:single_memory_arcsine}
\end{equation}
where $\Lambda_r(t)=\frac{2\gamma}{r}\ln(rt)$.

\subsection{Average density profile}

In this section, we study the average density $\langle\hat n_x(t)\rangle_r$ in the presence of resetting with uniform-memory. We show that, at large times, the density takes the scaling form $\langle\hat n_x(t)\rangle_r\simeq g\!\left(x/\Lambda_r(t)\right)$, where $\Lambda_r(t)=2\gamma\ln(rt)/r$ is the logarithmically growing front edge position and $g(z)$ is the same scaling function as in the reset-free problem, given in Eq.~\eqref{eq:g_bulk}. Thus, uniform-memory resetting preserves the shape of the bulk density profile, while replacing the ballistic front edge position $2\gamma t$ by the much slower logarithmic scale $\Lambda_r(t)$.

As in the reset-free case discussed in Sec.~\ref{sec:domainwallnoreset}, the average density is defined as the expectation value of the local occupation operator $\hat n_x$, introduced in Eq.~\eqref{eq:mx_def}. In the presence of resetting, this expectation value is taken with respect to the reset-averaged many-body density matrix $\hat\rho_r^{(N)}(t)$, namely
\begin{equation}
    \langle\hat n_x(t)\rangle_r
    =
    \operatorname{Tr}_{1,\ldots,N}
    \left[
    \hat\rho_r^{(N)}(t)\hat n_x
    \right].
\end{equation}
Since the derivation leading to Eq.~\eqref{eq:mx_rho1} relies only on the definition of the partial trace and on permutation symmetry, it remains valid in the presence of resetting. We therefore have
\begin{equation}
    \langle\hat n_x(t)\rangle_r
    =
    N\langle x|\hat\rho_r^{(1)}(t)|x\rangle,
    \label{eq:mx_memory_rho1}
\end{equation}
where we recall that $\hat\rho_r^{(1)}(t):=
\mathrm{Tr}_{2,\ldots,N}[\hat\rho_r^{(N)}(t)]$
is the one-particle reduced density matrix.
Hence, determining the average density requires only the one-body reduced density matrix.

As shown in Eq.~\eqref{eq:subordination_p}, although resetting
acts on the full $N$-body state, the one-body reduced density
matrix admits the same effective-time representation as the
full density matrix in Eq.~\eqref{eq:subordination}, namely
\begin{equation}
    \hat\rho_r^{(1)}(t)
    =
    \int_0^t\mathrm{d}q\,P(q,t)\hat\rho_0^{(1)}(q),
    \label{eq:rho1_memory_subordination}
\end{equation}
where $\hat\rho_0^{(1)}(q)$ is the one-body reduced density
matrix of the reset-free system evolved for a time $q$
from the same initial condition. Projecting Eq.~\eqref{eq:rho1_memory_subordination} onto the momentum basis and using Eq.~\eqref{eq:fdefinition} gives
\begin{equation}
    \rho_r^{(1)}(k,k';t)
    =
    \rho_0^{(1)}(k,k';0)
    f\left(i[E(k)-E(k')],t\right),
    \label{eq:rho1_memory_momentum}
\end{equation}
where $f(s,t)$ is given in Eq.~\eqref{eq:f_characteristic}.

Substituting Eq.~\eqref{eq:rho1_memory_subordination} into Eq.~\eqref{eq:mx_memory_rho1}, we recover the effective time representation for the average density,
\begin{equation}
    \langle\hat n_x(t)\rangle_r
    =
    \int_0^t\mathrm d q\,P(q,t)
    \langle\hat n_x(q)\rangle_0.
    \label{eq:mx_subordination_1}
\end{equation}
We can therefore calculate the average density in two equivalent ways: either by averaging the previously obtained reset-free density $\langle\hat n_x(q)\rangle_0$, given in Eq.~\eqref{eq:mx_bessel}, over the effective-time distribution $P(q,t)$, or by computing the one-body reduced density matrix directly in the presence of resetting and extracting its diagonal matrix elements. For completeness, we present both approaches below. 

We begin with the second approach, computing the average density directly from the one-body reduced density matrix in the presence of resetting, using Eq.~\eqref{eq:mx_memory_rho1}.
We consider the same initial condition as in the reset free case, i.e. a domain-wall initial condition, in which the occupied single-particle states are $|0\rangle,|-1\rangle,\ldots,|-(N-1)\rangle$. The corresponding initial density matrix is given in Eq.~\eqref{eq:rho1_result} and it reads, after re-labelling $n\to n-1$
\begin{equation}
    \hat\rho_0^{(1)}(0)
    =
    \frac{1}{N}
    \sum_{n=0}^{N-1}
    |-n\rangle\langle-n|,
    \label{eq:rho1_initial_domainwall}
\end{equation}
where $|-n\rangle$ is the single particle state corresponding to one particle at position $x=-n$ along the chain.
Using the momentum eigenstates of $\hat H$ given in Eq.~\eqref{eq:momentum_normalization}, i.e. $\langle x|k\rangle=e^{ikx}/\sqrt{2\pi}$, the matrix elements of Eq.~\eqref{eq:rho1_initial_domainwall} in momentum space read
\begin{align}
    \rho_0^{(1)}(k,k';0)
    &=
    \langle k|\hat\rho_0^{(1)}(0)|k'\rangle
    =
    \frac{1}{2\pi N}
    \sum_{j=0}^{N-1}
    e^{i(k-k')j}.
    \label{eq:rho1_initial_momentum}
\end{align}
We now insert two resolutions of the identity in Eq.~\eqref{eq:mx_memory_rho1} and use Eqs.~\eqref{eq:rho1_memory_momentum} and \eqref{eq:rho1_initial_momentum}. This gives
\begin{align}
    \langle\hat n_x(t)\rangle_r
    &=
    N
    \int_{-\pi}^{\pi}\mathrm d k
    \int_{-\pi}^{\pi}\mathrm d k'\,
    \langle x|k\rangle
    \langle k|\hat\rho_r^{(1)}(t)|k'\rangle
    \langle k'|x\rangle=
    \sum_{j=0}^{N-1}
    \int_{-\pi}^{\pi}\frac{\mathrm d k}{2\pi}
    \int_{-\pi}^{\pi}\frac{\mathrm d k'}{2\pi}
    e^{i(k-k')(x+j)}
    f\left(i[E(k)-E(k')],t\right).
    \label{eq:mx_memory_momentum}
\end{align}
The double integral in Eq.~\eqref{eq:mx_memory_momentum} is precisely the single-particle probability $P_{x+j}^{(r)}(t)$, derived in Eq.~\eqref{eq:Px_memory_exact}, for a particle initially at the origin to be found at site $x+j$. Therefore Eq.~\eqref{eq:mx_memory_momentum} becomes
\begin{equation}
    \langle\hat n_x(t)\rangle_r
    =
    \sum_{j=0}^{N-1}
    P_{x+j}^{(r)}(t).
    \label{eq:mx_memory_finiteN}
\end{equation}
In the limit $N\to\infty$, setting $\ell=x+j$, we obtain
\begin{equation}
    \langle\hat n_x(t)\rangle_r
    =
    \sum_{\ell=x}^{\infty}
    P_{\ell}^{(r)}(t).
    \label{eq:mx_memory_sum}
\end{equation}
Thus the many-particle density profile is completely determined by the single-particle probability in the presence of uniform-memory resetting, exactly as in the reset-free problem.

At large times, we can use Eq.~\eqref{eq:single_memory_arcsine}, so that Eq.~\eqref{eq:mx_memory_sum} becomes
\begin{equation}
    P_{\ell}^{(r)}(t)
    \simeq
    \frac{1}{\Lambda_r(t)}
    \mathcal{\tilde V}\left(\frac{\ell}{\Lambda_r(t)}\right),
    \qquad
    \mathcal{\tilde V}(z)
    =
    \frac{\Theta(1-|z|)}
    {\pi\sqrt{1-z^2}},
    \label{eq:single_memory_scaling_dw}
\end{equation}
where we recall $\Lambda_r(t)=2\gamma T(t)=\frac{2\gamma }{r}\ln(rt)$. Since $\Lambda_r(t)$ becomes large at long times, the sum in Eq.~\eqref{eq:mx_memory_sum} can be replaced by an integral,
\begin{align}
    \langle\hat n_x(t)\rangle_r
    &\simeq
    \int_x^\infty\mathrm d y\,
    \frac{1}{\Lambda_r(t)}
    \mathcal{\tilde V}\left(\frac{y}{\Lambda_r(t)}\right)=
    \int_x^\infty\mathrm d y\,
    \frac{\Theta(\Lambda_r(t)-|y|)}
    {\pi\sqrt{\Lambda_r^2(t)-y^2}}.
    \label{eq:mx_memory_integral}
\end{align}
Introducing the scaling variable $z=x/\Lambda_r(t)$, for $|z|<1$ we obtain
\begin{align}
    \langle\hat n_x(t)\rangle_r
    &\simeq
    \int_x^{\Lambda_r(t)}
    \frac{\mathrm d y}
    {\pi\sqrt{\Lambda_r^2(t)-y^2}}
    =
    \frac{1}{\pi}
    \arccos\left(\frac{x}{\Lambda_r(t)}\right).
\end{align}
Taking also into account the regions outside the propagating front, the density assumes the scaling form
\begin{equation}
    \langle\hat n_x(t)\rangle_r
    \simeq
    g\left(\frac{x}{\Lambda_r(t)}\right),
    \label{eq:mx_memory_bulk}
\end{equation}
where
\begin{equation}
    g(z)
    =
    \begin{cases}
        1, & z<-1,\\
        \displaystyle\frac{1}{\pi}\arccos(z),
        & |z|<1,\\
        0, & z>1.
    \end{cases}
\end{equation}
Hence the scaling function is exactly the same as in the 
reset-free domain-wall problem (see Eq.~\eqref{eq:g_bulk}). 
At the macroscopic level, uniform-memory resetting changes only the characteristic length scale of the propagating front: the ballistic 
scale $2\gamma t$ is replaced by the much slower 
logarithmically growing scale $\Lambda_r(t)=\frac{2\gamma }{r}\ln(rt)$.

We now briefly derive the same result using the first approach, by averaging the reset-free density over the effective-time distribution $P(q,t)$ through Eq.~\eqref{eq:mx_subordination_1}. Substituting the Gaussian form of $P(q,t)$ derived in Eq.~\eqref{eq:P_gaussian_T}, into Eq.~\eqref{eq:mx_subordination_1} we obtain
\begin{align}
    \langle\hat n_x(t)\rangle_r
    &\simeq
    \int_0^t \mathrm{d}q\,
    \frac{1}{\sqrt{4\pi T(t)/r}}
    \exp\left[-\frac{(q-T(t))^2}{4T(t)/r}\right]
    g\left(\frac{x}{2\gamma q}\right).
\end{align}
It is clear that, at large times, the fluctuations of $q$, of order $\sqrt{T(t)/r}$, become negligible compared to its mean $T(t)$. Therefore, at large times $t$, the Gaussian distribution can be approximated by a Dirac delta function, $P(q,t)\simeq\delta(q-T(t))$. We thus obtain
\begin{align}
    \langle\hat n_x(t)\rangle_r
    &\simeq
    \int_0^t \mathrm{d}q\,
    \delta(q-T(t))
    g\left(\frac{x}{2\gamma q}\right)=
    g\left(\frac{x}{\Lambda_r(t)}\right),
\end{align}
where $\Lambda_r(t)=2\gamma T(t)=2\gamma\ln(rt)/r$ and $g(z)$ is given in Eq.~\eqref{eq:g_bulk}. This reproduces Eq.~(10) of the main text.

\subsection{Local density at the edge of the front}
\label{sec:localdensityfront_wreset}

We now study the local density close to the edge of the propagating front in the presence of resetting with uniform memory. The corresponding reset-free result is derived in Sec.~\ref{sec:finitetimecorrnoreset}. In the
previous subsection, we considered the bulk scaling regime, in which
$x/\Lambda_r(t)$ is kept fixed as $t\to\infty$. In this regime, the
fluctuations of the effective time $q$ around its mean $T(t)$ become
negligible, and the distribution $P(q,t)$ can be replaced, at leading
order, by $\delta(q-T(t))$. However, this approximation is no longer
valid when studying the local density on the scale of the fluctuations
around the front edge position, as we show in this section.

We start again from the effective time representation in
Eq.~\eqref{eq:mx_subordination_1},
\begin{equation}
    \langle\hat n_x(t)\rangle_r
    =
    \int_0^t\mathrm d q\,P(q,t)
    \langle\hat n_x(q)\rangle_0.
    \label{eq:mx_subordination}
\end{equation}
As shown in Eq.~\eqref{eq:edge_density}, for a fixed, large
effective time $q$, the reset-free density close to the right edge of the front
takes the scaling form
\begin{equation}
    \langle\hat n_x(q)\rangle_0
    \simeq
    \frac{1}{(\gamma q)^{1/3}}
    \mathcal{F}_0\left(
    \frac{x-2\gamma q}{(\gamma q)^{1/3}}
    \right).
    \label{eq:resetfree_edge_q}
\end{equation}
There are therefore two possible sources of broadening in the
presence of resetting. First, for a given value of $q$, the
reset-free front edge is at $2\gamma q$ and has an intrinsic
width of order $(\gamma q)^{1/3}$, as described by the reset-free
scaling function $\mathcal{F}_0$. Second, the effective time $q$
itself fluctuates according to $P(q,t)$, so that the position
$2\gamma q$ of the front edge also fluctuates. Averaging over $q$
can therefore produce an additional broadening of the density
profile. To determine the resulting scaling behavior, we need
to compare these two contributions.

At large physical times $t$, the distribution $P(q,t)$ is
asymptotically Gaussian, with mean $T(t)=\ln(rt)/r$ and variance
$2T(t)/r$, as derived in Eq.~\eqref{eq:q_moments}. Thus,
the typical fluctuations of $q$ are of order
$\sqrt{T(t)/r}$, which induce fluctuations of the front
position of order $\gamma\sqrt{T(t)/r}$. On the other hand,
since $q$ is typically of order $T(t)$, the intrinsic
reset-free broadening is of order $(\gamma T(t))^{1/3}$.
For fixed $r$ and $\gamma$, the former grows as $T^{1/2}(t)$,
whereas the latter grows only as $T^{1/3}(t)$. Hence,
the fluctuations of the effective time dominate at large
$t$ and determine the scaling of the front edge.

We therefore focus on positions close to the average front edge
$\Lambda_r(t)=2\gamma T(t)$ and introduce the scaling variables
\begin{equation}
    x=\Lambda_r(t)+\gamma\sqrt{\frac{T(t)}{r}}\,y,
    \qquad
    q=T(t)+\sqrt{\frac{T(t)}{r}}\,u,
    \qquad
    y,u=O(1).
    \label{eq:memory_edge_variables}
\end{equation}
Using the Gaussian form of $P(q,t)$ in
Eq.~\eqref{eq:P_gaussian_T}, we obtain
\begin{equation}
    P(q,t)\mathrm d q
    \simeq
    \frac{\mathrm d u}{2\sqrt{\pi}}
    e^{-u^2/4}.
    \label{eq:P_u}
\end{equation}
In the large-time limit, the integration interval can be
extended to the whole real axis in $u$, since the Gaussian
weight is concentrated around $u=O(1)$.
We now substitute the scaling variables in
Eq.~\eqref{eq:resetfree_edge_q}. The prefactor becomes
\begin{equation}
    (\gamma q)^{-1/3}
    =
    [\gamma T(t)]^{-1/3}
    \left(
    1+\frac{u}{\sqrt{rT(t)}}
    \right)^{-1/3},
    \label{eq:Jq_prefactor}
\end{equation}
while the argument of $\mathcal{F}_0$ reads
\begin{align}
    \frac{x-2\gamma q}{(\gamma q)^{1/3}}
    =
    \frac{
    \gamma\sqrt{T(t)/r}(y-2u)
    }{
    [\gamma T(t)]^{1/3}
    \left(1+u/\sqrt{rT(t)}\right)^{1/3}
    }=
    \frac{\gamma^{2/3}T^{1/6}(t)}{\sqrt r}
    \frac{y-2u}{
    \left(1+u/\sqrt{rT(t)}\right)^{1/3}
    }.
    \label{eq:edge_argument_memory}
\end{align}
Substituting these expressions into
Eq.~\eqref{eq:mx_subordination}, we find
\begin{align}
    \langle\hat n_x(t)\rangle_r
    &\simeq
    \frac{[\gamma T(t)]^{-1/3}}{2\sqrt{\pi}}
    \int_{-\infty}^{\infty}\mathrm d u\,
    e^{-u^2/4}
    \left(
    1+\frac{u}{\sqrt{rT(t)}}
    \right)^{-1/3}
    \nonumber\\
    &\qquad\times
    \mathcal{F}_0\left[
    \frac{\gamma^{2/3}T^{1/6}(t)}{\sqrt r}
    \frac{y-2u}{
    \left(1+u/\sqrt{rT(t)}\right)^{1/3}
    }
    \right].
    \label{eq:memory_edge_integral_before}
\end{align}

We can now take the large-time limit. Since
$u/\sqrt{rT(t)}\to0$, the factors
$(1+u/\sqrt{rT(t)})^{-1/3}$ tend to one.
More importantly, the argument of $\mathcal{F}_0$ contains
the factor $T^{1/6}(t)$, which becomes very large for
$rt\gg1$. Consequently, for fixed $y$ and $u$, its
argument tends to $+\infty$ when $u<y/2$, and to
$-\infty$when $u>y/2$.

From the asymptotic behaviors in
Eq.~\eqref{eq:F_airy_asympt}, we know that
$\mathcal{F}_0(s)$ is exponentially small for large
positive $s$, whereas
$\mathcal{F}_0(s)\simeq\sqrt{-s}/\pi$ for large
negative $s$. Therefore, at leading order, only
the region $u>y/2$ contributes to the integral.
Using the asymptotic form
$\mathcal{F}_0(s)\simeq\sqrt{-s}/\pi$, we obtain
\begin{align}
    \langle\hat n_x(t)\rangle_r
    &\simeq
    \frac{[\gamma T(t)]^{-1/3}}{2\pi^{3/2}}
    \left(
    \frac{\gamma^{2/3}T^{1/6}(t)}{\sqrt r}
    \right)^{1/2}
    \int_{y/2}^{\infty}
    \mathrm d u\,
    e^{-u^2/4}\sqrt{2u-y}=\left(\frac{2\gamma}{r\Lambda_r(t)}\right)^{1/4}\mathcal{F}_r(y),
    \label{eq:memory_edge_scaling_derivation}
\end{align}
where we have introduced the scaling function
\begin{equation}
    \mathcal{F}_r(y)
    =
    \frac{1}{2\pi^{3/2}}
    \int_{y/2}^{\infty}
    \mathrm d u\,
    e^{-u^2/4}\sqrt{2u-y}.
    \label{eq:Fmemory_integral}
\end{equation}
We thus obtain the large-time scaling form of the
local density near the right edge of the front,
\begin{equation}\label{eq:memory_edge_final}
    \langle \hat n_x(t)\rangle_r
    \simeq
    \left(\frac{2\gamma}{r\Lambda_r(t)}\right)^{1/4}
    \mathcal{F}_r\left(
    \frac{x-\Lambda_r(t)}
    {\sqrt{\gamma\Lambda_r(t)/(2r)}}
    \right),
\end{equation}
thus reproducing Eq.~(12) and~(13) of the main text. The front edge, at late times, is therefore centered around the logarithmically growing position $\Lambda_r(t)=2\gamma T(t)\approx (2\gamma/r) \ln(rt )$, while
its width grows as
$\sqrt{[\gamma/(2r)]\Lambda_r(t)}\sim \sqrt{\ln (rt)}$.

The scaling function $\mathcal{F}_r(y)$ can also be expressed
in terms of a parabolic cylinder function [see (12.5.1) in Ref.~\cite{NIST}]. Setting
$s=2u-y$ in Eq.~\eqref{eq:Fmemory_integral}, we obtain
\begin{align}
    \mathcal{F}_r(y)
    &=
    \frac{1}{4\pi^{3/2}}
    \int_0^\infty
    \mathrm d s\,s^{1/2}
    \exp\left[
    -\frac{(s+y)^2}{16}
    \right]=
    \frac{e^{-y^2/32}}{2^{3/4}\pi}
    D_{-3/2}\left(
    \frac{y}{2\sqrt2}
    \right),
    \label{eq:Fmemory_D}
\end{align}
where $D_\nu(y)=\frac{e^{-y^2/4}}{\Gamma(-\nu)}\int_0^\infty dt \,t^{-\nu-1}e^{-t^2/2-yt}$ denotes the parabolic cylinder
function. Its asymptotic behaviors are
\begin{equation}
    \mathcal{F}_r(y)
    \simeq
    \begin{cases}
        \displaystyle
        \frac{\sqrt{-y}}{\pi},
        & y\to-\infty,\\[3mm]
        \displaystyle
        \frac{2^{3/2}}{\pi y^{3/2}}
        e^{-y^2/16},
        & y\to+\infty.
    \end{cases}
    \label{eq:Fmemory_asympt}
\end{equation}
In particular, the $y\to-\infty$ behavior of
$\mathcal{F}_r(y)$ coincides with that of the reset-free
scaling function $\mathcal{F}_0(y)$, given in
Eq.~\eqref{eq:F_airy_asympt}. Consequently, the edge
density in the presence of resetting also matches smoothly
with the bulk profile in Eq.~\eqref{eq:mx_memory_bulk},
confirming the consistency of the two scaling forms in
their common regime of validity.

\subsection{Extreme-value statistics}

We now investigate the extreme-value statistics of the rightmost fermion in the presence of uniform-memory resetting. As shown in Sec.~\ref{sec:onepartnomemory}, in the absence of resetting, the position of the rightmost fermion follows the Tracy--Widom distribution at large times, with fluctuations of order $(\gamma t)^{1/3}$ around the ballistic front edge $2\gamma t$ [see Eq.~\eqref{eq:TW_density}]. For a reset-free evolution of duration $q$, we can therefore write
\begin{equation}
    X_{\max}(q)
    \simeq
    2\gamma q+(\gamma q)^{1/3}\chi,
    \label{eq:Xmax_fixedq}
\end{equation}
where $\chi$ is an $O(1)$ random variable distributed according to the Tracy--Widom density $f_0$ given in Eq.~\eqref{eq:TW_density}.

At large physical time $t$, the effective time is asymptotically Gaussian. Using Eq.~\eqref{eq:q_moments}, we can write
\begin{equation}
    q
    =
    T(t)+\sqrt{\frac{2T(t)}{r}}\, Z,
    \qquad
    T(t)=\frac{\ln(rt)}{r}=\frac{\Lambda_r(t)}{2\gamma},
    \label{eq:q_random}
\end{equation}
where $Z$ is an $O(1)$ Gaussian random variable with zero mean and unit variance.
Substituting this expression for $q$ into Eq.~\eqref{eq:Xmax_fixedq} gives
\begin{align}
    X_{\max}(t)
    &\approx
    \Lambda_r(t)
    +
    2\sqrt{\frac{\gamma\Lambda_r(t)}{r}}\,Z
    +
    \left(\frac{\Lambda_r(t)}{2}\right)^{1/3}
    \left(
    1+2\sqrt{\frac{\gamma}{r\Lambda_r(t)}}\,Z
    \right)^{1/3}\chi.
    \label{eq:Xmax_substitute_q}
\end{align}
Since $\Lambda_r(t)\to\infty$, the factor 
$\left(1+2\sqrt{\gamma/[r\Lambda_r(t)]}\,Z\right)^{1/3}$ 
on the right-hand side of Eq.~\eqref{eq:Xmax_substitute_q} 
tends to one at large $t$. Hence, retaining only the leading 
fluctuations,
\begin{equation}
    X_{\max}(t)
    \approx
    \Lambda_r(t)
    +
    2\sqrt{\frac{\gamma\Lambda_r(t)}{r}}\,Z.
    \label{eq:Xmax_random_time}
\end{equation}
The fluctuations of the front edge position induced by the randomness of $q$
are therefore of order $\Lambda_r^{1/2}(t)$, whereas the intrinsic
Tracy--Widom fluctuations remain of order $\Lambda_r^{1/3}(t)$. Thus the
memory-induced fluctuations dominate asymptotically.

We now derive the corresponding probability distribution explicitly.
Since probabilities are linear in the density matrix,
Eq.~\eqref{eq:subordination} implies
\begin{equation}
    \mathbb P_r[X_{\max}(t)=x]
    =
    \int_0^t\mathrm d q\,P(q,t)
    \mathbb P_0[X_{\max}(q)=x].
    \label{eq:Xmax_subordination}
\end{equation}
Inserting Eqs.~\eqref{eq:P_gaussian_T} and~\eqref{eq:TW_density} into Eq.~\eqref{eq:Xmax_subordination}, we obtain
\begin{equation}
    \mathbb P_r[X_{\max}(t)=x]
    \simeq
    \int_0^t\mathrm d q\,
    \frac{
    \exp\left[-\frac{\gamma r}{2\Lambda_r(t)}
    \left(q-\frac{\Lambda_r(t)}{2\gamma}\right)^2\right]
    }
    {\sqrt{2\pi\Lambda_r(t)/(\gamma r)}}
    \frac{1}{(\gamma q)^{1/3}}
    f_0\left(
    \frac{x-2\gamma q}{(\gamma q)^{1/3}}
    \right),
    \label{eq:Xmax_convolution}
\end{equation}
The fluctuations generated by the effective time have spatial width
$2\sqrt{\gamma\Lambda_r(t)/r}$.
We therefore introduce the scaling variables
\begin{equation}
    x
    =
    \Lambda_r(t)+2\sqrt{\frac{\gamma\Lambda_r(t)}{r}}\,y,
    \qquad
    q
    =
    \frac{\Lambda_r(t)}{2\gamma}
    +\sqrt{\frac{\Lambda_r(t)}{\gamma r}}\,z.
    \label{eq:Xmax_scaling_variables}
\end{equation}
With these scaling variables, the numerator of the argument of 
$f_0$ in Eq.~\eqref{eq:Xmax_convolution} becomes
\begin{align}
    x-2\gamma q
    &=
    2\sqrt{\frac{\gamma\Lambda_r(t)}{r}}(y-z).
    \label{eq:Xmax_numerator}
\end{align}
Moreover,
\begin{equation}
    (\gamma q)^{1/3}
    =
    \left(\frac{\Lambda_r(t)}{2}\right)^{1/3}
    \left(
    1+2\sqrt{\frac{\gamma}{r\Lambda_r(t)}}\,z
    \right)^{1/3},
    \label{eq:Xmax_denominator}
\end{equation}
and $\mathrm d q=\sqrt{\Lambda_r(t)/(\gamma r)}\,\mathrm d z$. 
The Gaussian factor also simplifies as
\begin{equation}
    \frac{\gamma r}{2\Lambda_r(t)}
    \left(q-\frac{\Lambda_r(t)}{2\gamma}\right)^2
    =
    \frac{z^2}{2}.
\end{equation}
Therefore Eq.~\eqref{eq:Xmax_convolution} becomes
\begin{align}
    \mathbb P_r[X_{\max}(t)=x]
    &\simeq
    \left(\frac{2}{\Lambda_r(t)}\right)^{1/3}
    \int_{-\infty}^{\infty}
    \frac{\mathrm d z}{\sqrt{2\pi}}\,
    e^{-z^2/2}
    \left(
    1+2\sqrt{\frac{\gamma}{r\Lambda_r(t)}}\,z
    \right)^{-1/3}
    \nonumber\\
    &\qquad\times
    f_0\left[
    2^{4/3}\sqrt{\frac{\gamma}{r}}\Lambda_r^{1/6}(t)
    \frac{y-z}
    {\left(1+2\sqrt{\frac{\gamma}{r\Lambda_r(t)}}\,z\right)^{1/3}}
    \right],
    \label{eq:Xmax_scaled_integral}
\end{align}
where the integration limits have been extended to the whole real axis,
since the relative fluctuations of $q$ around its mean
$\Lambda_r(t)/(2\gamma)$ vanish at large $t$.

We now define
\begin{equation}
    \lambda(t)
    \equiv
    \frac{2\sqrt{\gamma\Lambda_r(t)/r}}
    {(\Lambda_r(t)/2)^{1/3}}
    =
    2^{4/3}\sqrt{\frac{\gamma}{r}}\Lambda_r^{1/6}(t).
    \label{eq:lambda_TW}
\end{equation}
Hence $\lambda(t)\to\infty$, while 
$2\sqrt{\gamma/[r\Lambda_r(t)]}\to0$. 
At leading order, Eq.~\eqref{eq:Xmax_scaled_integral} therefore reduces to
\begin{equation}
    \mathbb P_r[X_{\max}(t)=x]
    \simeq
    \left(\frac{2}{\Lambda_r(t)}\right)^{1/3}
    \int_{-\infty}^{\infty}
    \frac{\mathrm d z}{\sqrt{2\pi}}\,
    e^{-z^2/2}
    f_0\left[\lambda(t)(y-z)\right].
    \label{eq:Xmax_scaled_integral_leading}
\end{equation}
Since $f_0$ is normalized, in the limit $\lambda\to\infty$, $\lambda f_0(\lambda s)\longrightarrow\delta(s)$.
Equivalently,
\begin{equation}
    f_0\left[\lambda(t)(y-z)\right]
    \simeq
    \frac{1}{\lambda(t)}
    \delta(y-z).
\end{equation}
Using $\frac{1}{(\Lambda_r(t)/2)^{1/3}\lambda(t)}
=\frac{1}{2\sqrt{\gamma\Lambda_r(t)/r}}$,
Eq.~\eqref{eq:Xmax_scaled_integral_leading} becomes
\begin{align}
    \mathbb P_r[X_{\max}(t)=x]
    &\simeq
    \frac{1}{2\sqrt{\gamma\Lambda_r(t)/r}}
    \int_{-\infty}^{\infty}
    \frac{\mathrm d z}{\sqrt{2\pi}}\,
    e^{-z^2/2}
    \delta(y-z)=
    \frac{1}{2\sqrt{\gamma\Lambda_r(t)/r}}
    \frac{e^{-y^2/2}}{\sqrt{2\pi}}.
\end{align}
Recalling that $y=[x-\Lambda_r(t)]/[2\sqrt{\gamma\Lambda_r(t)/r}]$, 
we finally obtain
\begin{equation}
    \mathbb P_r[X_{\max}(t)=x]
    \simeq
    \frac{1}{2\sqrt{\gamma\Lambda_r(t)/r}}
    f_r\left(
    \frac{x-\Lambda_r(t)}
    {2\sqrt{\gamma\Lambda_r(t)/r}}
    \right),
    \qquad
    f_r(y)
    =
    \frac{e^{-y^2/2}}{\sqrt{2\pi}}.
    \label{eq:Xmax_gaussian}
\end{equation}
This reproduces Eq.~(16) of the main text.
Thus the leading fluctuations of the rightmost fermion become Gaussian
under uniform-memory resetting, with width
$2\sqrt{\gamma\Lambda_r(t)/r}\sim\Lambda_r^{1/2}(t)$, 
whereas the intrinsic reset-free Tracy--Widom fluctuations 
at the typical effective time have width of order $\Lambda_r^{1/3}(t)$.

\subsection{Integrated current}

We finally investigate the mean and variance of the integrated current in the presence of uniform-memory resetting. As in the reset-free case discussed in Sec.~\ref{sec:onepartnomemory}, the integrated current counts the net number of fermions crossing the bond between sites $0$ and $1$ up to time $t$, and is defined in Eq.~\eqref{eq:C0_def}. Using Eq.~\eqref{eq:observable_subordination}, its expectation value can be obtained as
\begin{equation}
    \langle\hat C(t)\rangle_r
    =
    \int_0^t\mathrm d q\,P(q,t)
    \langle\hat C(q)\rangle_0.
    \label{eq:C_memory_mean_general}
\end{equation}
At large $t$, the distribution $P(q,t)$ is sharply concentrated around its mean $\langle q\rangle\simeq T(t)=\ln(rt)/r$, which itself becomes large as $t\to\infty$. We can therefore use the large-$q$ asymptotic expression for the reset-free mean current, $\langle\hat C(q)\rangle_0\simeq 2\gamma q/\pi$, obtained in Eq.~\eqref{eq:C_mean_asympt}. Substituting this result into Eq.~\eqref{eq:C_memory_mean_general}, we find
\begin{align}
    \langle\hat C(t)\rangle_r
    &\simeq
    \frac{2\gamma}{\pi}
    \int_0^t\mathrm d q\,qP(q,t)
    =
    \frac{2\gamma}{\pi}\langle q\rangle
    \simeq
    \frac{2\gamma T(t)}{\pi}
    =
    \frac{\Lambda_r(t)}{\pi},
    \label{eq:C_memory_mean}
\end{align}
where we recall $\Lambda_r(t)=\frac{2\gamma}{r}\ln(rt)$.
Thus the linear growth of the reset-free mean current is replaced by a logarithmic growth in physical time.

We now turn to the variance of the integrated current, defined as
\begin{equation}
    \operatorname{Var}_r[\hat C(t)]
    =
    \langle\hat C^2(t)\rangle_r
    -
    \langle\hat C(t)\rangle_r^2.
    \label{eq:C_memory_var_def}
\end{equation}
As for the mean, the second moment can be obtained thanks to Eq.~\eqref{eq:observable_subordination}.
Writing the reset-free second moment $\langle\hat C^2(q)\rangle_0$ as $\langle\hat C^2(q)\rangle_0=\operatorname{Var}_0[\hat C(q)]+\langle\hat C(q)\rangle_0^2$, we have
\begin{align}
    \langle\hat C^2(t)\rangle_r
    &=
    \int_0^t\mathrm d q\,P(q,t)
    \langle\hat C^2(q)\rangle_0=
    \int_0^t\mathrm d q\,P(q,t)
    \left[
    \operatorname{Var}_0[\hat C(q)]
    +
    \langle\hat C(q)\rangle_0^2
    \right].
    \label{eq:C_memory_secondmoment}
\end{align}
Substituting this expression into
Eq.~\eqref{eq:C_memory_var_def}, and using
Eq.~\eqref{eq:C_memory_mean_general} for the mean, we obtain
\begin{align}
    \operatorname{Var}_r[\hat C(t)]
    &=
    \int_0^t\mathrm d q\,P(q,t)
    \operatorname{Var}_0[\hat C(q)]+
    \int_0^t\mathrm d q\,P(q,t)
    \langle\hat C(q)\rangle_0^2
    -
    \left[
    \int_0^t\mathrm d q\,P(q,t)
    \langle\hat C(q)\rangle_0
    \right]^2.
    \label{eq:C_memory_totalvariance}
\end{align}
This expression separates two distinct sources of fluctuations.
The first term is the reset-free current variance averaged
over the effective time $q$, while the remaining two terms
account for the fluctuations of the reset-free mean current
induced by the randomness of $q$.

To determine which contribution dominates at large times,
we use the reset-free asymptotic results obtained in
Eqs.~\eqref{eq:C_mean_asympt} and \eqref{eq:C_var_log},
namely $\langle\hat C(q)\rangle_0\simeq 2\gamma q/\pi$
and
$\operatorname{Var}_0[\hat C(q)]
\simeq[\ln(2\gamma q)+\bar c]/(2\pi^2)$.
Since $P(q,t)$ is concentrated around the large effective
time $T(t)$, these expressions can be substituted into
Eq.~\eqref{eq:C_memory_totalvariance}, giving
\begin{align}
    \operatorname{Var}_r[\hat C(t)]
    &\simeq
    \frac{1}{2\pi^2}
    \int_0^t\mathrm d q\,P(q,t)
    \left[
    \ln(2\gamma q)+\bar c
    \right]+
    \frac{4\gamma^2}{\pi^2}
    \left(
    \langle q^2\rangle-\langle q\rangle^2
    \right).
    \label{eq:C_memory_var_asympt}
\end{align}

We now compare the large-time behavior of these two terms.
Since the fluctuations of $q$ are small compared to its
mean $T(t)$, the first term on the right-hand side of 
Eq.~\eqref{eq:C_memory_var_asympt} is asymptotically
\begin{equation}
    \frac{1}{2\pi^2}
    \int_0^t\mathrm d q\,P(q,t)
    \left[\ln(2\gamma q)+\bar c\right]
    \simeq
    \frac{\ln\Lambda_r(t)+\bar c}{2\pi^2},
\end{equation}
which grows only as $\ln\Lambda_r(t)$.
On the other hand, using Eq.~\eqref{eq:q_moments}, the second term 
on the right-hand side of Eq.~\eqref{eq:C_memory_var_asympt} grows as
\begin{equation}
    \frac{4\gamma^2}{\pi^2}
    \left(
    \langle q^2\rangle-\langle q\rangle^2
    \right)
    =
    \frac{4\gamma^2}{\pi^2}
    \operatorname{Var}(q)
    \simeq
    \frac{4\gamma\Lambda_r(t)}{\pi^2r}.
\end{equation}
Since $\Lambda_r(t)$ grows faster than $\ln\Lambda_r(t)$,
this contribution dominates over the intrinsic
reset-free fluctuations. We therefore obtain
\begin{align}
    \operatorname{Var}_r[\hat C(t)]
    &\simeq
    \frac{4\gamma^2}{\pi^2}
    \operatorname{Var}(q)
    \simeq
    \frac{4\gamma\Lambda_r(t)}{\pi^2r}.
    \label{eq:C_memory_var_final}
\end{align}
Together with Eq.~\eqref{eq:C_memory_mean}, this reproduces the asymptotic results for the mean and variance of the integrated current reported in the main text.
Thus, both the mean and the variance of the integrated
current grow linearly with $\Lambda_r(t)$, i.e., logarithmically
with the physical time.
However, unlike in the reset-free system, the leading
current fluctuations are no longer determined by the
intrinsic quantum fluctuations, but by the randomness
of the effective time introduced by uniform-memory
resetting.

\section{Numerical simulations}
\label{sec:numerical_simulations}

In this section, we compare our analytical predictions with numerical simulations. As discussed in the main Letter, our framework allows us to decouple the unitary quantum dynamics, governed by the Hamiltonian $\hat{\mathcal H}$, from the classical stochastic process describing the effective time $q(t)$. This considerably simplifies the numerical simulations, since instead of simulating the full quantum many-body dynamics, it is sufficient to simulate the one-dimensional classical stochastic process $q(t)$ and use the relation
\begin{equation}
\label{eq:observablerelation_APP}
   \langle\hat O(t)\rangle_r
   =\int_0^t\mathrm{d}q\,P(q,t)
   \langle\hat O(q)\rangle_0.
\end{equation}
In practice, for a fixed physical time $t$, we generate independent realizations of $q(t)$ and evaluate the corresponding reset-free expectation value $\langle\hat O(q(t))\rangle_0$, which can be computed either analytically or numerically. In our case, we exploit known exact analytical results for the reset-free observables, but we emphasize that the same approach remains applicable when only numerical results are available.
Averaging these values over the realizations then provides a numerical estimate of $\langle\hat O(t)\rangle_r$, without explicitly simulating the quantum many-body dynamics in the presence of resetting.

As a first check, we numerically simulate the stochastic process $q(t)$, whose dynamics is governed by
\begin{equation}
q(t+\mathrm{d}t)=
\begin{cases}
q(t)+\mathrm{d}t,
& \text{with probability }1-r\,\mathrm{d}t,\\[2mm]
q(\tau),
& \text{with probability }r\,\mathrm{d}t,
\end{cases}
\label{eq:q_forward_dynamics}
\end{equation}
where $\tau$ is chosen uniformly from $[0,t]$, with the initial condition $q(0)=0$.
A direct way to simulate this process, which we refer to as the forward method, consists in generating successive resetting times from an exponential distribution with rate $r$. Between resets, $q(t)$ increases linearly with unit slope, while at each reset it jumps to the value $q(\tau)$ attained at a uniformly selected past time $\tau$, as illustrated in Fig.~\ref{fig:q_dynamics}. This requires keeping track of the previously visited values of $q(t)$ along the trajectory.
\begin{figure}[htbp]
    \centering
    \includegraphics[width=0.9\textwidth]{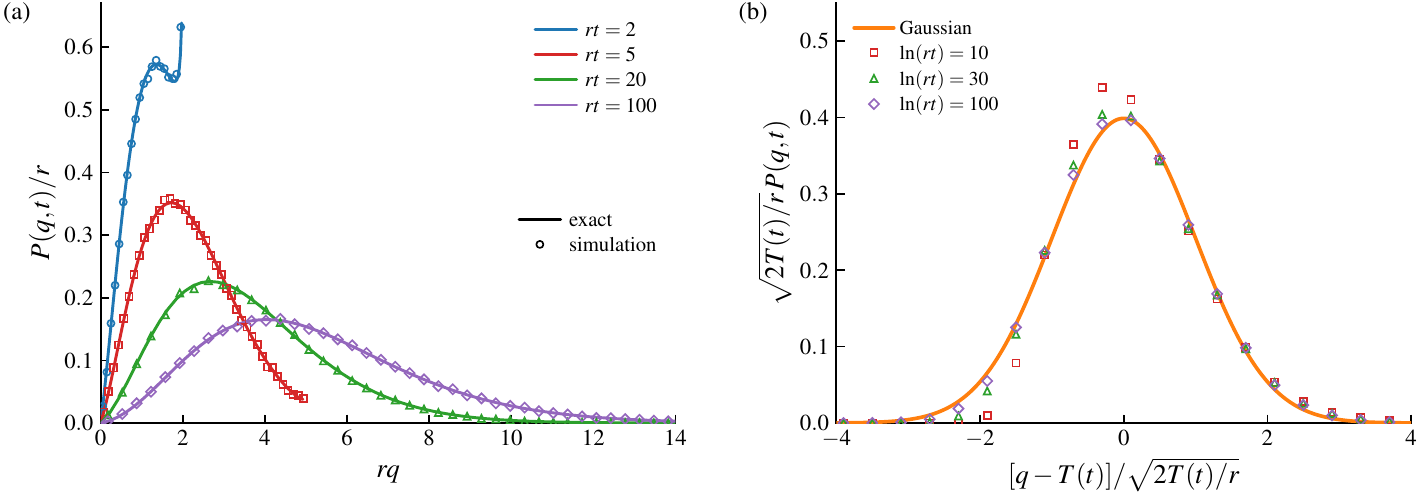}
    \caption{
    Probability distribution $P(q,t)$ of the effective time.
    (a) Comparison between the continuous part of the exact
    distribution, obtained by numerically inverting the
    Laplace transform in Eq.~\eqref{eq:f_characteristic}
    after subtracting the singular contribution in Laplace
    space (solid lines), and numerical simulations performed
    using the forward method (symbols), for $rt=2,5,20,100$.
    The singular contribution $e^{-rt}\delta(q-t)$
    is not shown.
    (b) Rescaled effective-time distribution for
    $\ln(rt)=10,30,100$, obtained using the backward method
    (symbols), compared with the asymptotic Gaussian prediction
    in Eq.~\eqref{eq:P_gaussian_T} (solid line).
    Here $T(t)=\ln(rt)/r$.
}
    \label{fig:Pq_numerical}
\end{figure}

A more efficient approach is provided by the backward method, which allows us to sample $q(t)$ at a given physical time $t$ without generating its entire history. Starting from $t$, we draw the time $\Delta$ elapsed since the last (past) reset from an exponential distribution with rate $r$. If $\Delta\geq t$, no reset has occurred and $q(t)=t$. Otherwise, the last reset occurred at time $t-\Delta$, when the process returned to a value attained at a uniformly chosen past time $\tau\in[0,t-\Delta]$. Since $q$ subsequently increases at unit speed, we obtain the recursive relation
\begin{equation}
q(t)=
\begin{cases}
t, & \Delta\geq t,\\[2mm]
\Delta+q(\tau), & \Delta<t,
\end{cases}
\label{eq:q_backward_dynamics}
\end{equation}
where $\tau\sim U[0,t-\Delta]$ is chosen uniformly in $[0,t-\Delta]$ and
where the recursion is repeated until the first case is reached. Unlike the forward method, which generates all resetting events up to time $t$, the backward method follows only the sequence of past times needed to determine $q(t)$. It is therefore computationally more efficient, particularly at large times, when the number of resetting events in the forward method grows as $rt$.
\begin{figure}[htbp]
    \centering
    \includegraphics[width=0.65\textwidth]{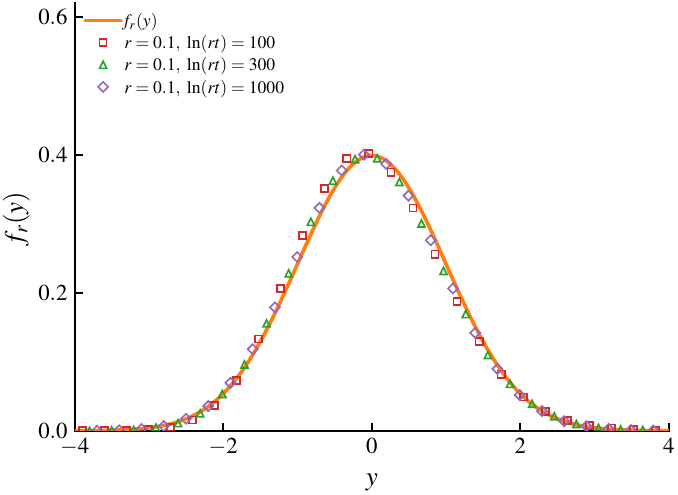}
\caption{
    Rescaled probability density of the rightmost fermion
    for $r=0.1$ and $\ln(rt)=100,300,1000$.
    The scaling variable is
    $y=[x-\Lambda_r(t)]/[2\sqrt{\gamma\Lambda_r(t)/r}]$,
    where $\Lambda_r(t)=2\gamma\ln(rt)/r$.
    Symbols represent numerical simulations obtained
    using the backward method, while the solid line
    shows the analytical Gaussian scaling function
    $f_r(y)=e^{-y^2/2}/\sqrt{2\pi}$ given in
    Eq.~\eqref{eq:Xmax_gaussian}.}
    \label{fig:evs_numerical}
\end{figure}
Using the methods described above, we numerically investigate 
the different observables studied in this work. The numerical 
results for the macroscopic density profile and the local density 
at the propagating front edge, both with and without resetting, 
are presented in Fig.~1 of the main text. Here, we present 
additional numerical results for the effective-time distribution, 
the extreme-value statistics of the rightmost fermion, and the 
mean and variance of the integrated current.

In Fig.~\ref{fig:Pq_numerical}, we compare numerical
simulations of the effective-time distribution $P(q,t)$
with our analytical predictions. Note that, at finite times,
$P(q,t)$ contains a singular contribution
$e^{-rt}\delta(q-t)$, corresponding to trajectories
with no resetting events. In panel (a), we use the
forward method and compare the results with the continuous
part of the exact distribution, obtained by numerically
inverting the Laplace transform in
Eq.~\eqref{eq:f_characteristic} after subtracting
the contribution $e^{-(r+s)t}$ associated with the
delta function. Both the analytical and numerical
distributions shown in panel (a) are thus normalized
to $1-e^{-rt}$, before restricting the plotted range.
In panel (b), we use the backward method to explore
much longer times and compare the rescaled distributions
with the asymptotic Gaussian prediction in
Eq.~\eqref{eq:P_gaussian_T}.

We now turn to the extreme-value statistics of the rightmost fermion. As shown in the main Letter, uniform-memory resetting changes the asymptotic distribution of $X_{\max}(t)$ from Tracy--Widom to Gaussian. In Fig.~\ref{fig:evs_numerical}, we compare the analytical scaling function $f_r(y)$, given in Eq.~\eqref{eq:Xmax_gaussian}, with numerical simulations obtained using the backward method for different values of $t$. The numerical results show good agreement with the theoretical prediction, confirming the Gaussian scaling of the extreme-value statistics at large times.
\begin{figure}[htbp]
    \centering
    \includegraphics[width=0.9\textwidth]{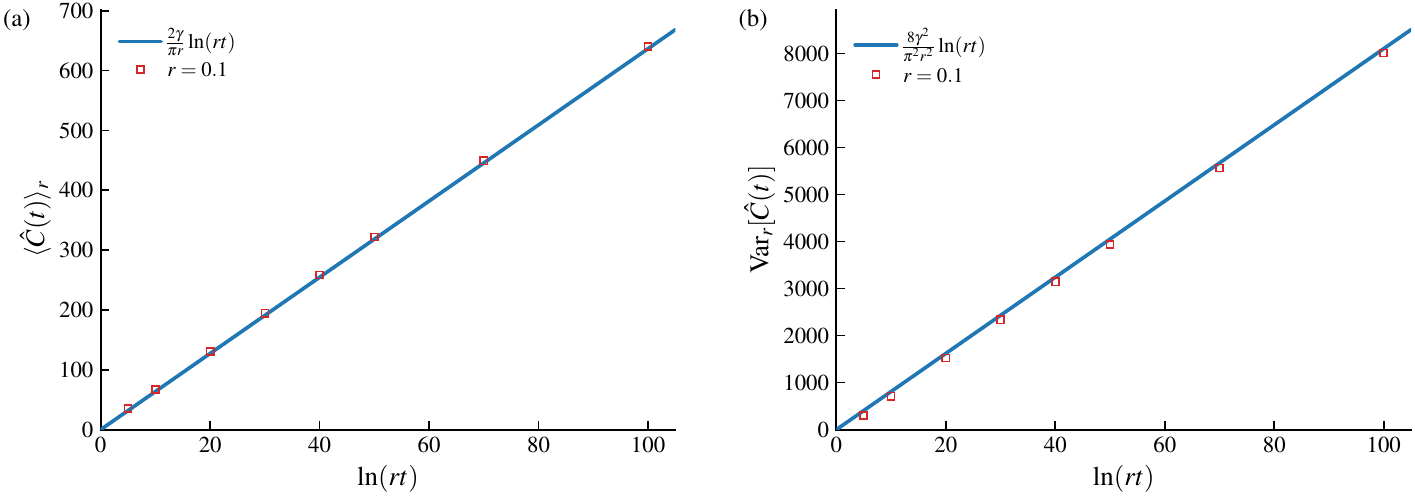}
    \caption{
    Mean and variance of the integrated current
    as functions of $\ln(rt)$ for $r=0.1$.
    (a) Mean integrated current
    $\langle\hat C(t)\rangle_r$.
    (b) Variance $\operatorname{Var}_r[\hat C(t)]$.
    Symbols represent numerical simulations obtained
    using the backward method, while solid lines
    show the analytical predictions
    $\langle\hat C(t)\rangle_r\simeq\Lambda_r(t)/\pi$ and
    $\operatorname{Var}_r[\hat C(t)]\simeq
    4\gamma\Lambda_r(t)/(\pi^2r)$,
    respectively.}
    \label{fig:current_numerical}
\end{figure}

Finally, we consider the mean and variance of the integrated current. As derived in Eqs.~\eqref{eq:C_memory_mean} and \eqref{eq:C_memory_var_final}, both quantities grow logarithmically with time. In Fig.~\ref{fig:current_numerical}, we compare these analytical predictions with numerical simulations obtained using the backward method for $r=0.1$. The numerical results show excellent agreement with the theoretical predictions for both the mean and the variance, confirming the logarithmic growth of the integrated current and its fluctuations.

\FloatBarrier